%% file: ms.tex
\documentclass[12pt,preprint]{aastex}

\usepackage{natbib}
  \renewenvironment{thebibliography}[1]{%
    \begin{oldthebibliography}{#1}%
      \setlength{\parskip}{0ex}%
      \setlength{\itemsep}{0ex}%
  }%
  {%
    \end{oldthebibliography}%
  }

\newcommand{\kms}{km s$^{-1}$}
\newcommand{\kmsmpc}{{\rm km\ s\ }^{-1}{\rm\ Mpc}^{-1}}
\newcommand{\Kms}{{\rm km\ s}^{-1}}
\newcommand{\Bband}{B}
\newcommand{\Vband}{V}
\newcommand{\BminusV}{({\Bband}{\rm -}{\Vband})}
\newcommand{\bminusv}{[{\Bband}{\rm -}{\Vband}]}

\newcommand{\EBV}{E\bminusv}
\newcommand{\Ebv}{E\BminusV}

\newcommand{\bvri}{\protect\hbox{$BV\!RI$}}

\newcommand{\gtrsi}{\mathrel{\hbox{\rlap{\hbox{\lower4pt\hbox{$\sim$}}}\hbox{$>$}}}}

\shorttitle{Spectropolarimetric Diversity of SNe Ia} \shortauthors{Leonard et al.}

\received{2005 March 17}
\begin{document}

\title{Evidence for Spectropolarimetric Diversity in Type Ia Supernovae}

\vspace{2cm}

\author{Douglas C. Leonard\footnote{NSF Astronomy and Astrophysics
        Postdoctoral Fellow.}}
\affil{Astronomy Department, MS 105-24, California Institute of
       Technology, Pasadena, CA 91125}
\email{leonard@astro.caltech.edu}

\author{Weidong Li,
        Alexei V. Filippenko,
        Ryan J. Foley, and
        Ryan Chornock}
\affil{Department of Astronomy, University of California, Berkeley,
 CA 94720-3411}
\email{weidong@astro.berkeley.edu, alex@astro.berkeley.edu, chornock@astro.berkeley.edu, rfoley@astro.berkeley.edu}

\vspace{1cm}

\begin{abstract}

We present single-epoch, post-maximum spectropolarimetry of four Type Ia
supernovae (SNe~Ia) that span a range of spectral and photometric properties:
SN~2002bf and SN~2004dt exhibit unusually high-velocity (HV) absorption lines.
SN~1997dt is probably somewhat subluminous, and SN~2003du is slightly
overluminous.  We detect polarization modulations across strong lines in all
four objects, demonstrating that all are intrinsically polarized.  However, the
nature and degree of the polarization varies considerably.  Including all
SNe~Ia studied thus far, the following order emerges in terms of increasing
strength of line-polarization features: ordinary/overluminous $<$ subluminous
$<$ HV~SNe~Ia, with the strength of the line-polarization features increasing
from $0.2\%$ in the slightly overluminous SN~2003du to $2\%$ in both of the
HV~SNe~Ia in our study.  The most convincing explanation for the line
polarization of all objects is partial obscuration of the photosphere by clumps
of intermediate-mass elements (IMEs) forged in the explosion; the polarization
characteristics of the HV~SNe~Ia in particular effectively rule out a simple
ellipsoidal asphericity as the root cause of their line polarization.  That SNe
Ia are separable into different groups based on their {\it spectropolarimetric}
characteristics may help narrow down progenitor possibilities and/or explosion
physics.  In particular, the recently proposed gravitationally confined
detonation model may provide an attractive explanation for many of the observed
polarization characteristics of HV~SNe~Ia.

\end{abstract}

\medskip

\section{Introduction}
\label{sec:1}

The substantial homogeneity of Type~Ia supernovae (SNe~Ia), together with the fact
that the peak luminosity of individual objects can be observationally determined
through comparisons with well-calibrated samples of nearby SNe~Ia, has propelled 
them to ``gold standard'' status among extragalactic distance indicators. Precise 
distance measurements out to $z \approx 1.7$ have been made \citep{Riess01,Riess04},
revealing the surprising cosmological result that the expansion rate of the
universe is currently accelerating \citep{Riess98,Perlmutter99}; see
\citet{Filippenko04a,Filippenko05} for extensive reviews. Although the
progenitor systems have not yet been conclusively identified, the general 
consensus is that SNe~Ia arise from carbon-oxygen white dwarfs (CO WDs) that 
accrete matter through some mechanism until they achieve a density of $\sim 3 
\times 10^9 {\rm\ g\ cm}^{-3}$ in their centers, leading to a runaway thermonuclear 
reaction that incinerates the star \citep{Woosley86}.  This occurs, coincidentally, 
when a CO WD's mass is nearly the Chandrasekhar limit of $\sim$1.4~$M_\odot$.
Simulations of exploding CO WDs are able to reproduce the main spectral and
photometric characteristics of SNe~Ia (e.g., \citealt{Leibundgut00}, and
references therein), lending theoretical support to this scenario.

However, many questions remain concerning the SN~Ia progenitors and 
explosion mechanism.  What is the nature of the ``donor'' that is
responsible for providing the material that the WD accretes?  Is there more
than one channel by which the accretion can take place?  Where does the
thermonuclear runaway begin inside the WD, and how does it propagate throughout
the star?  As with so many things, the ``devil is in the details,'' and there
is a growing sentiment that the answers to these fundamental questions may come
from careful study of the {\it differences} seen among SNe~Ia, rather than from
the similarities alone.

Spectropolarimetry offers the only direct probe of early-time SN geometry, and
thus is an important diagnostic tool for discriminating among SN~Ia progenitor
systems and theories of the explosion physics.  The essential idea is this: A
hot, young SN atmosphere is dominated by electron scattering, which by its
nature is highly polarizing.  For an unresolved source that has a spherical
distribution of scattering electrons, the directional components of the
electric vectors of the scattered photons cancel exactly, yielding zero net
linear polarization.  Any asymmetry in the distribution of the scattering
electrons, or of absorbing material overlying the electron-scattering
atmosphere, results in incomplete cancellation, and produces a net polarization
\citep[see, e.g.,][]{Leonard11}.

Initial broad-band polarimetry studies found SNe~Ia to possess zero or, at
most, very weak ($< 0.2\%$) intrinsic polarization \citep{Wang96}, suggesting a
high degree of symmetry for their scattering atmospheres.  More recent
spectropolarimetric studies capable of resolving individual line features,
however, are revealing a complex picture, with both continuum and line
polarization now convincingly established for at least a subset of the SN~Ia
population.

To date, three SNe~Ia have been examined in detail with spectropolarimetry at
early times: SN~1999by \citep{Howell01}, SN~2001el \citep{Kasen03,Wang03}, and,
most recently, SN~2004dt \citep{Wang05}.  In this paper we present single-epoch
spectropolarimetry of four SNe~Ia: SN~1997dt, SN~2002bf, SN~2003du, and
SN~2004dt, obtained about $21$, 3, 18, and 4 days (respectively) after maximum
light. 

A particular motivation for this multi-object investigation is to attempt to
link spectral and photometric peculiarities of individual SNe~Ia with their
spectropolarimetric characteristics.  Accordingly, the four events span a range
of properties: SN~1997dt is likely somewhat subluminous, SN~2002bf and
SN~2004dt exhibit unusually high-velocity absorption lines (in the case of
SN~2002bf, the highest ever seen in an SN~Ia for the epochs considered), and
SN~2003du is slightly overluminous.

This paper is organized as follows.  We briefly review the present photometric,
spectroscopic, and spectropolarimetric state of knowledge of SNe~Ia in
\S~\ref{sec:2}, focusing particular attention on objects sharing
characteristics with the SNe~Ia included in our spectropolarimetric survey.  We
describe and present the spectropolarimetry in \S~\ref{sec:3}; to assist in the
classification, we also include optical photometry and additional spectroscopy
for two of the events, SN~2002bf and SN~2003du.  We analyze the data in
\S~\ref{sec:4}, and present our conclusions in \S~\ref{sec:5}.  Note that
preliminary discussions of the spectropolarimetry of SN~1997dt and SN~2002bf
have been given by \citet{Leonard1} and \citet{Filippenko04}, respectively.

\section{Background}
\label{sec:2}
\subsection{Photometric Properties of SNe~Ia}
\label{sec:2.1}

Type Ia SNe typically rise to peak $B$-band brightness in about 20 days,
decline by about $3$ mag over the next $35$ days, and then settle into a nearly
constant descent of $\sim$1.55 mag (100\ day)${^{-1}}$ for the next year.
However, it is now confirmed beyond doubt that deviations exist from the
central trend; see, for example, \citet{Phillips99}.  Some events rise and fall
more slowly, producing ``broader'' light curves (e.g., SN~1991T), whereas
others rise and fall more quickly, yielding ``narrower'' light curves (e.g.,
SN~1991bg).  The width of the light curves near peak correlates strongly with
luminosity, in the sense that broader light curves generally indicate
intrinsically brighter objects (but see \citealt*{Jha05}, for exceptions), with
a total spread of about a factor of sixteen in absolute peak $B$-band
brightness among the population \citep{Altavilla04}.

The ability to individually determine the luminosity of SNe~Ia through
examination of their light-curve shapes has been the primary driver for
cosmological applications.  There are several photometric calibration
techniques in common use that correlate the observed light-curve shape, or
supernova color, with luminosity (see \citealt{Wang05a}, and references
therein). The simplest calibration technique involves measuring the decline in
$B$ during the first 15 days after maximum $B$-band brightness, $\Delta
m_{15}(B)$. A typical value is $\Delta m_{15}(B) \approx 1.1$ mag, which
corresponds to $M_B = -19.26 \pm 0.05$ mag according to the calibrations of
\citet{Hamuy96}.  Underluminous, SN 1991bg-like objects yield values up to
$\Delta m_{15}(B) = 1.94$ mag, while overluminous, SN 1991T-like objects 
go as low as $\Delta m_{15}(B) = 0.81$ mag \citep{Altavilla04}.  There is a
continuum of values in between the two extremes.

For the two SNe with optical photometry presented in this paper, we shall use
the ``multicolor light-curve shape'' method \citep[MLCS; e.g.,][]{Riess96},
which has been revised (and hereafter referred to as MLCS2k2) by \citet{Jha02}
and S. Jha et al. (in preparation) to include $U$-band light curves from
\citet{Jha05}, a more self-consistent treatment of extinction, and an improved
determination of the unreddened SN~Ia color.  This method is based on
determining $\Delta$ (a dimensionless number related to magnitude that 
parameterizes the light-curve shape), $t_0$ (the time of $B$-band maximum light), 
$\mu_0$ (the distance modulus for an assumed value of $H_0$, here taken to be 
$65\ \kmsmpc$), and $A_V^0$ (the visual extinction at $t_0$).

The mechanism that produces the dispersion in SN~Ia luminosity is not known;
present speculations range from different progenitor systems to differences in
the explosion mechanism and flame propagation (e.g., deflagration, detonation,
delayed-detonation, or, most recently, gravitationally confined detonation; see,
respectively, \citealt*{Nomoto84,Arnett69,Khokhlov91,Plewa04}).
Some have further speculated that global asymmetries in the expanding ejecta
may result in viewing-angle dependent luminosity \citep{Wang03}.  One
observational fact that all theories must confront is that the most luminous
SNe~Ia have thus far only been seen in late-type galaxies; the faintest
objects tend to prefer elliptical galaxies, but have also been found in spirals
\citep[e.g., ][and references therein]{Howell01a,Benetti05}.

\subsection{Spectroscopic Properties of SNe~Ia}
\label{sec:2.2}

Early-time spectra of SNe~Ia (see \citealt{Filippenko97} for a review)
typically exhibit lines of intermediate-mass elements (IMEs), such as
magnesium, silicon, sulfur, and calcium, with some contribution from iron-peak
elements, especially at near-ultraviolet wavelengths.  As time progresses, and
the photosphere recedes deeper into the ejecta, lines of Fe come to dominate
the spectrum.  This spectral evolution suggests a burning front that
incinerates some of the progenitor's carbon and oxygen all the way to the
iron peak deep inside the ejecta, but then leaves the outer layers only
partially burned.  It is interesting to note that overluminous SNe~Ia show
enhanced Fe features at early times, whereas subluminous ones show weak
early-time Fe features.

At early times (e.g., $-20 {\rm\ d} < t < 20 {\rm\ d}$ from the date of maximum
$B$ brightness, $B_{\rm max}$), the spectra of typical SNe~Ia evolve rapidly
and with such uniformity that it is possible to determine the age of an event
relative to the date of $B_{\rm max}$ to within $\sim$2 days from a single
spectrum alone \citep{Riess97,Foley05}. The spectroscopic peculiarities of subluminous
and overluminous events currently preclude their ``spectral feature ages'' from
being derived accurately through comparison with average SN~Ia spectra.  For
instance, pre-maximum spectra of overluminous events lack the strong
\ion{Si}{2} $\lambda 6355$ absorption that is so prominent in normal and
underluminous SN~Ia spectra \citep{Filippenko92a}.

Another point of distinction among SN~Ia spectra comes from the blueshifts of
the spectral lines.  Increasing attention is being paid to a small but growing
group of ``high-velocity'' (HV) SNe~Ia, whose spectra around maximum light are
characterized by unusually broad and highly blueshifted absorption troughs in
many line features, indicating optically thick ejecta moving about
4000--5000~\kms\ faster than is typically seen for SNe~Ia
\citep{Branch87,Benetti05}. Well-studied examples of HV~SNe~Ia include
SN~1983G \citep[][and references therein]{Branch93}, SN~1984A
\citep{Branch87,Barbon89}, SN~1997bp \citep{Anupama97}, SN~1997bq \citep[][and
references therein]{Lentz01}, SN~2002bo \citep{Benetti04}, SN~2002dj
\citep{Benetti04}, and SN~2004dt \citep{Wang05}.  \citet{Wang05} note that in a
spectrum of one HV~SN~Ia, SN~2004dt, some line features do not possess
abnormally high velocity, such as those identified with \ion{S}{2}.  Since two
of our objects, SN~2002bf and SN~2004dt, are HV~SNe~Ia, we briefly review the
salient features that are known about this class of objects.

The most thorough and recent study of the diversity of SN~Ia expansion
velocities is that of \citet{Benetti05}, who apply a statistical treatment to
data from 26 SNe~Ia and find that HV SNe~Ia indeed make up a kinematically
distinct group.  Its members have normal peak luminosity (based on their
$\Delta m_{15}(B)$ values: $1.09 \lesssim \Delta m_{15}(B) \lesssim 1.37$ mag),
reside in all types of host galaxies (e.g., both ellipticals and spirals, but
with a preference for later types), and have very strong \ion{Si}{2} features.
They are further distinguished by a large temporal velocity gradient,
$\dot{v}$\ $ > 70\ \Kms {\rm\ day}^{-1}$, where $\dot{v}$ is defined to be the
average daily rate of decrease of the expansion velocity between maximum light
and the time the \ion{Si}{2} $\lambda 6355$ feature disappears.  HV~SNe~Ia
appear to have similar photometric characteristics to ordinary Type~Ia events,
although there is some indication of subtle differences in their color
evolution \citep{Benetti04}.

Whether HV SNe~Ia represent the extreme end of a continuum of more typical
SNe~Ia, or require a different explosion mechanism, progenitor system, or
explosion physics, remains controversial.  What is certain is that any model
must have significant optical depth in the IMEs at high velocities at early
times.  \citet{Branch93} originally proposed that HV~SNe~Ia may simply result
from more energetic explosions.  However, \citet{Benetti04} show that the Fe
nebular lines have velocities comparable to those of normal SNe~Ia, which tends
to weaken the argument for higher overall kinetic energy.  There is also
evidence from early-time spectra of SN~2002bo that HV~SNe~Ia possess very
little unburned carbon in their outer layers, in contrast with more typical
SNe~Ia like SN~1994D \citep{Benetti04}.  The high velocities of the IMEs
coupled with the lack of primordial carbon may indicate that burning to Si
penetrates to much higher layers in HV~SNe~Ia than it does in more normal
events.  This is a feature of some delayed-detonation models, such as those
studied by \citet{Lentz01}.  However, \citet{Wang05} argue that a strong
detonation wave is unlikely to generate the clumpy and asymmetrically
distributed silicon layer that is inferred from pre-maximum spectropolarimetric
observations of SN~2004dt.  As another alternative, \citet{Benetti04} propose
that IMEs produced at deeper layers may simply be more efficiently mixed
outward in HV~SNe~Ia than is typical.  A prediction of this mechanism is that
while the IME products of explosive carbon burning will be mixed in with the
primordial C and O in the outer layers, the C and O should also be mixed inward
within these events, and exist at lower velocities than are normally seen.

It is appropriate at this point to mention the recently proposed
gravitationally confined detonation (GCD) model of \citet{Plewa04}, which
stands as an intriguing alternative to the standard deflagration or
delayed-detonation scenarios.  The mechanism involves the slightly off-center
ignition of a deflagration that produces a buoyancy-driven bubble of material
that reaches the stellar surface at supersonic speeds, where it laterally
accelerates the outer stellar layers.  This material, gravitationally confined
to the white dwarf, races around the star and, in $\sim$2~s, converges at a
point opposite to the location of the bubble's breakout, creating conditions
capable of igniting the nuclear fuel and triggering a detonation that can
incinerate the white dwarf and result in an energetic explosion.  

Of interest to the present study is the recent work of \citet{Kasen05}, who
investigate the spectral, and spectropolarimetric, consequences of the GCD
model.  They focus their investigation on the interaction of the expanding
ejecta with an ellipsoidal, metal-rich extended atmosphere formed from the
bubble of deflagration products (taken to be 57\% Si, 27\% S, 7.1\% Fe, and
2.7\% Ca, plus smaller amounts of other metals; see \citealt{Khokhlov93}), and
find that a dense, optically thick pancake of metal-rich material is formed at
potentially large velocity on the side of the ejecta where the bubble emerged.
For low atmosphere masses (e.g., resulting from a bubble of mass $0.008
M_\odot$), the pancake of material spans the velocity range 17,000--28,000
\kms\ and is geometrically detached from the bulk of the SN ejecta.  This might
explain the detached, high-velocity \ion{Ca}{2} near-infrared (IR) triplet
absorption seen in pre-maximum spectra of some SNe~Ia (e.g., SN~2001el, see
\S~\ref{sec:2.3.3}).  For larger bubble and, hence, atmosphere masses ($m_{\rm
atm} \gtrsim 0.016 M_\odot$), the absorbing pancake moves at lower velocities
(e.g., 10,000--21,000 \kms\ for $m_{\rm atm} = 0.08 M_\odot$) and could blend
with the region of IMEs in the SN ejecta and potentially increase the blueshift
of several of the spectral features.  This could provide an
orientation-dependent explanation for the origin of HV~SNe~Ia, whereby sight
lines in which the pancake more completely blocks the photosphere produce the
anomalously large velocities.  The spectropolarimetric consequences of this
model are discussed in \S~\ref{sec:2.3.1}.

\subsection{Spectropolarimetric Properties of SN~Ia}
\label{sec:2.3}
\subsubsection{Supernova Polarization Mechanisms}
\label{sec:2.3.1}

The first definitive proof that some SNe are polarized came from
observations of the Type II SN~1987A \citep{Cropper88}, which exhibited a
modest temporal increase in continuum polarization during the photospheric
phase, as well as sharp polarization modulations across strong P-Cygni flux
lines \citep{Jeffery91a}.  While intrinsic polarization has now been
established in over a dozen SNe (for recent reviews, see \citealt{Wheeler00};
\citealt{Filippenko04}; \citealt{Leonard11}), including at least three SNe~Ia, 
the exact origin of both continuum and line polarization remains controversial.

There are essentially two mechanisms by which supernova continuum polarization
is thought to be produced: (1) a globally aspherical photosphere and
electron-scattering atmosphere
\citep[e.g.,][]{Shapiro82,Hoflich91,Jeffery91,Leonard2,Wang01}, and (2)
ionization asymmetry produced by the decay of asymmetrically distributed
radioactive $^{56}{\rm Ni}$, perhaps flung out into \citep{Chugai92,Hoflich01}
or beyond \citep{Kawabata02,Leonard8} the expanding ejecta in clumps.  The
simplest, and most well-studied, globally aspherical geometry is that of an
ellipsoid, and we shall make frequent reference to the ``ellipsoidal model'' in
the following discussion.  In the second model, in which an ionization asymmetry
is present, continuum polarization is generated by light from the (either
spherical or aspherical) photosphere scattering off of asymmetrically
distributed free electrons that exist in clouds surrounding clumps of
radioactive $^{56}{\rm Ni}$.  Variations on both of these polarization
mechanisms are also possible.  For instance, an aspherical distribution of
$^{56}{\rm Ni}$ could also result in SN polarization by providing an asymmetry
in a source of optical photons (produced by the thermalization of
$\gamma$-rays) relative to the scattering medium.

To explain polarization modulations seen across spectral lines, it is important
to differentiate between the emission peaks and blueshifted absorption troughs
that are characteristic of P-Cygni profiles in total-flux spectra of SNe.  Line
peaks have usually been assumed to consist of intrinsically unpolarized
photons.  This is because although resonance scattering by a line is an
inherently polarizing process \citep{Jeffery91}, directional information for
scattered photons is lost in an SN atmosphere since the timescale for
randomizing collisional redistribution of the relative level populations within
the fine structure of the atomic levels of a line transition is much shorter
than the characteristic timescale for absorption and reemission in a line
\citep{Hoflich96}.  The assumption of intrinsically unpolarized emission lines
is also commonly used to derive the interstellar polarization (ISP;
\S~\ref{sec:2.3.2}).  Note, though, that if ionization asymmetry exists above
the photosphere (from, e.g., clumps of radioactive $^{56}{\rm Ni}$), then even
intrinsically unpolarized emission-line photons may become polarized by
electron scattering within the SN atmosphere.

For SNe~Ia, line-blanketing, due largely to Fe, is particularly severe at
wavelengths below $\sim$5000~\AA, where theoretical models \citep{Howell01}
suggest that nearly complete depolarization of any ``continuum'' light may be
assumed.  Conversely, the broad spectral region $6800 \lesssim \lambda
\lesssim 7800$~\AA\ is largely devoid of line opacity \citep{Kasen04,Howell01},
at least near maximum light, and thus may give a good indication of true
continuum polarization level.

The explanation of polarization changes in absorption troughs is controversial.
One well-studied hypothesis, which follows logically from the ellipsoidal
model, is that selective blocking of more forward-scattered and, hence, less
polarized, light in P-Cygni absorption troughs results in trough polarization
increases \citep{Jeffery91,Leonard1,Leonard3}, potentially producing
``inverted'' P-Cygni polarization profiles when coupled with emission peak
depolarization.  Some SN modelers have claimed that this ``geometrical
dilution'' mechanism, however, is rather poor at polarizing SN~Ia light, and
that polarization {\it decreases} should actually be seen in absorption
troughs, since much of the light reaching the observer in those spectral
regions has been absorbed and reemitted by the line \citep{Howell01}.  

A key point is that, in either case, to produce line trough polarization changes
under the ellipsoidal model, some continuum polarization must exist.  In fact,
\citet{Leonard4} show that the strength of a line trough polarization feature
can be used to place a lower bound on the true intrinsic continuum polarization
level under the ellipsoidal model according to

\begin{equation}
p_{cont} \geq \frac{\Delta p_{tot}} {(I_{cont}/I_{trough}) - 1},
\label{eqn:1}
\end{equation}

\noindent 
where $I_{cont}$ is the interpolated value of the continuum flux
at the location of the line trough, $I_{trough}$ is the total flux at the
line's flux minimum, and $\Delta p_{tot}$ is the total polarization change
observed in the line trough.  Thus, a critical test of the ellipsoidal model is
to see whether the continuum polarization observed in the spectral region
6800--7800~\AA\ is sufficient to explain observed line trough polarization
changes.  Lack of significant continuum polarization in an object with strong
line features argues against the ellipsoidal model.  An additional prediction
is that {\it no} rotation of the polarization angle (PA) should be seen across
the line, since the continuum and line-forming regions share the same geometry.

A final way to produce a polarization change in a line trough is through
asymmetry in the distribution of elements in the ejecta material located above
the photosphere along the line-of-sight (l-o-s), hereafter referred to as the
``clumpy ejecta'' model.  Asymmetry in the distribution of material with
significant optical depth may unevenly block the underlying photospheric light,
thereby producing a net polarization change and/or PA rotation in a line
trough, even when the photosphere is spherical. The GCD model discussed in
\S~\ref{sec:2.2} provides a natural mechanism by which to generate line-trough
polarization through this model, since it predicts an optically thick pancake
of high-velocity, metal-rich material overlying the photosphere on the side in
which the bubble emerged.  In particular, if the GCD model provides the correct
mechanism by which to produce HV~SNe~Ia, then polarization changes in the
absorption troughs of the strong, HV metal lines in these objects should be
particularly pronounced since the l-o-s necessarily intercepts a substantial
fraction of the pancake.

For SNe~Ia, an interesting alternative to the simple ellipsoidal model for
producing {\it both} line and continuum polarization is presented by
\citet{Kasen04}, who explore the polarization consequences of a conical hole in
the ejecta due to the interaction with a companion star, which we shall refer
to as the ``ejecta-hole'' model.  By considering various viewing angles and
hole sizes, \citet{Kasen04} demonstrate that both continuum and line
polarization and PA changes can be generated.  In general, viewing angles
almost directly down the hole yield low continuum polarization with
polarization increases in strong line troughs, whereas sight lines more nearly
perpendicular to the hole result in larger continuum polarization (prominently
seen in the spectral region 6800--7800~\AA) and ``inverted P-Cygni'' line
polarization profiles; this latter case is nearly indistinguishable from the
predictions of the simple ellipsoidal model.

Ionization asymmetry is also capable of generating both continuum polarization
as well as PA and polarization level changes through line features, since both
continuum and line photons will scatter off of concentrations of free electrons
in (or beyond) the ejecta.  An impressive example of this mechanism potentially
being at work is given by \citet{Chugai92} for the case of SN~1987A, in which
the spectropolarimetry data, as well as asymmetries in the flux line profiles,
are convincingly reproduced by the effects of two clumps of $^{56}{\rm Ni}$ in
the far (receding) hemisphere of the ejecta.

\subsubsection{Removing Interstellar Polarization}
\label{sec:2.3.2}

A problem that plagues interpretation of all SN polarization measurements is
proper removal of the ISP.  Since directional extinction resulting from
aspherical interstellar dust grains aligned by some mechanism along the l-o-s
to an SN can contribute a large polarization to the observed signal, an attempt
must be made to remove it prior to analyzing SN spectropolarimetry data. This
is notoriously difficult, although a number of different techniques have been
advanced over the years, which we here summarize.

An excellent way to derive Galactic ISP is through observations of distant,
intrinsically unpolarized, ``probe stars'' close to (within $\sim$0.5$\arcdeg$)
the l-o-s to the SN \citep[e.g.,][]{Leonard8,Leonard7}.  Deriving the total
ISP, which includes the contribution from dust in the host galaxy, however, is
more difficult.

The most basic technique is to {\it place limits on the ISP from reddening
considerations.}  Since the same dust that polarizes starlight should redden it
as well, it seems logical to expect a correlation between reddening and ISP.
Because the alignment of dust grains is not total (or has multiple preferred
orientations due to non-uniformity of the magnetic field along the l-o-s), and
grains are probably only moderately elongated particles, it is not surprising
that, through the analysis of thousands of reddened Galactic stars, only an
upper bound on the polarization efficiency of Galactic dust can be derived
\citep{Serkowski75}: ${\rm ISP} / E_{B-V} < 9.0\%\ {\rm mag}^{-1}$.  However,
the polarization efficiency of the dust in external galaxies is not well
studied, and in one of the few investigations carried out to date,
\citet{Leonard7} find compelling evidence for polarization efficiency well in
excess of the empirical Galactic limit for dust in NGC~3184 along the l-o-s to
the Type II-P SN~1999gi.  It is not clear at this point whether meaningful
constraints can thus be placed on the ISP of extragalactic SNe that are
significantly reddened by host-galaxy dust.  Nonetheless, total reddening
arguments are still often used to set ``reasonable'' limits on the ISP.

Another method to get a handle on the ISP is to assume axisymmetry for any SN
asphericity, a situation that reveals itself through a straight-line
distribution of points when the spectropolarimetry is plotted in the $q$--$u$
plane.  If axisymmetry exists, the ISP is constrained to lie along the axis
defined by the line \citep{Howell01}.  Its absolute value, however, is
uncertain without additional input.

A more precise technique relies on the theoretical expectation that {\it
unblended emission lines consist of unpolarized light}, and that any
polarization observed in emission-line photons comes from the ISP
\citep{Jeffery91a,Tran97,Wang96,Leonard3}.  This assumption is thought to be
most valid at early times, when any $^{56}{\rm Ni}$ concentrations are likely
to be below the photosphere.  Note, though, that to use this method, care must
be taken to isolate the emission-line photons from the underlying continuum
light \citep[e.g.,][]{Tran97}, which may be intrinsically polarized.

A related method is to assume that a particular spectral region is
intrinsically {\it completely} unpolarized, with all of the observed
polarization therefore coming from ISP.  Some empirical support exists that, in
some objects, this may be the case for very strong emission lines
\citep{Kawabata02,Leonard8,Wang04a}.  For SNe~Ia in particular, it has
sometimes been assumed that the far blue spectral region (e.g., typically $<
5000$~\AA) satisfies this criterion due to the effect of the heavy
line-blanketing and multiple, depolarizing, line scatters, largely due to
iron-group elements.  This qualitative expectation is demonstrated
quantitatively by \citet{Howell01}, who present model polarization spectra
resulting from delayed-detonation models in realistic SN Ia atmospheres.  More
recently, \citet{Wang05} update this approach by choosing only specific
spectral regions at blue wavelengths ($\lambda < 5000$~\AA) that are not
dominated by obvious individual line features in either flux or polarization
for the ISP determination, rather than the arbitrary blue edge of the spectrum, as 
adopted by \citet{Howell01}. The central idea here is that while strong individual line 
features might impart their own polarization signature (e.g., through geometrical 
dilution in ellipsoidal models, or through blocking of the parts of the photosphere 
in clumpy-ejecta models), the spectral regions in between specific features
probably decrease any effective continuum polarization by the numerous
overlapping spectral lines at blue wavelengths.

Finally, if polarimetry is obtained after the electron-scattering
optical depth of the atmosphere has dropped well below unity, then it
may be adequate to assume the observed polarization to be due entirely
to ISP across the whole spectrum \citep[e.g.,][]{Wang03}.

Improperly removed, ISP can increase or decrease the derived intrinsic
polarization, and it can change ``valleys'' into ``peaks'' (or vice
versa) in the polarization spectrum. Since it is so difficult to be
certain of accurate removal of ISP, it is generally safest to focus on
(a) temporal changes in the polarization with multiple-epoch data, (b)
distinct line features in spectropolarimetry having high
signal-to-noise ratio (S/N), and (c) continuum polarization unlike the
characteristic ``Serkowski-law'' wavelength dependence imparted by
dust \citep[e.g.,][]{Whittet92}.

\subsubsection{Previous SN~Ia Polarimetry Studies}
\label{sec:2.3.3}

Evidence for intrinsic polarization in SNe~Ia was initially difficult to find,
as the first investigations detected only marginally significant polarization
among normal-brightness events observed near maximum, $p < 0.2$\%
\citep[e.g.,][]{Wang96,Wang97}.  Broad-band polarimetry of one overluminous
event observed over a month after maximum also found no intrinsic polarization
down to a level of $p \approx 0.3\%$ \citep{Wang96}.  Significant advances have
been made in the last few years. \citet{Leonard1} reported the first
convincing, albeit weak, features in the polarization of an SN~Ia, SN~1997dt,
which was likely a subluminous event; these data are presented and analyzed in
more detail in the present study.

The first thorough spectropolarimetric study of an SN~Ia is that by
\citet{Howell01}.  In spectropolarimetry obtained at maximum light, the
subluminous SN~1999by exhibits a polarization change of $\sim$0.8\% from
4800~\AA\ to 7100~\AA, and a sharp polarization modulation of $\sim$0.4\%
across the strong \ion{Si}{2} $\lambda6355$ absorption.  These features are
explained within the context of an ellipsoidal model with a global asphericity
of $\sim$20\%, observed equator-on.  This physical picture was achieved by
assuming the ISP to be the observed polarization at the blue edge of the
spectrum, which was about $0.2\% {\rm\ at\ } 4800$~\AA.  With this choice of
ISP, the inferred intrinsic polarization rises from $0\%$ to about $0.8\%$ from
blue to red, with a sharp depolarization from $0.4\%$ to near $0\%$ across the
\ion{Si}{2} $\lambda 6355$ feature.  An argument supporting both the
ellipsoidal model as well as the ISP choice is that after ISP removal the PA
becomes nearly independent of wavelength.  \citet{Howell01} note, on the other hand, 
that a wavelength-independent PA also results if one assumes the ISP to be the
observed polarization of the far {\it red} edge of the spectrum; in this case,
the polarization modulation across the \ion{Si}{2} $\lambda 6355$ becomes a
polarization increase.  Such a scenario could result from selective blocking of
forward-scattered light, as described in \S~\ref{sec:2.3.1}.  However,
\citet{Howell01} find the theoretical arguments supporting the former ISP to be
more compelling.  With this ISP, the redward rise in intrinsic polarization is
attributed to the decreasing importance of line opacities, and the increased
influence of continuum electron scattering at longer wavelengths.  The
depolarization across the \ion{Si}{2} $\lambda6355$ absorption is also
attributed to the depolarizing effect of line scattering.  Despite the
detection of intrinsic polarization for SN~1997dt and SN~1999by, their unusual
(subluminous) nature nevertheless left some doubt about intrinsic polarization in {\it
normal} SNe~Ia.

   That doubt has recently been put to rest with the work of \citet{Kasen03} on
SN~2001el.  For this normal-luminosity event, the percent polarization changed
from blue to red by $\sim$0.4\% in spectropolarimetry obtained 1 week before
maximum brightness.  However, the extraordinary feature here is the existence
of distinct high-velocity \ion{Ca}{2} near-infrared (IR) triplet absorption
($v$ = 18,000--25,000 km s$^{-1}$) in addition to the usual, lower-velocity
\ion{Ca}{2} feature.  A similar, but much weaker, high-velocity feature had
been previously observed in SN~1994D, and perhaps in other SNe~Ia as well; the
number of pre-maximum spectra covering the near-IR spectral range is small.
The polarization is seen to change dramatically in this feature, by $\sim$0.4\%.  
The \ion{Ca}{2} near-IR feature is examined by \citet{Kasen03} in an
elegant study, which concludes that it is likely due to photospheric
obscuration by a clumped shell of high-velocity material.  Using multi-epoch
data, \citet{Wang03} demonstrate that the nature of the polarization changes
over the course of two weeks following this early epoch, becoming nearly
undetectable a week after maximum brightness, further solidifying the case for
intrinsic polarization at early times.  Although not commented on by either
study, it appears that the \ion{Si}{2} $\lambda 6355$ line also shows a
polarization modulation in the earliest epoch, and with a different PA from the
\ion{Ca}{2} near-IR feature.  Although it does not affect the main results, a
cautionary note on the difficulty of ISP determination is set by the fact that
the two studies arrive at quite different values: \citet{Kasen03} derive an ISP
of very nearly $0\%$ by assuming the blue edge of the spectrum to be
unpolarized, whereas \citet{Wang03} obtain an ISP of $\sim$0.6\% by
attributing the observed polarization $38$ days after maximum light, when
SN~2001el is argued to be in the nebular phase, entirely to ISP.

Most recently, \citet{Wang05} analyze a single epoch of pre-maximum
spectropolarimetry of SN~2004dt, an HV~SN~Ia for which an ISP of $q_{\rm ISP} =
0.2 \pm 0.1\%$ and $u_{\rm ISP} = -0.2 \pm 0.1\%$ is derived from the observed
polarization of a handful of narrow, blue spectral regions in which no single
spectral feature dominates in the total-flux spectrum.  This results in rather
low continuum polarization, $p \lesssim 0.4\%$, but very strong modulations
across spectral lines.  The polarization spikes reach $2\%$ in the deep troughs
due to \ion{Si}{2} $\lambda 4130$ and $\lambda 6355$; lesser peaks are observed
in features identified with \ion{Mg}{2} $\lambda 4471$, a blend of \ion{Si}{2}
$\lambda\lambda 5041, 5056$ and \ion{Fe}{2} $\lambda\lambda 4913, 5018, 5169$,
and the \ion{Ca}{2} near-IR triplet.  All line polarization has similar directional
behavior in the $q$--$u$ plane, suggesting a common origin.  Interestingly,
whereas other strong line features in the total-flux spectrum are characterized
by strong polarization modulations, \ion{O}{1} $\lambda7774$ shows no
polarization signature.  These features are explained in terms of optically
thick bubbles of IMEs, the result of partial burning, that are asymmetrically
distributed within an essentially spherical oxygen substrate that remains from
the progenitor. Note that \citet{Wang97a} find polarization
variation at a level of $> 0.5\%$ across the strong \ion{Si}{2}
$\lambda 6355$ feature of another HV~SN~Ia, SN~1997bp, in unpublished
spectropolarimetry obtained near maximum light.

From this small sample, the observations thus far suggest that normal and,
perhaps, overluminous events are weakly polarized ($p \lesssim 0.4\%$), with
subluminous ones possessing somewhat larger values ($p \approx 0.8\%$).  HV~SNe~Ia
have modest continuum polarization, but possess the highest line polarization
of all, achieving polarizations of up to $2\%$ in the strongest features.  We
assess the robustness of these tentative trends with the present study of
four SNe~Ia.

\subsection{The Type Ia Supernovae 1997dt, 2002bf, 2003du, and 2004dt}
\label{sec:2.4}

SN~1997dt was discovered \citep{Qiao97} by the Beijing Astronomical Observatory
Supernova Survey \citep{Li96} on 1997 November 22.44 (UT dates are used
throughout this paper) at an unfiltered magnitude of $\sim$15.3 in the Sbc
galaxy NGC~7448. Images of the same field taken eight days earlier show no star
at the position of the SN to a limiting unfiltered magnitude of about 18.5.  An
optical spectrum obtained immediately (0.06 days) after discovery showed it to
be a Type~Ia event \citep{Qiao97}; a subsequent examination by \citet{Li01a}
estimates the age at discovery to be $-7 \pm 5$ days relative to maximum light,
which places the date of maximum at very roughly 1997 November 29.
\citet{Tonry03} report that SN~1997dt suffers from a host extinction of $A_V =
0.46 {\rm\ mag}$, the median of the values derived from MLCS and ``Bayesian
Adapted Template Match Method'' \citep[J. L. Tonry et al. 2005, in
preparation;][]{Tonry03} analyses of its unpublished light curves.  An MLCS2k2
fit (S. Jha et al., in preparation) suggests that it is subluminous, with
$\Delta = 0.94$ (corresponding to a $\Delta m_{15}(B) \approx 1.8$ mag; see
\S~\ref{sec:2.1}), although there is a long shoulder of probability to lower
$\Delta$ values, indicating that there is a wide range of light-curve
shapes that can fit the sparse number of photometric points (2 in $B$, 3 in
$V$, and 2 in $I$).  From a pre-maximum spectrum posted at the Center for
Astrophysics' Recent Supernovae
Page,\footnote{\url{http://cfa-www.harvard.edu/cfa/oir/Research/supernova/RecentSN.html}}
we estimate $\cal R$(\ion{Si}{2}) $\approx 0.3$ (see \citealt{Nugent95}), which
implies $\Delta m_{15}(B) \approx 1.33$ mag from the correlation derived by
\citet{Benetti04}, also consistent with a somewhat subluminous classification.

SN~2002bf was discovered \citep{Martin02} by the Lick Observatory and Tenagra
Observatory Supernova Searches \citep[LOTOSS;][]{Schwartz00} on 2002 February
22.30 at an unfiltered magnitude of $\sim$17, very close to the nucleus of the
Sb galaxy PGC~029953.  An image of the same field taken twenty days earlier
showed nothing at the position of SN~2002bf to a limiting unfiltered magnitude
of $\sim$19. Optical spectra obtained on 2002 March 6.21 by \citet{Matheson02}
and on 2002 March 7.41 by \citet{Filippenko02} identified it as a Type~Ia
event.  Both groups noted that the expansion velocity derived from the
absorption minimum of the \ion{Si}{2} $\lambda 6355$~\AA\ line was
significantly greater than normal for an SN~Ia near maximum light, indicating
that it may be an HV~SN~Ia.  A ``spectral feature age'' \citep{Riess97} of $0
\pm 2$ days was derived from the March 6 spectrum \citep{Matheson02}.  The
MLCS2k2 analysis of the light curves presented in \S~\ref{sec:4.2.1} yields
2002 March $4.37 \pm 0.50$ as the date of maximum $B$ light, consistent with
the age on March 6 derived by \citet{Matheson02}.

SN~2003du was discovered \citep{Schwartz03} by LOTOSS on 2003 April 22.4 at an
unfiltered magnitude of $\sim$15.9 in the SBd galaxy UGC~9391.  An image of the
same field taken fifteen days before discovery showed nothing at the position
of SN~2003du to a limiting unfiltered magnitude of $\sim$19. An optical
spectrum obtained shortly thereafter, on 2003 April 24.06, identified it as a
Type~Ia event roughly two weeks before maximum light \citep{Kotak03}.  The
MLCS2k2 analysis of the light curves presented in \S~\ref{sec:4.2.1} yields
2003 May $6.12 \pm 0.50$ as the date of maximum $B$ light, consistent with the
earlier spectral age estimate.  It also agrees with the epochs of maximum
determined by the recent photometric studies of SN~2003du by \citet{Anupama05}
and \citet{Gerardy04}.

SN~2004dt was discovered \citep{Moore04} by the Lick Observatory Supernova
Search \citep{Filippenko03a} at an unfiltered magnitude of $\sim$16.1 in the
SBa galaxy NGC 799 on 2004 August 11.48.  An image of the same field taken ten
days earlier showed nothing at the position of SN~2004dt to a limiting
unfiltered magnitude of $\sim$18.  Spectra taken within 2 days of discovery by
\citet{Galyam04}, \citet{Patat04}, and \citet{Salvo04} confirmed it to be an
SN~Ia before maximum light.  \citet{Patat04} noted that several absorption
lines showed high expansion velocities, a result confirmed by the recent study
by \citet{Wang05}, suggesting that, like SN~2002bf, SN~2004dt is an HV~SN~Ia.
A preliminary analysis of the light curves of SN~2004dt shows that maximum
light occurred near 2004 August 20 (W. Li, personal communication).  A series
of {\it Hubble Space Telescope (HST)} UV spectral observations was obtained as
part of program GO-10182 (P.I. Filippenko), and will be analyzed in a future
paper.

\section{Observations and Reductions}
\label{sec:3}

We obtained single-epoch spectropolarimetry of SN~1997dt, SN~2002bf, SN~2003du,
and SN~2004dt on days 21, 3, 18, and 4 (respectively) after maximum light.  We
also obtained additional optical spectroscopy and \bvri\ photometry of
SN~2002bf and SN~2003du.  For SN~2002bf, our photometry samples $-10$ to $57$
days from the time of maximum light, with one additional flux spectrum taken on
day 9 after maximum.  For SN~2003du, our ground-based photometry covers $-5$ to
$113$ days from maximum, with one additional epoch on day 436 taken using the
High Resolution Channel (HRC) of the Advanced Camera for Surveys (ACS) on board
{\it HST}.  Five additional spectral epochs sample its development from days
$24$ to $82$ after maximum.

\subsection{Photometry}
\label{sec:3.1}

\subsubsection{Ground-Based Photometry of SN 2002bf and SN 2003du}
\label{sec:3.1.1}

\begin{figure}
\scalebox{1.0}{
\plotone{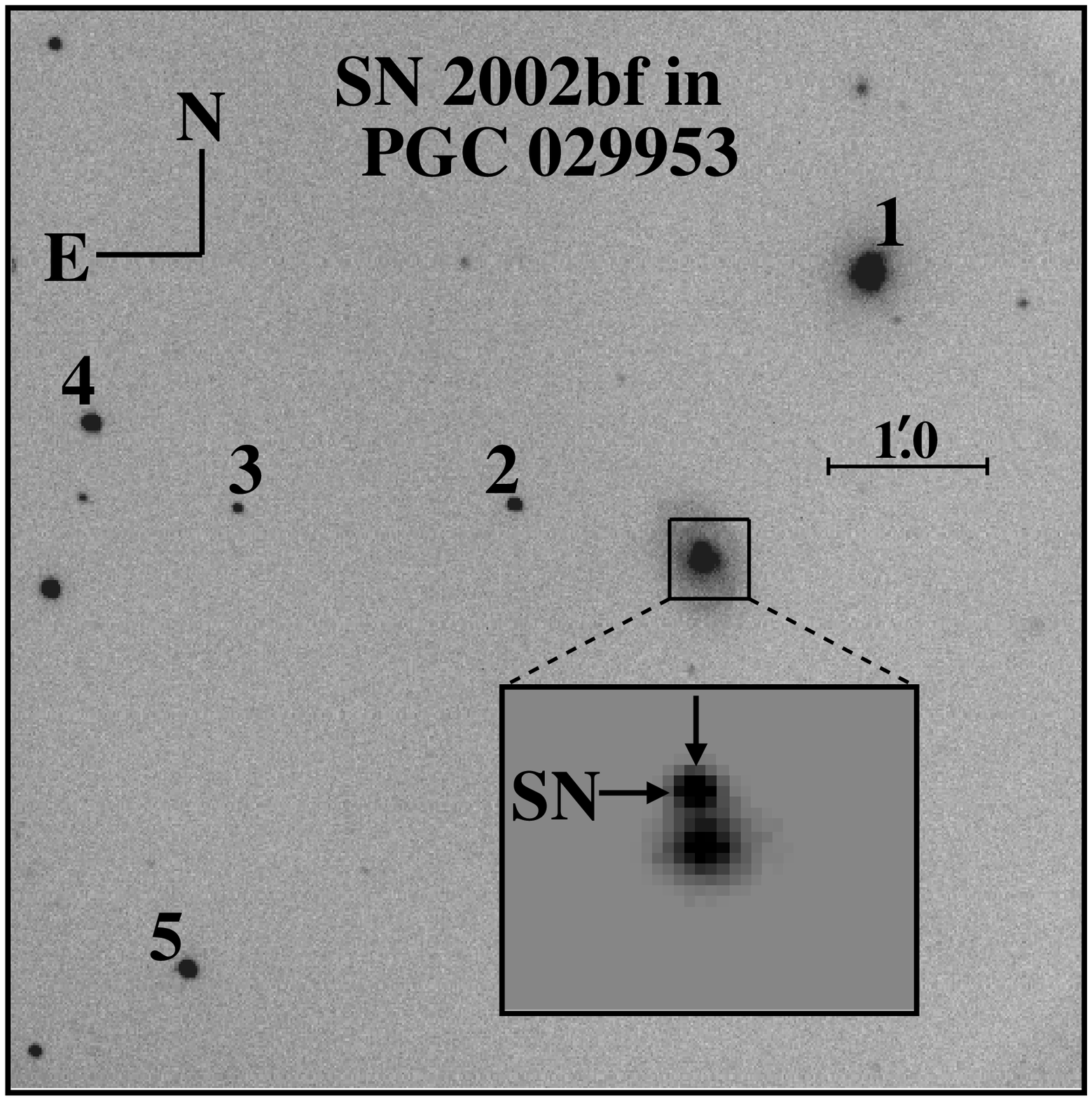}
}
\vskip 0.2in
\caption{$B$-band image of PGC~029953 taken on 2002 March 15 with KAIT, with the
local standards listed in Table~\ref{tab:1} marked.  For clarity the inner
$0\farcm5$ of PGC~029953 is expanded with the contrast adjusted to clearly show
the SN ({\it inset}).  SN~2002bf (SN) is measured to be $0\farcs5$ east 
and $4\farcs1$ north of the low-surface-brightness center of PGC~029953.
\label{fig:1} }
\end{figure}

\begin{figure}
\scalebox{1.0}{
\plotone{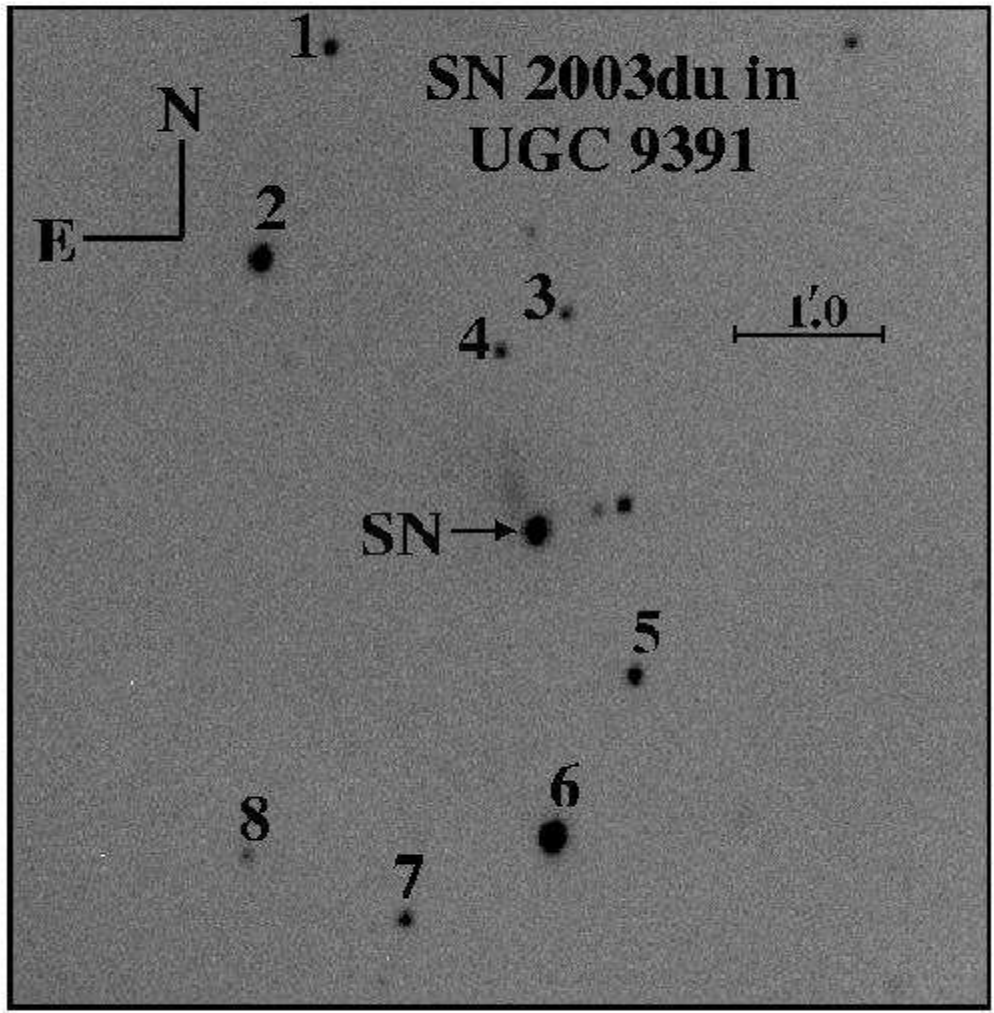}
}
\vskip 0.2in
\caption{$V$-band image of UGC 9391 taken on 2003 May 19 with KAIT, with the
local standards listed in Table~\ref{tab:2} marked.  SN~2003du (SN) is
measured to be $9\farcs3$ west and $14\farcs9$ south of the center of UGC~9391.
\label{fig:2} }
\end{figure}

All ground-based photometric data were obtained using either the 0.76-m Katzman
Automatic Imaging Telescope \citep[KAIT; ][]{Filippenko01,Li03} or the Nickel 1
m reflector \citep{Li01}, both located at Lick Observatory.
Figures~\ref{fig:1} and \ref{fig:2} show KAIT images of PGC~029953 and
UGC~9391, the host galaxies of SN~2002bf and SN~2003du, respectively.  Also
labeled in the KAIT images are the ``local standards'' in both fields that were
used to measure the relative SN brightness on non-photometric nights.  We
obtained 12 epochs of Johnson-Cousins \bvri\ photometry (\citealt[][for
$BV$]{Johnson66}; \citealt[][for $RI$]{Cousins81}) for SN 2002bf, all taken
with KAIT, and 38 epochs of \bvri\ photometry for SN~2003du, 33 of them taken
with KAIT and five with the Nickel telescope.  We also obtained three
pre-maximum unfiltered observations of SN~2002bf with KAIT (approximating the
$R$ band, see \citealt{Li03}), and one additional epoch of $RI$ photometry with
KAIT of SN~2003du.  

For the photometry we employed the usual techniques of galaxy ``template''
subtraction \citep[][and references therein]{Li00}, point-spread function
fitting \citep[][and references therein]{Stetson91}, and using ``local
standards'' to determine the \bvri\ brightnesses of the SNe on non-photometric
nights; in general, we closely followed the technique detailed by
\citet{Leonard6}.  We note that the galaxy subtraction procedure for SN~2002bf
was particularly challenging since it is only $4\farcs1$ from its host galaxy's
center.

The absolute calibration of the SN~2002bf field was accomplished on the
photometric nights of 2002 May 14 and 2004 March 17 with the Nickel telescope,
and 2003 February 3 and 2004 March 18 with KAIT, by observing several fields of
Landolt (1992) standards over a range of airmasses in addition to the SN~2002bf
field.  The absolute calibration of the SN~2003du field was similarly derived
from data taken on the photometric nights of 2003 May 31, June 1, June 26, and
August 27 with the Nickel telescope, and of 2003 May 22 and 2004 March 18 with
KAIT.  The color terms used to transform the filtered instrumental magnitudes
to the standard Johnson-Cousins system are those of \citet{Foley03}.  We list
the measured \bvri\ magnitudes and the $1\sigma$ uncertainties, taken as the
quadrature sum of a typical photometric error and the $1\sigma$ scatter of the
photometric measurements from all of the photometric nights, of the local
standard stars in Tables~\ref{tab:1} and \ref{tab:2}.

\begin{figure}
\plotone{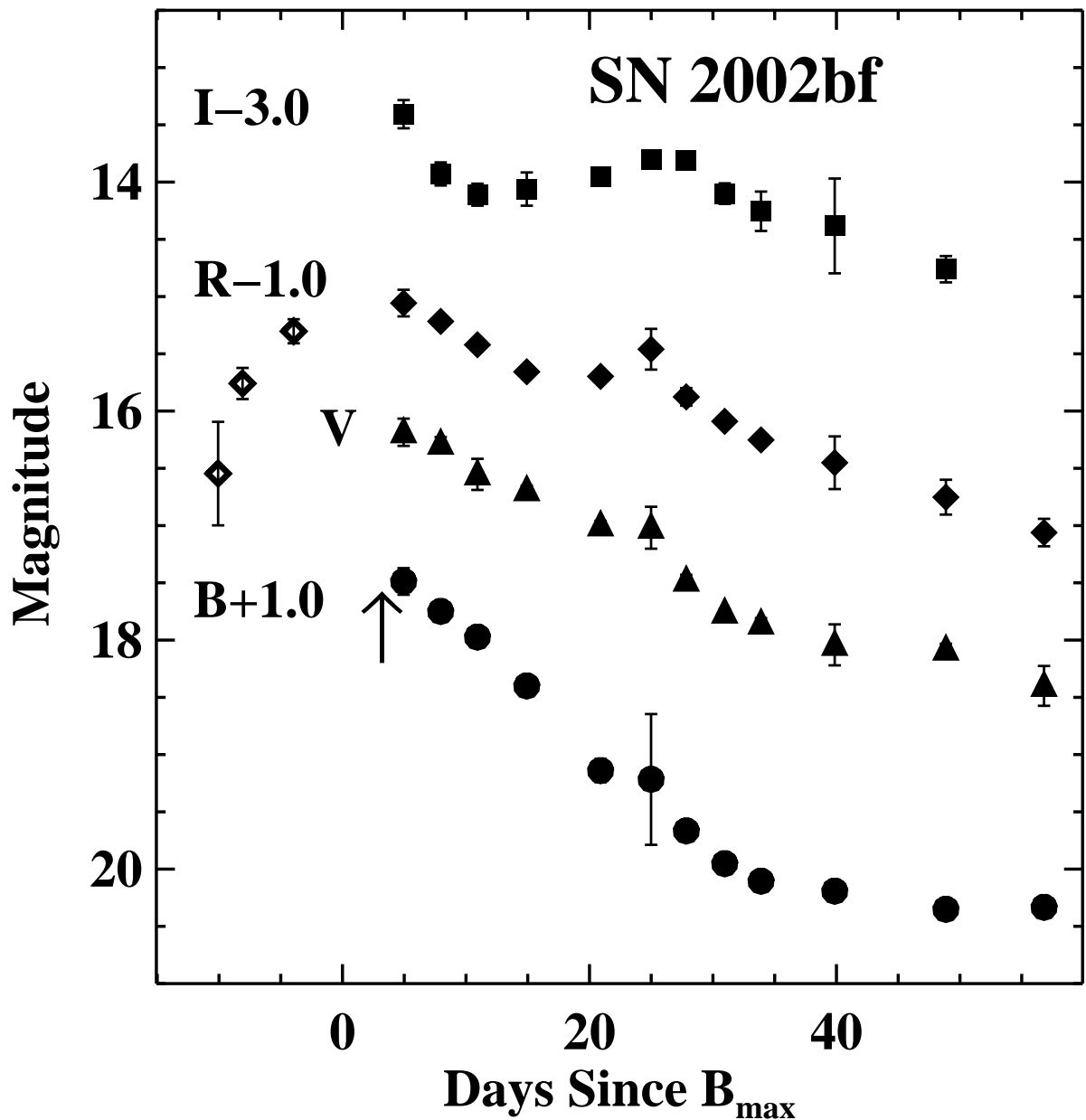}
\vskip 0.2in
\caption{\bvri\ light curves for SN 2002bf from Table~\ref{tab:3}.  For
clarity, the magnitude scales for {\it BRI} have been shifted by the amounts
indicated.  The open symbols on days $-10$, $-8$, and $-4$ represent KAIT
unfiltered magnitudes translated to the $R$ band, as described by \citet{Li03}.
{\it Arrow} indicates the epoch of our spectropolarimetric observation.
\label{fig:3} }
\end{figure}

\begin{figure}
\plotone{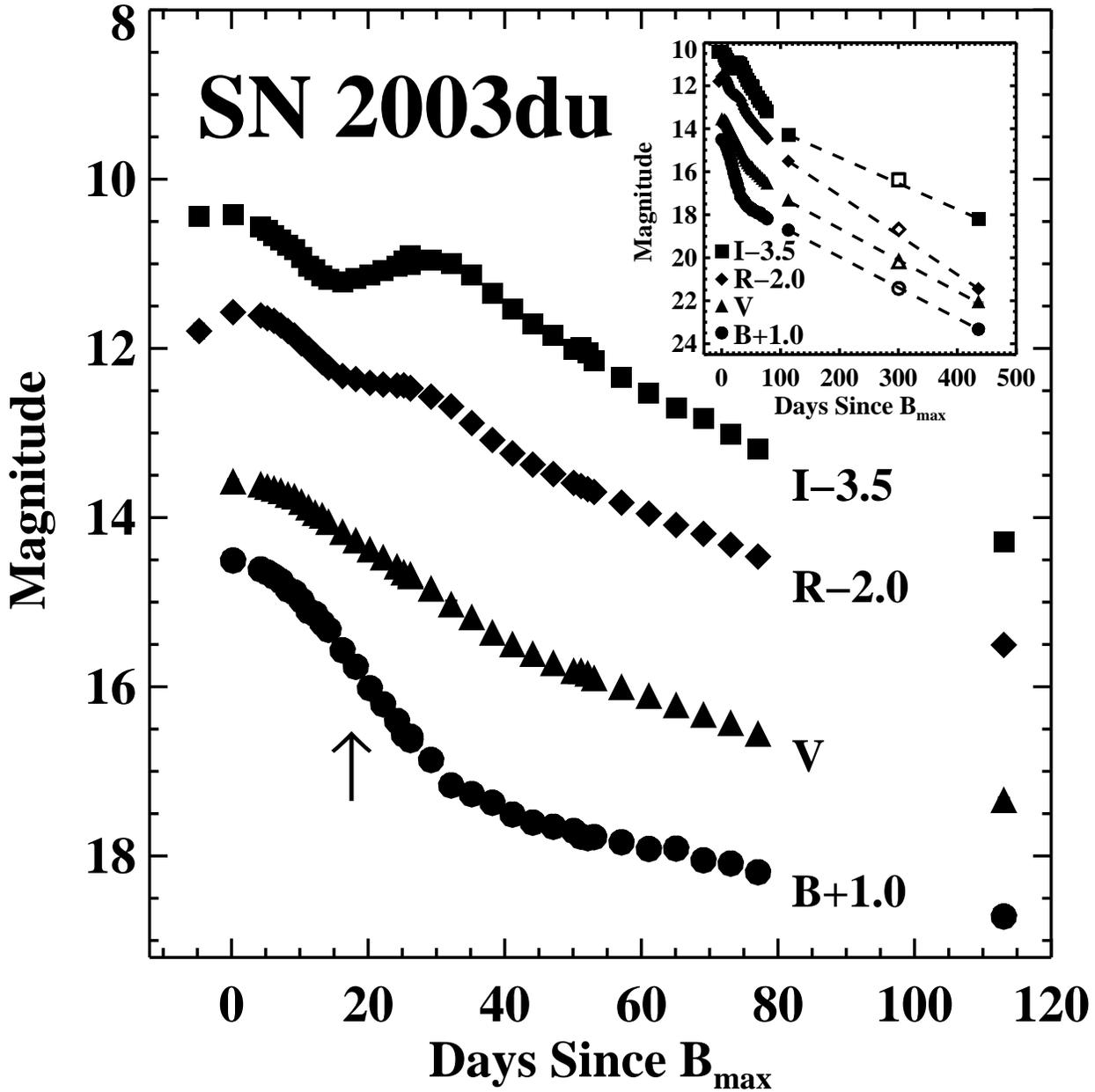}
\vskip 0.2in
\caption{\bvri\ light curves for SN 2003du from Table~\ref{tab:4}.  For
clarity, the magnitude scales for {\it BRI} have been shifted by the amounts
indicated.  In most cases the error bar is smaller than the plotted symbol.
{\it Arrow} indicates the epoch of our spectropolarimetric observation.  The
{\it inset} shows the complete light curve, including the {\it HST} ACS/HRC
observations on day 436.  {\it Dotted lines} connect the day 113 and day 436
data points, and {\it open symbols} are data from \citet{Anupama05} on day
301 (2004 January 3).
\label{fig:4} }
\end{figure}

After deriving the \bvri\ magnitudes of the SNe based on a comparison with each
of the local standards, we took the weighted mean of the individual estimates
as the final standard magnitude of the SNe at each epoch in each filter.  The
results of our ground-based photometric observations are given in
Tables~\ref{tab:3} and \ref{tab:4} and shown in
Figures~\ref{fig:3} and \ref{fig:4}.  The reported uncertainties come from the
quadratic sum of the photometric errors (reported by DAOPHOT) and the
transformation errors.  For SN~2002bf, the uncertainty produced by the
difficult galaxy-subtraction process contributed the majority of the error on
nights with low S/N.

\subsubsection{{\it Hubble Space Telescope} Photometry of SN 2003du}
\label{sec:3.1.2}

We obtained {\it HST}\ images during the course of two orbits of a
$29^{\prime\prime} \times 26^{\prime\prime}$ field of view centered on
SN~2003du on 2004 July 15, 434 days after $B_{\rm max}$, with the ACS/HRC
detector through filters F435W, F555W, F625W, and F814W (hereafter referred to
as $B, V, R, {\rm\ and\ } I$, respectively), as part of our Snapshot survey
program (GO-10272; P.I. Li) to investigate the late-time photometric behavior and
environment of nearby SNe.  SN~2003du was detected in all images.  Total
exposure times in \bvri\ were, respectively, $1680 {\rm\ s}$ (data archive
designation j8z441011/3011), $960 {\rm\ s}$ (j8z442011/4011), $720 {\rm\ s}$
(j8z441021/3021), and $1440 {\rm\ s}$ (j8z442021/4021).  

To derive the {\it HST} photometry, we followed as closely as possible the
procedure detailed by \citet{Sirianni05}, including correction for the effects
of SN light contaminating the background region, aperture corrections, the
``red-halo'' effect (for the $I$-band), and CTE degradation \citep{Riess03}.
We translated the resulting instrumental magnitudes to the standard
Johnson-Cousins \bvri\ system by using the coefficients and color corrections
tabulated by \citet{Sirianni05}.  The final results of our {\it HST} photometry
are included in Table~\ref{tab:4} and displayed in Figure~\ref{fig:4}.

\subsection{Spectropolarimetry and Spectroscopy}
\label{sec:3.2}

We obtained single epochs of spectropolarimetry for SN~2002bf and SN~2003du on
2002 March 7 and 2003 May 24, respectively, with the Low-Resolution Imaging
Spectrometer \citep{Oke95} in polarimetry mode (LRISp; Cohen
1996)\footnote{Instrument manual available at
\url{http://www2.keck.hawaii.edu/inst/lris/pol\_quickref.html}.} at the
Cassegrain focus of the Keck-I 10-m telescope.  We observed SN~1997dt on 1997
December 20 with LRISp using the Keck II 10-m telescope, and SN~2004dt on 2004
August 24 with the Kast double spectrograph \citep{Miller93} with polarimeter
at the Cassegrain focus of the Shane 3-m telescope at Lick Observatory.  We
reduced the polarimetry data according to the methods outlined by
\citet{Miller88} and detailed by \citet{Leonard3} and \citet{Leonard4}.  

The polarization angle offset between the half-wave plate and the sky
coordinate system was determined by observing the following polarized standard
stars from the list of \citet{Schmidt92a} and setting the observed $V$-band
polarization position angle (i.e., $\theta_V$, the debiased, flux-weighted
average of the polarization angle over the wavelength range 5050--5950~\AA; see
\citealt{Leonard3}) equal to the cataloged value: BD $+64^\circ106$ (1997
December 20), BD $+59^\circ389$ (2002 March 7), and HD 161056 (2003 May 24).
On the night of 2004 August 24, we averaged the polarization angle offsets
derived from observations of three polarized standards from the
\citet{Schmidt92a} list, HD 204827, BD $+59^\circ389$, and HD 19820; the
individual offsets were internally consistent to within $1^\circ$.  To check
for instrumental polarization, the following null standards taken from the
lists of \citet{Turnshek90}, \citet{Mathewson70}, \citet{Schmidt92a}, and
\citet{Berdyugin95}, were also observed: HD 94851 (1997 December 20), HD 57702
(2002 March 7), HD 109055 and BD $+32^\circ3739$ (2003 May 24), and HD 212311
(2004 August 24).  All stars were measured to be null to within $0.1\%$, which
is also our estimate of the systematic uncertainty of a continuum polarization
measurement made with either the Keck or Lick spectropolarimeters
\citep[e.g.,][]{Leonard3}.

Additional specifics of the observations of SN~2002bf taken on 2002 March 7,
including an investigation of the potential impact of second-order light
contamination and instrumental polarization (both shown to be minimal) in the
setup used on this night, are given by \citet{Leonard8}.  Second-order light
contamination is not a concern for our observation of SN~1997dt due to its
limited spectral range. For SN~2003du, the use of a dichroic (D560) to split
the beam near 5600~\AA\ eliminates second-order light contamination on the
red side.  Our spectropolarimetric observation of SN~2004dt was taken in a
setting that included the use of an order-blocking filter (GG455) to prevent
contamination by second-order light at red wavelengths.  Beyond $\sim$9000~\AA, 
however, second-order contamination may exist, but for reasons similar to
those discussed by \citet{Leonard8} we believe it to have minimal
impact for this particular object.

\begin{figure}
\hspace{-0.5in}
\rotatebox{90}{
\scalebox{0.9}{
\plotone{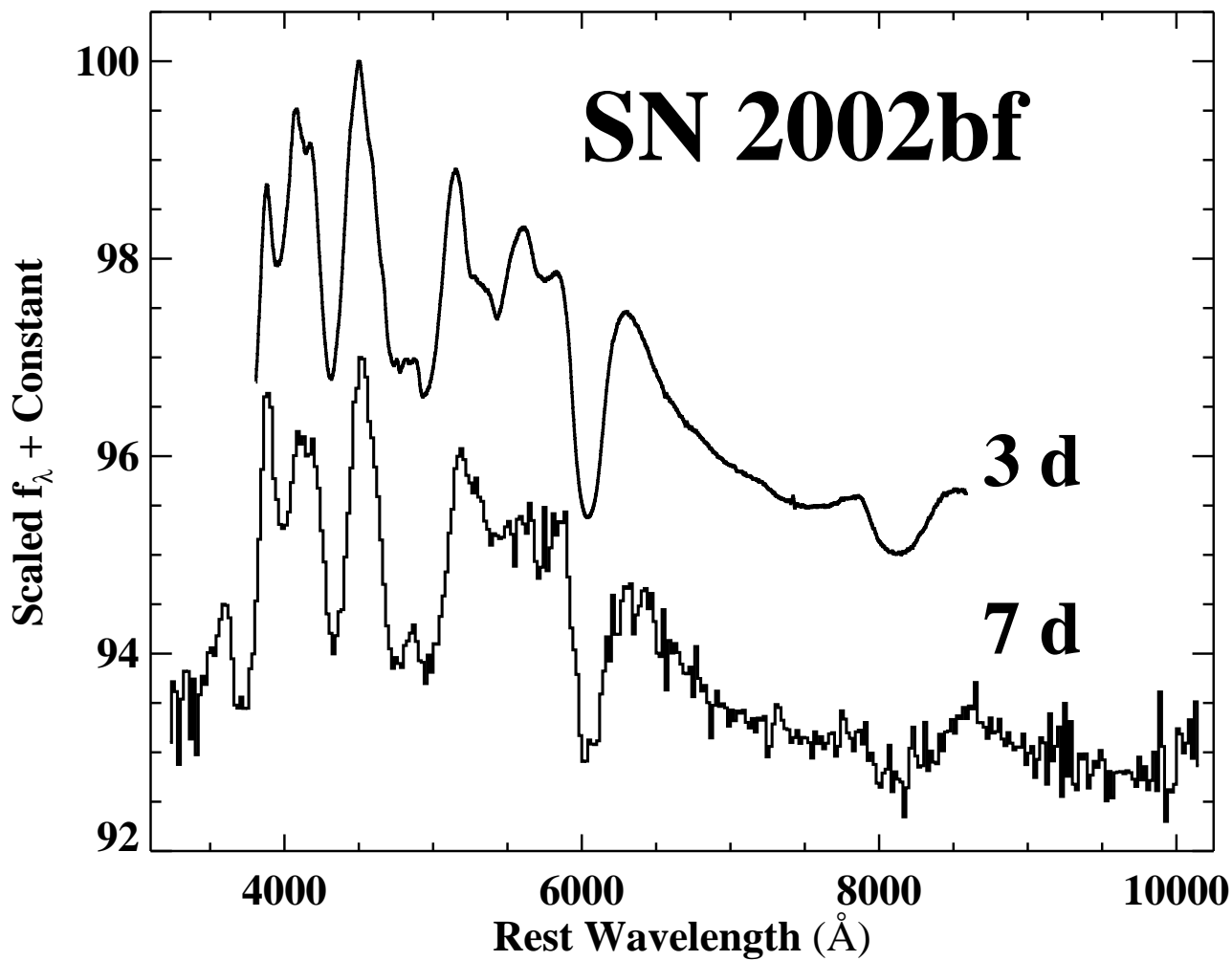}
}}
\vskip 0.2in
\caption{Optical flux spectra of SN~2002bf labeled by epoch with respect to $B$
  maximum.  For clarity, the spectra have been scaled and shifted vertically by
  arbitrary amounts.  Note that while the spectrum at $t = 3$ days is binned 2~\AA\ 
  bin$^{-1}$, the spectrum at 7 days has been binned at 20~\AA\ 
  bin$^{-1}$ in order to increase the S/N.
\label{fig:9} }
\end{figure}

\begin{figure}
\hspace{-0.5in}
\rotatebox{90}{
\scalebox{0.9}{
\plotone{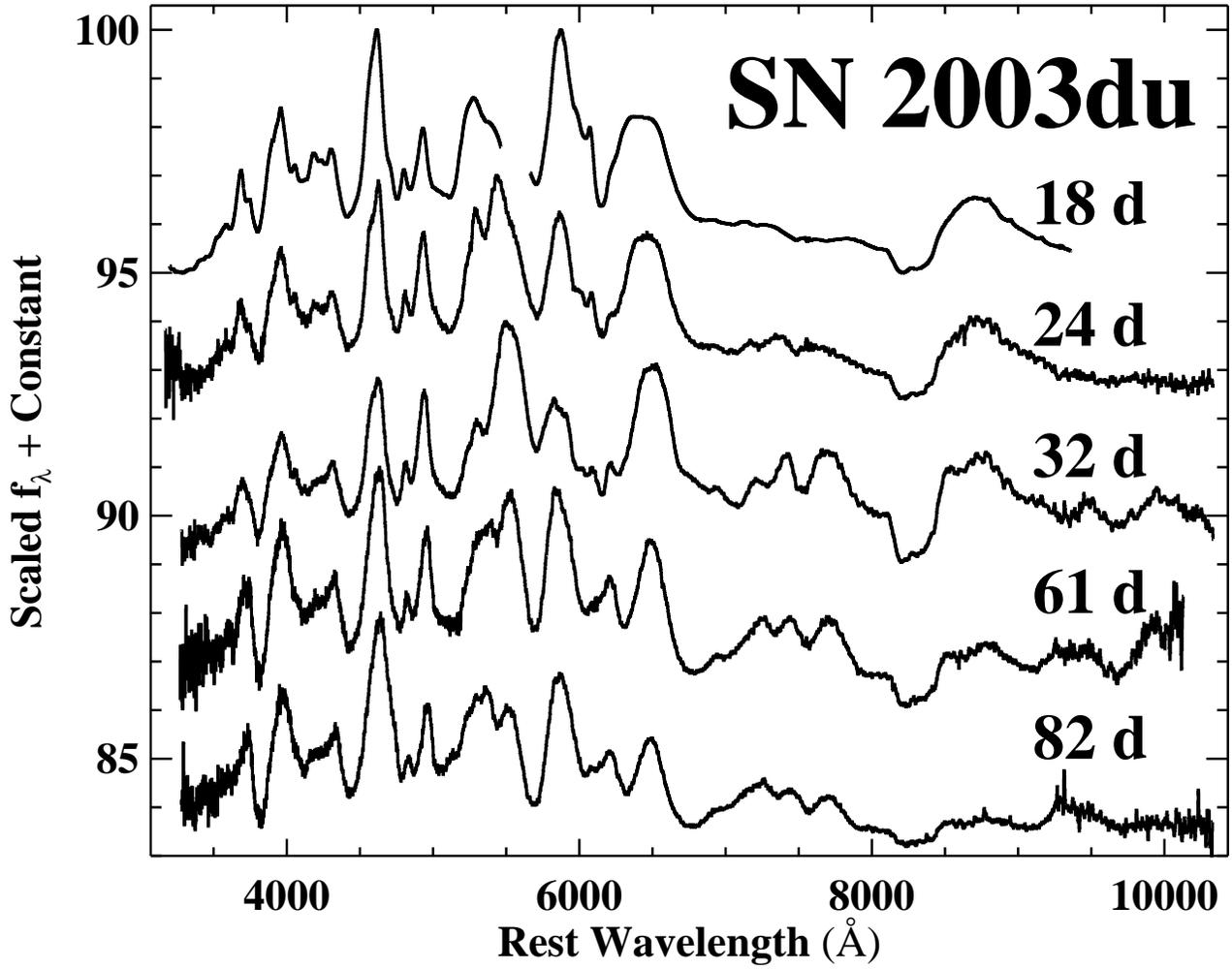}
}}
\vskip 0.2in
\caption{Optical flux spectra of SN~2003du labeled by epoch with respect to
  $B$ maximum.  For clarity, the spectra have been scaled and shifted
  vertically by arbitrary amounts.  
\label{fig:10} }
\end{figure}

To derive the total-flux spectra, we extracted all one-dimensional
sky-subtracted spectra optimally \citep{Horne86} in the usual manner.  Each
spectrum was then wavelength and flux calibrated, and was corrected for
continuum atmospheric extinction and telluric absorption bands
\citep{Wade88,Bessell99,Matheson00}.  With the exception of the
spectropolarimetric observations of SN~1997dt and SN~2002bf, all spectra were
taken near the parallactic angle \citep{Filippenko82}, so the spectral shape
should be quite accurate.  Table~\ref{tab:5} lists the spectropolarimetric and
spectral observations for all four SNe.  Figures~\ref{fig:5}--\ref{fig:8}
show the observed spectropolarimetry data of the four objects, and
Figures~\ref{fig:9} and \ref{fig:10} show the complete series of spectra
obtained for SN~2002bf and SN~2003du, respectively.

\section{Analysis}
\label{sec:4}

\subsection{Spectroscopy}
\label{sec:4.1}

Our spectrum of SN~1997dt, taken $\sim$21 days after maximum, shows typical
features for an SN~Ia at this phase (Fig.~\ref{fig:5}a).  Similarly, our
spectral sequence of SN~2003du (Fig.~\ref{fig:10}) closely follows the
evolution of the normal-luminosity SN~Ia~1994D \citep{Patat96,Filippenko97} in
terms of the strengths and blueshifts of line features.  This is consistent
with the analyses of \citet{Anupama05} and \citet{Gerardy04}, in which spectra
taken before and shortly after maximum light were also examined.  This
convincingly establishes SN~2003du as a spectroscopically ``typical'' SN~Ia
\citep{Branch93a}.

As discussed in \S~\ref{sec:2.4}, the spectra of both SN~2002bf and SN~2004dt
are peculiar in one regard: the blueshifts of many of the spectral lines, most
noticeably \ion{Si}{2} $\lambda 6355$, occur at significantly higher velocity
than is typical for an SN~Ia at this phase, and indicate that these are both
HV~SNe~Ia.

\begin{figure}
\hspace{0.5in}
\rotatebox{0}{
\scalebox{0.9}{
\plotone{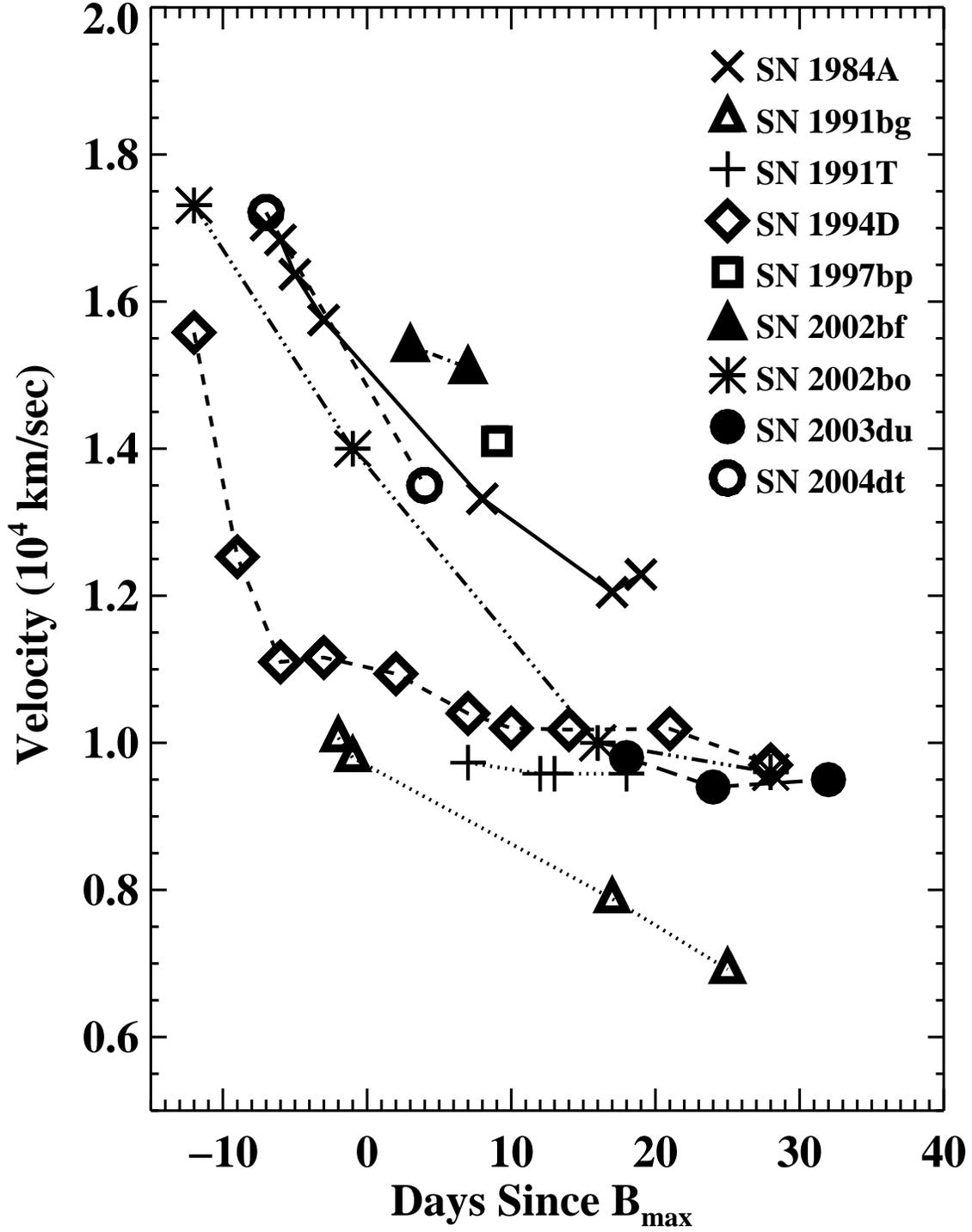}
}}
\vskip 0.2in
\caption{Blueshifts of \ion{Si}{2} $\lambda 6355$ for nine SNe~Ia, from
  Table~\ref{tab:6}.  Lines connecting data points for each SN are shown 
  to help guide the eye, and are not the results of formal fits.
\label{fig:11} }
\end{figure}

For comparison, we have measured the velocities of the \ion{Si}{2}
$\lambda6355$ line in a number of other HV~SNe~Ia from our database, along with
the spectroscopically normal SN~1994D, the subluminous SN~1991bg
\citep[e.g.,][]{Filippenko92b}, and the overluminous SN~1991T
\citep[e.g.,][]{Filippenko92a}.  We present the results in Table~\ref{tab:6}
and Figure~\ref{fig:11}, from which it is clear that both SN~2002bf and
SN~2004dt belong to the class of HV SNe~Ia; indeed, SN~2002bf is the most
extreme HV~SN~Ia yet observed for its epochs.  The figure also suggests that
line velocity is not strongly correlated with luminosity, although the
expansion velocity of the subluminous SN~1991bg is somewhat lower than typical
values.

\begin{figure}
\hspace{-1.0in}
\rotatebox{90}{
\scalebox{0.9}{
\plotone{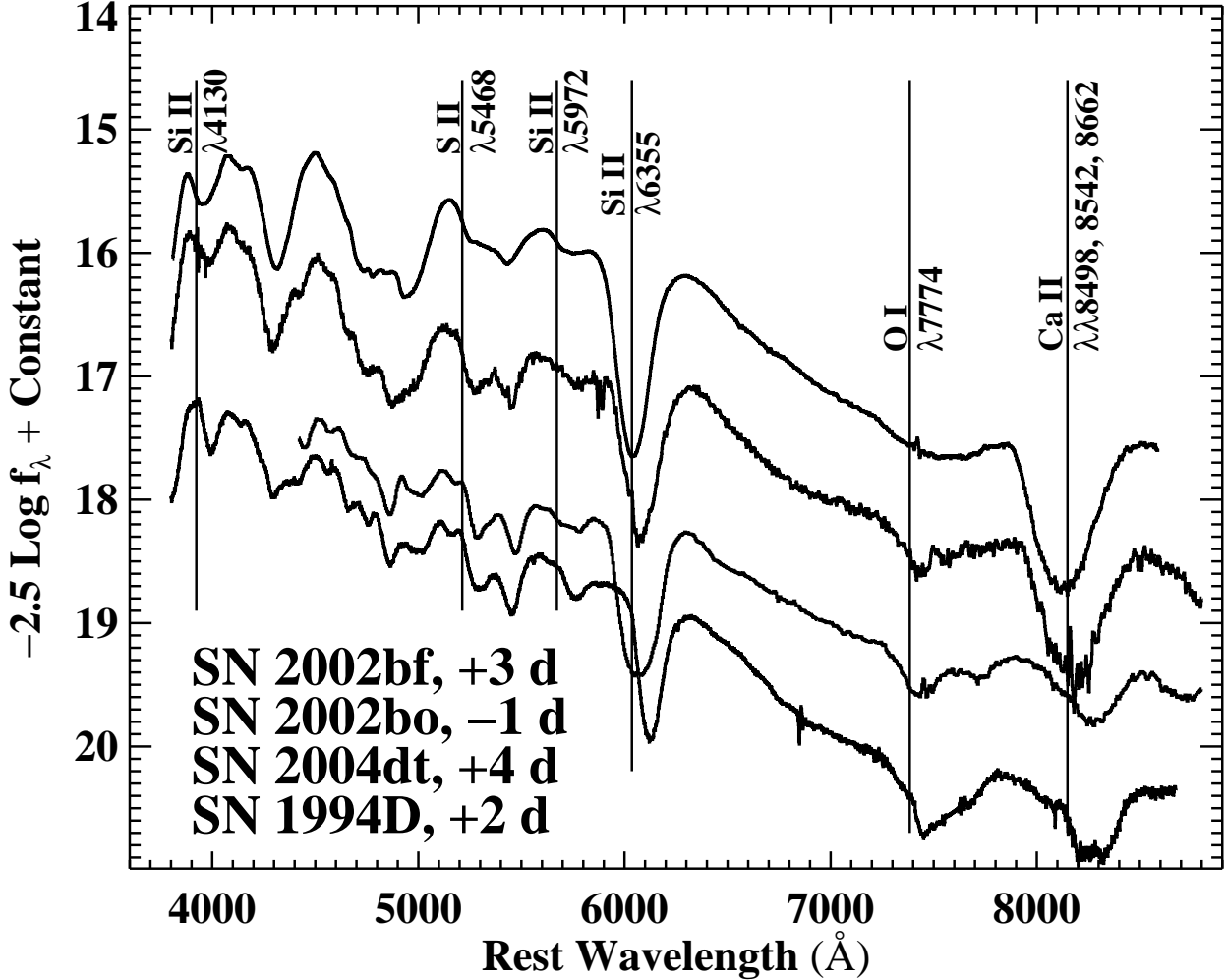}
}}
\vskip 0.2in
\caption{Spectra of four SNe~Ia (from top down: SN~2002bf, SN~2002bo,
  SN~2004dt, SN~1994D) with time from $B_{\rm max}$ indicated, demonstrating
  the diversity of P-Cygni absorption-line blueshifts.  All but SN~1994D are
  considered to be ``high-velocity'' SNe~Ia, due to their unusually broad and
  highly blueshifted \ion{Si}{2} $\lambda6355$ \AA\ line.  The expected
  locations of several additional line features, if blueshifted by $15,400\
  \Kms$ (the blueshift of the \ion{Si}{2} $\lambda6355$ line for SN~2002bf)
  from their rest wavelengths, are indicated by {\it vertical lines}.  The
  spectra have been dereddened by the following amounts: $\Ebv = 0.08$ mag
  (SN~2002bf; \S~\ref{sec:4.3.2}), $\Ebv = 0.43$ mag (SN~2002bo;
  \citealt{Benetti04}), $\Ebv = 0.11$ mag (SN~2004dt; \S~\ref{sec:4.3.2}), and
  $\Ebv = 0.06$ mag (SN~1994D; \citealt{Patat96}).
\label{fig:12} }
\end{figure}

Figure~\ref{fig:12} presents a spectral comparison near maximum light of three
HV~SNe~Ia (SN~2002bf, SN~2002bo, and SN~2004dt) with the spectroscopically
normal SN~1994D.  The excessive blueshift of the \ion{Si}{2} $\lambda 6355$
trough is obvious for the three HV~SN~Ia compared with SN~1994D.  While it must
be cautioned that the spectra span a range of ages of about five days near
maximum light, a time when significant spectral development occurs, the
blueshift differences are much greater than can be explained by age differences
alone.  In addition, the \ion{Si}{2} line is significantly stronger in the
HV~SNe~Ia compared with SN~1994D, with equivalent widths of $\gtrsim 140$~\AA\
compared with $\sim$100~\AA\ for SN~1994D.  Conversely, the \ion{O}{1}
$\lambda 7774$ absorption is relatively weaker in the HV~SN~Ia events, with
equivalent widths of $\lesssim 100$~\AA\ in the HV~SNe~Ia compared with
$125$~\AA\ for SN~1994D.  The fact that the \ion{Si}{2} line is significantly
stronger in the HV~SNe~Ia sample, and the \ion{O}{1} line relatively weaker, is
consistent with the scenario in which a greater fraction of C and O is burned
to IMEs in HV~SN~Ia than in more typical events; it also follows naturally from
the GCD model, since the obscuring pancake is formed from the products of
oxygen burning.

While HV~SNe~Ia share many spectral characteristics, it is clear that they also
exhibit spectral diversity.  For instance, whereas the absorption trough of the
\ion{Ca}{2} near-IR triplet is significantly blueshifted for SN~2002bf and
SN~2002bo relative to SN~1994D ($16,300\ \Kms$ and $14,500\ \Kms$ for SN~2002bf
and SN~2002bo, respectively, compared with $\sim$10,900~\kms\ for SN~1994D,
where we have assumed $\lambda_0 = 8579$ for the \ion{Ca}{2} near-IR triplet, a
value derived using the prescription given by \citealt{Leonard5}), it has a
more normal blueshift ($v = 10,900$~\kms) in SN~2004dt.  The equivalent widths
of the \ion{Ca}{2} near-IR lines also show a suggestive trend, with both
SN~2002bf and SN~2002bo having widths of $> 200$~\AA, while both SN~2004dt and
SN~1994D have equivalent widths of $\sim$100~\AA.  This may indicate that the
explosive nucleosynthesis did not proceed up to Ca as far out in the atmosphere
(or in the bubble in the GCD scenario) of SN~2004dt as it did in the other two
HV~SNe.

Finally, it is clear from Figure~\ref{fig:12} that not all lines share the
extreme velocities seen in the \ion{Si}{2} $\lambda 6355$ feature. For
instance, as first noticed by \citet{Wang05} in a pre-maximum spectrum of
SN~2004dt, the \ion{S}{2} ``W'' feature in the spectra of SN~2002bo and SN~2004dt
indicates velocities comparable to those in SN~1994D; for SN~2002bf they
are somewhat larger, but still not near the velocity of the \ion{Si}{2}
$\lambda 6355$ line.  \citet{Wang05} propose that this may indicate that sulfur
is more confined to the lower-velocity, inner region.

\subsection{Photometry}
\label{sec:4.2}

\subsubsection{Ground-Based Photometry}
\label{sec:4.2.1}

The results of the MLCS2k2 application to the photometry of SN~2002bf and
SN~2003du are given in Table~\ref{tab:7}.  For details of the MLCS2k2 procedure
used, see S. Jha et al. (in preparation); an overview of the technique is
provided by \citet{Riess05}.  The MLCS2k2 analysis finds SN~2002bf to be of
typical luminosity.  SN~2003du is slightly overluminous, although its
pre-maximum spectral evolution \citep{Anupama05} demonstrates that it is {\it
not} a SN~1991T-like event, as the strength of the \ion{Si}{2} $\lambda 6355$
feature is comparable to that seen in spectra of normal SNe~Ia.

For SN~2002bf, $B_{\rm max}$ occurred on 2002 March $4.37 \pm 0.50$, which is
$10$ days after discovery and 5 days before our filtered observations
commenced.  For SN~2003du, $B_{\rm max}$ occurred on 2003 May $6.12 \pm 0.50$,
which is $14$ days after discovery, and the same day that our \bvri\
observations began.  Our derived date of maximum light for SN~2003du agrees
with those found by \citet{Anupama05} and \citet{Gerardy04} using independent
data sets.

\subsubsection{{\it HST} Photometry of SN~2003du}
\label{sec:C1c}

Late-time photometry (e.g., $t > 200 {\rm\ d}$) of SNe~Ia exists for only a
handful of objects \citep[see, e.g., ][and references therein]{Milne01}.  From
the small sample, there are two main features of note.  First, SN~Ia decline
rates are typically much faster than the decay slope of $^{56}{\rm Co}
\rightarrow$ $^{56}{\rm Fe}$ of $0.98\ {\rm mag\ (100\ d)}^{-1}$ predicts.
This decay mechanism is thought to be primarily responsible for powering the
luminosity from the early nebular phase out to $\sim$1000 days for SNe of all
types.  Essentially, the $^{56}{\rm Co}$ decays release most of their energy in
the form of $\gamma$-rays, which, given enough optical depth, can become
trapped in the ejecta and Compton scatter off free electrons.  The energetic
electrons generate optical photons primarily through ionization and excitation
of atoms, and the ejecta are transparent to these photons.  A small portion
($\sim$3.5\%, see \citealt*{Arnett79}) of the total $^{56}{\rm Co}$ decay
energy comes in the form of positrons, which may deposit their kinetic energy
in the ejecta and then annihilate with electrons, producing two $\gamma$-ray
photons of energy $E_\gamma = m_e c^2$.  The steeper decline seen in the
late-time photometry of SNe~Ia has been explained by significant transparency
of the ejecta to $\gamma$-ray photons \citep{Milne99} and positron escape
\citep{Milne01}.

A quantitative measure of the late-time decline is given by
\citet{Cappellaro97}, who investigate how the decline from maximum to 300 days
after peak $V$ brightness, denoted $\Delta m_{300}(V)$, correlates with
intrinsic SN brightness as derived from the $\Delta m_{15}(B)$ parameter.
Although limited by small sample size (5 objects), they find a convincing
correlation, with $\Delta m_{300}(V)$ going from 6.7 mag to 8.4 mag as the
sample runs from overluminous (SN~1991T) to subluminous (SN~1991bg) events,
with the normal-brightness SN~1994D characterized by $\Delta m_{300}(V) = 7.3$
mag.  A second feature that has thus far been seen in only two SN~Ia events
(SN~1991T and SN~1998bu; see \citealt*{Schmidt94a} and \citealt*{Cappellaro01},
respectively) is a sudden flattening of the late-time optical light curves,
which has been attributed to the contribution of a light echo from foreground
dust clouds.

Our photometric data from {\it HST}, taken $436$ days after $B_{\rm max}$,
allow us to investigate the late-time photometric behavior of SN~2003du.  From
the inset of Figure~\ref{fig:4}, it is clear that SN~2003du, like all SNe~Ia
analyzed before it, declines significantly faster than the $^{56}{\rm Co}
\rightarrow$ $^{56}{\rm Fe}$ decay rate.  In fact, we measure the average decay
rate in $V$ (shown by \citealt*{Milne01} to track the bolometric luminosity of
an SN~Ia quite accurately) of SN~2003du between our last two photometric epochs
on days $113$ and $436$ to be $\Delta V = 1.47 \pm 0.02 {\rm\ mag\ (100\
d)}^{-1}$.  The slope also appears to have been rather constant throughout the
period between our two last epochs, as indicated by the good agreement of the
late-time data taken from \citet{Anupama05} near day 300 and the decay slope
determined from our data alone.  Using the \citet{Anupama05} data point and our
estimate of $V_{\rm max}$ (Table~\ref{tab:7}), we derive $\Delta m_{300}(V) =
6.74$, which is most consistent with the values found by \citet{Cappellaro97}
for overluminous events.  That SN~2003du may be somewhat overluminous was also
suggested by its peak $V$ magnitude of $-19.67 \pm 0.02$ mag, which is 0.17 mag
brighter than the fiducial template used in the MLCS2k2 procedure
(Table~\ref{tab:2}).  There is no evidence from our data of any contribution to
the SN brightness from a light echo, although we note that the major
indications of additional contributions to the apparent brightness of SNe~1991T
and 1998bu did not become obvious until epochs $\gtrsim 500 {\rm\ d}$.

\subsection{Reddening}
\label{sec:4.3}

\subsubsection{Techniques to Estimate SN Ia Reddening}
\label{sec:4.3.1}

Accurate determination of SN reddening is crucial both for deriving intrinsic
SN properties as well as interpreting spectropolarimetry, since the same dust
that reddens SN light can also polarize it as discussed in \S~\ref{sec:2.3.2}.
When multi-band photometry is available, the MLCS2k2 technique can accurately
estimate the total extinction (see \citealt{Riess05} for discussion).  When
this is lacking, other methods must be used.

Galactic extinction along the l-o-s is accurately estimated by the dust maps of
\citet[][hereafter SFD]{Schlegel98} to an estimated precision of $\sim$15\%.
For host-galaxy extinction, a rather crude approach, to which many
investigators resort in the absence of photometry, is to employ the rough
correlation found between the total equivalent width ($W_\lambda$) of the
interstellar (IS) \ion{Na}{1} D doublet ($\lambda\lambda 5890, 5896$) and
reddening \citep{Barbon90}.  The \citeauthor{Barbon90} correlation has
been subsequently improved upon by \citet{Munari97}, who derive a more precise
relation that uses just the equivalent width of the \ion{Na}{1} D2 ($\lambda
5890$) line.  Both relations, however, warrant healthy degrees of skepticism
since sodium is known to be only a fair tracer of the hydrogen gas column
(especially in dense environments, where sodium may be heavily depleted; e.g.,
\citealt*{Cohen73}), from which the dust column is then estimated.  The
dust-to-gas ratio also varies significantly among galaxies (e.g.,
\citealt*{Issa90}), and it could well be that in the case of SNe,
circumstellar, rather than interstellar, dust is present, for which the
dust-to-gas ratio (or the extinguishing properties of the dust itself) could be
unusual.  A final complication is that, at the resolution typical of most
optical spectra, individual absorption components along the l-o-s, whose
contributions should be considered separately to determine the total reddening,
are blended into a single profile.  In such situations, the \citet{Munari97}
relation formally yields an {\it upper limit} to the reddening, and caution
must be used since the derived value could seriously overestimate the actual
reddening.  Nonetheless, the \citet{Munari97} relation is still frequently used
to get approximate reddening values, or upper limits, especially in the case of
null detections of IS \ion{Na}{1} D lines.

\subsubsection{The Reddening of the Four SNe Ia}
\label{sec:4.3.2}

For SN~2002bf and SN~2003du we shall adopt the MLCS2k2 values of $\Ebv = 0.08
\pm 0.04$ mag and $\Ebv = 0.01 \pm 0.01$ mag (respectively) reported in
Table~\ref{tab:7} (for $R_V = 3.1$).  For SN~1997dt, \citet{Tonry03} report a
host-galaxy reddening value of $\Ebv_{\rm Host} = 0.15$ mag (assuming $R_V =
3.1$), the median of the values derived from MLCS2k2 and ``Bayesian Adapted
Template Match Method'' analyses.  The SFD dust maps predict $\Ebv_{\rm
SFD}^{\rm MW} = 0.06$ mag for SN~1997dt, for a total estimate of $\Ebv = 0.21$
mag.

Our spectrum of SN~1997dt also exhibits strong \ion{Na}{1} D IS absorption at
both the redshift of the host galaxy and the MW.  The resolution of our
spectrum, $\sim$5~\AA\ (Table~\ref{tab:5}), while sufficient to deblend the
\ion{Na}{1} D doublet, is not fine enough to resolve the individual absorption
components that likely contribute to the D1 and D2 profiles.  For host-galaxy
absorption, we measure $W_\lambda^{\rm tot} = 0.77$~\AA, with $W_\lambda^{\rm
D2} = 0.42$~\AA\ and $W_\lambda^{\rm D1}= 0.35$~\AA.  This translates to
$\Ebv_{\rm NaID}^{\rm Host} \leq 0.21$ mag from the \citet{Munari97} relation.
For the MW, we measure $W_\lambda^{\rm tot} = 0.76$~\AA, with $W_\lambda^{\rm
D2} = 0.44$~\AA\ and $W_\lambda^{\rm D1} = 0.32$~\AA, yielding $\Ebv_{\rm
NaID}^{\rm MW} \leq 0.23$ mag from the \citeauthor{Munari97} relation.  The
values given by SFD and \citet{Tonry03} are consistent with the upper limits
derived from the sodium relation.  The purported accuracy of the SFD Galactic
reddening value and the upper limit set by the \citeauthor{Munari97} relation
for the host reddening would indicate an upper reddening limit of $\Ebv_{\rm
total} \leq 0.27$ mag.  This limit will prove to have important implications
when we examine the spectropolarimetry of SN~1997dt in \S~\ref{sec:4.4.2}.  The
fact that the \citeauthor{Munari97} relation overpredicts the reddening for
both the host galaxy and, especially, the MW, may indicate that multiple,
unresolved components make up the IS line profiles.  Given the upper reddening
limit, we conclude $\Ebv = 0.21 \pm 0.06$ mag for SN~1997dt.

For SN~2004dt, $\Ebv_{\rm SFD}^{\rm MW} = 0.03$ mag, and \citet{Wang05} report
a range of $\Ebv_{\rm total} = 0.14$ to 0.2 mag from analysis of the color of
their pre-maximum spectrum.  However, they also note that SN~2004dt does not
show noticeable \ion{Na}{1} D IS lines at the redshift of NGC~799.  We confirm
the lack of \ion{Na}{1} D lines in our spectrum as well, although its poor
resolution ($\sim$18~\AA, see Table~\ref{tab:5}) makes deriving even an upper
limit to the strength of the \ion{Na}{1} D IS lines difficult.  Formally, our
procedure yields $W_\lambda {\rm (3\sigma)} = 0.1$~\AA\ for host-galaxy
\ion{Na}{1} D, which translates to a predicted upper limit of $\Ebv_{\rm
NaID}^{\rm Host} < 0.02$ mag.  We thus confirm the discrepancy noted by
\citet{Wang05} between the lack of IS sodium absorption and reddening inferred
from other methods. It could be that the spectroscopic peculiarities of HV
SNe~Ia produce photometric irregularities that affect the general reddening
relations.  As mentioned in \S~\ref{sec:2.2}, \citet{Benetti04} do find
photometric peculiarities, albeit minor, in their study of another HV~SN~Ia,
SN~2002bo. However, since the \ion{Na}{1} D relation, especially for the host
galaxy, is also prone to error, we have no convincing way to decide between the
two estimates.  We thus take the simple average of the low and high reddening
estimates, and incorporate the disparity into our final estimate's uncertainty.
This yields $\Ebv = 0.11 \pm 0.06$ mag as our best reddening estimate for
SN~2004dt.

\subsection{Spectropolarimetry of Four SNe Ia}
\label{sec:4.4}

To summarize: SN~1997dt is likely a somewhat subluminous event that is thought
to be reddened by $\Ebv = 0.21 \pm 0.06$ mag, with an upper limit of $0.27$
mag.  SN~2002bf and SN~2004dt are HV~SNe~Ia, with SN~2002bf being the most
extreme example yet observed.  Finally, SN~2003du is a slightly overluminous
event that is minimally reddened.  At maximum light, a previous subluminous
event has been found to be moderately polarized ($p \approx 0.8\%$), whereas
normal to overluminous examples have been less so ($p \lesssim 0.4\%$).  The
two HV~SNe~Ia that have reported polarization measurements show the greatest
polarization features yet observed for any SN~Ia type, with values approaching
$2\%$ in the strongest lines of SN~2004dt (\S~\ref{sec:2.3.3}).

We now examine our single-epoch spectropolarimetry of SN~1997dt, SN~2002bf, SN
2003du, and SN~2004dt obtained on days 21, 3, 18, and 4, respectively, after
maximum light.  For each SN, we shall first discuss the observed polarization
and then attempt to remove the ISP through the technique of \citet{Wang05},
which assumes that spectral regions lacking strong individual flux or
polarization features at blue wavelengths ($\lambda < 5000$~\AA) are
intrinsically unpolarized.  Since the choice of the specific spectral regions
to use for ISP determination is admittedly somewhat subjective, we shall be
careful to point out how our conclusions would change with other ISP choices.

Following ISP removal and examination of the data in the $q$--$u$ plane, we
calculate the intrinsic polarization of the SN by rotating the $q$--$u$ axes
through a single angle, $\theta$, that places the greatest degree of
polarization change across the spectrum along the rotated $q$ axis.  The angle
by which the axes are rotated is determined through a uniform-weight,
least-squares fit to the ISP-subtracted data in the $q$--$u$ plane.  The
polarization degree measured along the rotated $q$ axis is then referred to as
the rotated Stokes parameter \citep[RSP;][]{Tran95,Leonard3}, whereas that
measured along the rotated $u$ axis is denoted URSP.\footnote{Note that RSP
and URSP, as defined here, are identical to the \citet{Wang01} definitions of
the polarization along the ``dominant'' and ``orthogonal'' axes, $P_d$ and
$P_o$, respectively.  That is, by definition, RSP shows polarization strength
variation along the chosen polarization angle (i.e., the dominant axis),
whereas the URSP shows polarization strength variation along the axis that is
orthogonal to the dominant axis in the $q$--$u$ plane.  The main difference
between our construction of the RSP in the {\it observed} polarization plots
(Figures \ref{fig:5}--\ref{fig:8}) and the {\it ISP-subtracted} data
(Figures \ref{fig:15}, \ref{fig:17}, and \ref{fig:19}) is that in
the former the polarization angle about which the RSP is determined is a
smoothly varying function of wavelength (i.e., it is the polarization angle,
$\theta$, smoothed over many bins; see \citet{Leonard3} for more details on the
procedure), whereas in the latter, RSP and URSP are determined with respect to
a single, unchanging polarization angle.  For consistency with
spectropolarimetry work in other fields, as well as our own prior studies, we
shall continue to use the RSP and URSP designations, rather than the $P_d$ and
$P_o$ designations of \citet{Wang01}.}

We shall then analyze the resulting spectropolarimetry within the context of
the models described in \S~\ref{sec:2.3.1}.  By construction, the greatest
degree of polarization change, especially from the continuum regions, should
occur in the RSP.  Examination of URSP is especially important in the line
features, however, as it can be used to discriminate between the predictions of
the ellipsoidal and clumpy-ejecta models.  In particular, the simple
ellipsoidal model demands the existence of a single value of $\theta$ capable
of placing all line polarization changes along the RSP (i.e., URSP should show
no polarization features).

\subsubsection{Two HV~SNe~Ia:  SN 2002bf and SN 2004dt}
\label{sec:4.4.1}

We begin by considering the spectropolarimetry of the two HV~SNe~Ia, SN~2002bf
and SN~2004dt, which are displayed in Figures~\ref{fig:6} and \ref{fig:8},
respectively.  Both objects show very low levels of observed polarization, with
$p_V = 0.03\%$, $\theta_V = 62^\circ$ for SN~2002bf and $p_V = 0.25\%$,
$\theta_V = 146\arcdeg$ for SN~2004dt, where $p_V$ and $\theta_V$ approximate
the rest-frame $V$-band polarization and polarization angle derived by
calculating the debiased, flux-weighted averages of $q$ and $u$ over the
interval 5050--5950~\AA\ \citep[see][]{Leonard3}.  The polarization of
SN~2004dt may exhibit an upward trend with wavelength, increasing from about
$0.3\%$ at blue wavelengths to $\sim$1.0\% at the red edge ($\lambda =
9600$~\AA); a similar but smaller trend exists in the data for SN~2002bf.
Neither object shows significantly different polarization in the ``continuum''
region 6800--7800~\AA\ from that observed in the heavily line-blanketed
region below 5000~\AA\ (\S~\ref{sec:2.3.1}), suggesting that the intrinsic
continuum polarization is quite low for both objects.

The salient features of both data sets are the extraordinarily large
polarization modulations ($\sim$2\%) across certain P-Cygni lines, most
notably the \ion{Ca}{2} near-IR triplet for SN~2002bf and the \ion{Si}{2}
$\lambda 6355$ line for SN~2004dt.  Smaller features may be discerned across
other lines in both objects.

Before attempting ISP removal, it is instructive to compare our SN~2004dt data
with those obtained by \citet{Wang05} taken eleven days earlier, when the SN
was about seven days before maximum light.  The interval separating the two
observations is one marked by rapid spectral and photometric evolution.  For
SN~2004dt, the velocity of the minimum of the high-velocity \ion{Si}{2}
$\lambda 6355$ line recedes by about $3700\ \Kms$ (from $17,200$ to $13,500\
\Kms$), and the weaker, lower-velocity lines such as \ion{S}{2} $\lambda\lambda
5612, 5654$ ($\lambda_0 = 5635$~\AA) decrease by about $2100\ \Kms$ (from
$11,000$ to $8900\ \Kms$).  Remarkably, in spite of the great changes that
occur in the flux spectrum between the two epochs, we find the
spectropolarimetry to be virtually unchanged.  The overall polarization, in
both magnitude and polarization angle, is consistent with the earlier epoch,
and the detected line-polarization features, most notably those at \ion{Si}{2}
$\lambda 6355$ and the \ion{Ca}{2} near-IR triplet, are also very similar.  As
with the \citeauthor{Wang05} data, our observations do not show significant
polarization modulation across the important \ion{O}{1} $\lambda 7774$ line,
although this line is now much weaker in the total-flux spectrum than it was
earlier.  Weaker polarization features, which are clearly detected in the
earlier epoch, are difficult to confirm in our lower S/N data, but our data are
not inconsistent with the earlier measurements, within the errors.

As discussed in \S~\ref{sec:2.3.3}, \citet{Wang05} explain the
spectropolarimetry of SN~2004dt in terms of optically thick bubbles of IMEs
that are asymmetrically distributed within an essentially spherical oxygen
substrate that remains from the progenitor material.  \citet{Wang05} further
note that the polarization behavior of the high-velocity lines is similar to
the low-velocity lines in the early epoch, implying that the structures that
obscure the photosphere have great radial extent. Within this context, then, it
may not be surprising that the polarization features have not evolved
significantly between the two epochs, despite the great evolution of line
velocity.

In order to directly compare the spectropolarimetry of SN~2002bf and SN~2004dt
with each other, we must attempt to remove the ISP.  To establish the ISP, we
apply the technique of \citet{Wang05} and choose the spectral regions indicated
in Figures~\ref{fig:6}a and \ref{fig:8}a.  For SN~2002bf, this yields $(q_{\rm
ISP}, u_{\rm ISP}) = (0.01\%, 0.05\%)$, or ISP$_{\rm max} = 0.05\%$ at
$\theta_{\rm ISP} = 39\arcdeg$ at an assumed peak wavelength of the ISP of
$\lambda_{\rm max} = 5500$~\AA.  For SN~2004dt we derive $(q_{\rm ISP}, u_{\rm
ISP}) = (0.3\%, -0.2\%)$, which is within the uncertainty of the ISP found by
\citet{Wang05} through this same approach, $(q_{\rm ISP}, u_{\rm ISP}) =
(0.2\%, -0.2\%)$.  In order to facilitate direct comparisons between our data
and those of \citet{Wang05}, we shall adopt their ISP value for our study of
SN~2004dt as well.

After removal of the small amount of ISP for SN~2002bf, we derive a
best-fitting axis with ${\rm PA\ } = 123\arcdeg$ (Fig.~\ref{fig:13}), about
which we calculate the intrinsic RSP and URSP shown in Figure~\ref{fig:15}.  In
a similar way, for SN~2004dt we derive a best-fitting axis of ${\rm PA\ } =
146\arcdeg$, which may be compared with the PA of $\sim$150$\arcdeg$ found by
\citet{Wang05}.  Again, since the two values do not differ significantly, we
adopt the \citeauthor{Wang05} direction for ease of comparison between the two
data sets.  The axis and ISP choices for SN~2004dt are indicated in
Figure~\ref{fig:14}, and the resulting RSP and URSP are shown in
Figure~\ref{fig:15}.

From examination of Figure~\ref{fig:15}, it is clear that the two events have
spectropolarimetric similarities. Both show low overall polarization and
modulations across the \ion{Si}{2} $\lambda 6355$ and \ion{Ca}{2} near-IR
triplet absorptions.  The detailed character of the line polarizations do
differ, however.  For SN~2002bf there is a $2\%$ polarization change across the
\ion{Ca}{2} feature, with a more modest ($\sim$0.4\%) feature detected in the
\ion{Si}{2} line.  (When discussing overall ``polarization change'' across a
line feature without specific reference to either the RSP or URSP, we
effectively mean the quadrature sum of the changes seen in the two parameters.)
For SN~2004dt, the situation is reversed.  In the case of the \ion{Ca}{2} line,
this may be related to its relative strength and velocity in the two spectra,
as both the equivalent width and velocity of the feature in the flux spectrum
are much greater in SN~2002bf than in SN~2004dt.  As discussed earlier
(\S~\ref{sec:4.1}), it is possible that burning to Ca occurred more extensively
in SN~2002bf than it did in SN~2004dt.  The explanation for the \ion{Si}{2}
line disparity, however, is not so obvious, as the lines have similar strengths
and velocities.

Clearly, strong line and weak continuum polarization levels disfavor the simple
ellipsoidal model.  Applying Eq.~(\ref{eqn:1}) to the data for SN~2002bf and
SN~2004dt yields lower bounds on the expected continuum polarization of $p_{\rm
cont} \geq 0.7\%$ and $p_{\rm cont} \geq 1.3\%$, respectively. For both
objects, however, we measure $p < 0.4\%$ in both the observed and
ISP-subtracted data for the spectral region 6800--7800~\AA, which is
largely devoid of line opacity in SNe~Ia near maximum (\S~\ref{sec:2.3.1}).
The ellipsoidal model therefore appears to be ruled out as the explanation for
the polarization characteristics of these objects.

The clumpy-ejecta model, on the other hand, provides a natural explanation for
high line and low continuum polarization levels (\S~\ref{sec:2.3.1}).  In
particular, the GCD scenario can successfully explain (1) the high line
velocities, (2) the large polarization change in the line troughs, (3) the
great radial extent of the obscuring material (i.e., for $m_{\rm atm} = 0.08
M_\odot$, the obscuring pancake spans the range 10,000--21,000 \kms), and (4)
the lack of significant continuum polarization. A remaining challenge is to
explain the lack of any polarization change across the \ion{O}{1} $\lambda
7774$ line, especially in the earlier \citet{Wang05} data for SN~2004dt, when
this line is extremely strong in the total-flux spectrum.  Recent
nucleosynthesis calculations based on multi-dimensional (2D and 3D)
hydrodynamical simulations of the thermonuclear burning phase in SNe~Ia show
that as much as $40\%$ to $50\%$ of the ejected matter in SNe~Ia is unburned
carbon and oxygen \citep{Travaglio04}.  This has led \citet{Wang05} to propose
that the primordial oxygen is nearly spherically distributed, within which
asymmetrically distributed IME clumps, or an absorbing pancake in the GCD
model, are embedded.

With this in mind, an important prediction of the clumpy-ejecta model
(including the GCD scenario) is that the \ion{O}{1} $\lambda 7774$ line should,
in fact, show a polarization change with a polarization angle that differs by
$90^\circ$ from what is observed in the \ion{Si}{2} and \ion{Ca}{2}
lines since its distribution is essentially the inverse of these IMEs.
However, some oxygen is probably also contained in the clumps or pancake, since
it is produced by explosive carbon burning in small quantities and should be
present when such burning products as magnesium exist, whose spectral signature
is unequivocally seen in the spectra.  This would tend to reduce the
polarization level in the \ion{O}{1} $\lambda 7774$ line.  Neither our data,
nor those of \citet{Wang05}, are of sufficiently high S/N to detect such a
change, and it must be left to future, higher-S/N studies focused especially at
early times when the \ion{O}{1} $\lambda 7774$ feature is strong.  Finding such
a PA change would further strengthen the case for the clumpy-ejecta and/or GCD
scenario.

We note that the ejecta-hole model of \citet{Kasen04} is also capable of
producing large line polarization with weak continuum polarization for
sight-lines near to the hole (\S~\ref{sec:2.3.1}).  Arguing against this
model in these cases, though, is that it offers no natural explanation for why
high line velocities should be associated with sight lines near to the hole.
In fact, the \citeauthor{Kasen04} models predict {\it lower} absorption
velocities when viewing down the hole.

In all, then, our study finds that HV~SNe~Ia are, as a group, characterized by
much stronger line-polarization features than are seen in other SN~Ia
varieties.  The ellipsoidal model is incapable of explaining the polarization
characteristics of these objects, whereas the clumpy-ejecta and ejecta-hole
models are more successful.  On balance, the case for clumpy ejecta appears to
be the most convincing explanation, with the GCD model investigated by
\citet{Kasen05} able to reproduce many of the observed spectral and
spectropolarimetric features.

\subsubsection{SN 1997dt}
\label{sec:4.4.2}

\begin{figure}
\hspace{0.5in}
\rotatebox{0}{
\scalebox{0.8}{
\plotone{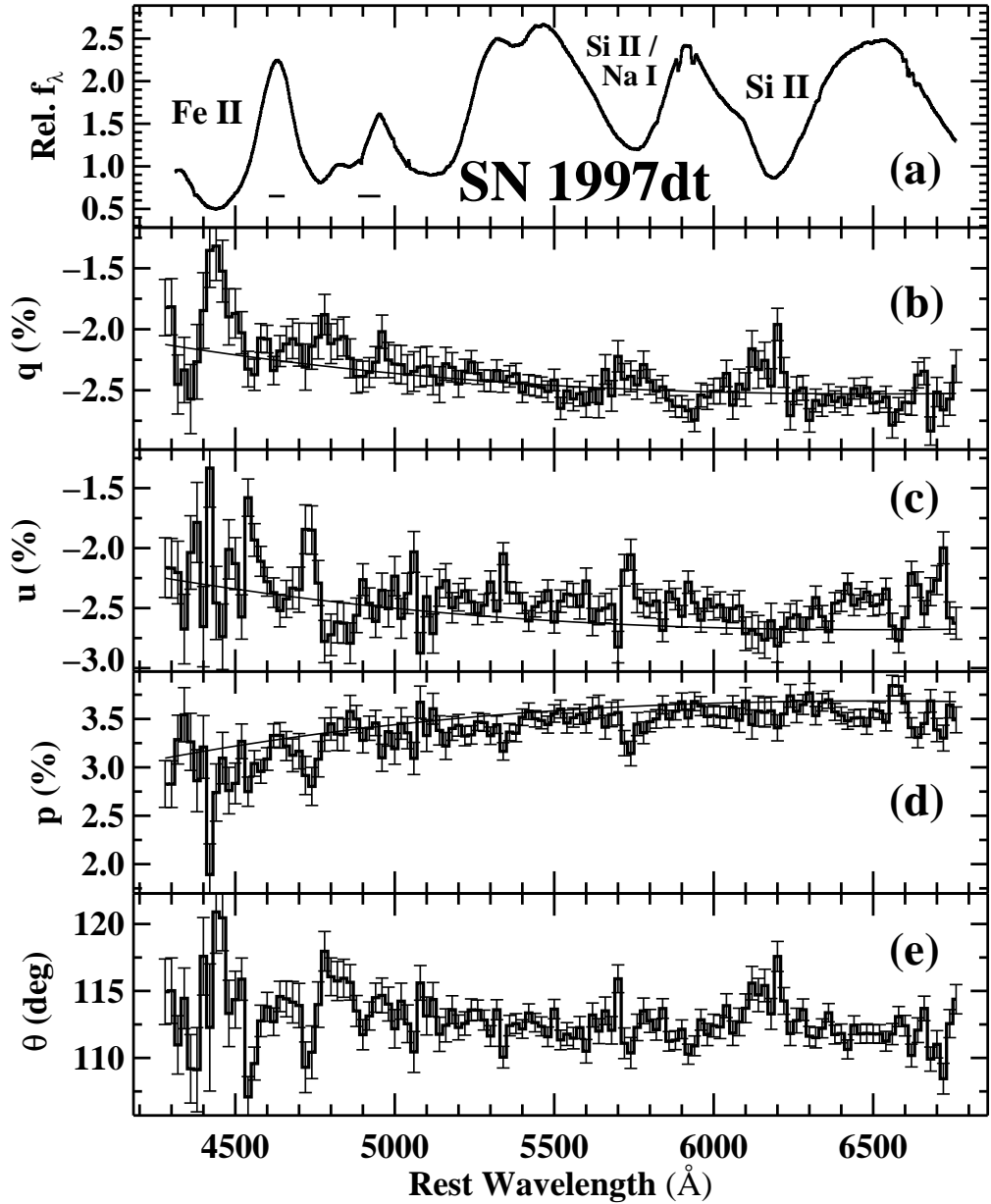}
}}
\vskip 0.2in
\caption{Observed polarization data for SN~1997dt obtained on 1997 Dec. 20,
  when the SN was about 21 days past maximum $B$ brightness.  ({\it a}) Total
  flux, in units of $10^{-15}$ ergs s$^{-1}$ cm$^{-2}$ \AA$^{-1}$, with
  prominent absorption features identified.  Line identifications in this and
  all figures are from \citet{Li01}, and references therein.  {\it Horizontal
  dashes} indicate spectral regions used to determine the ISP.  ({\it b}-{\it
  c}) Normalized $q$ and $u$ Stokes parameters. ({\it d}) Observed degree of
  polarization. ({\it e}) Polarization angle in the plane of the sky.  The
  total-flux spectrum is shown at 2~\AA\ bin$^{-1}$, whereas the polarization
  data are binned at 20~\AA\ bin$^{-1}$ to improve the S/N.  The displayed
  polarization $p$ is the ``rotated Stokes parameter'' in this and all figures;
  see discussion in \S~\ref{sec:4}.  The {\it thin lines} in {\it b, c}, and
  {\it d} show a Serkowski law ISP characterized by the parameters derived in
  \S~\ref{sec:4.4.2}.  Note that the NASA/IPAC Extragalactic Database (NED)
  recession velocity of $2194\ \Kms$ for NGC~7448 has been removed in this and
  all figures showing spectra of SN~1997dt.
\label{fig:5} }
\end{figure}

\begin{figure}
\hspace{0.5in}
\rotatebox{0}{
\scalebox{0.8}{
\plotone{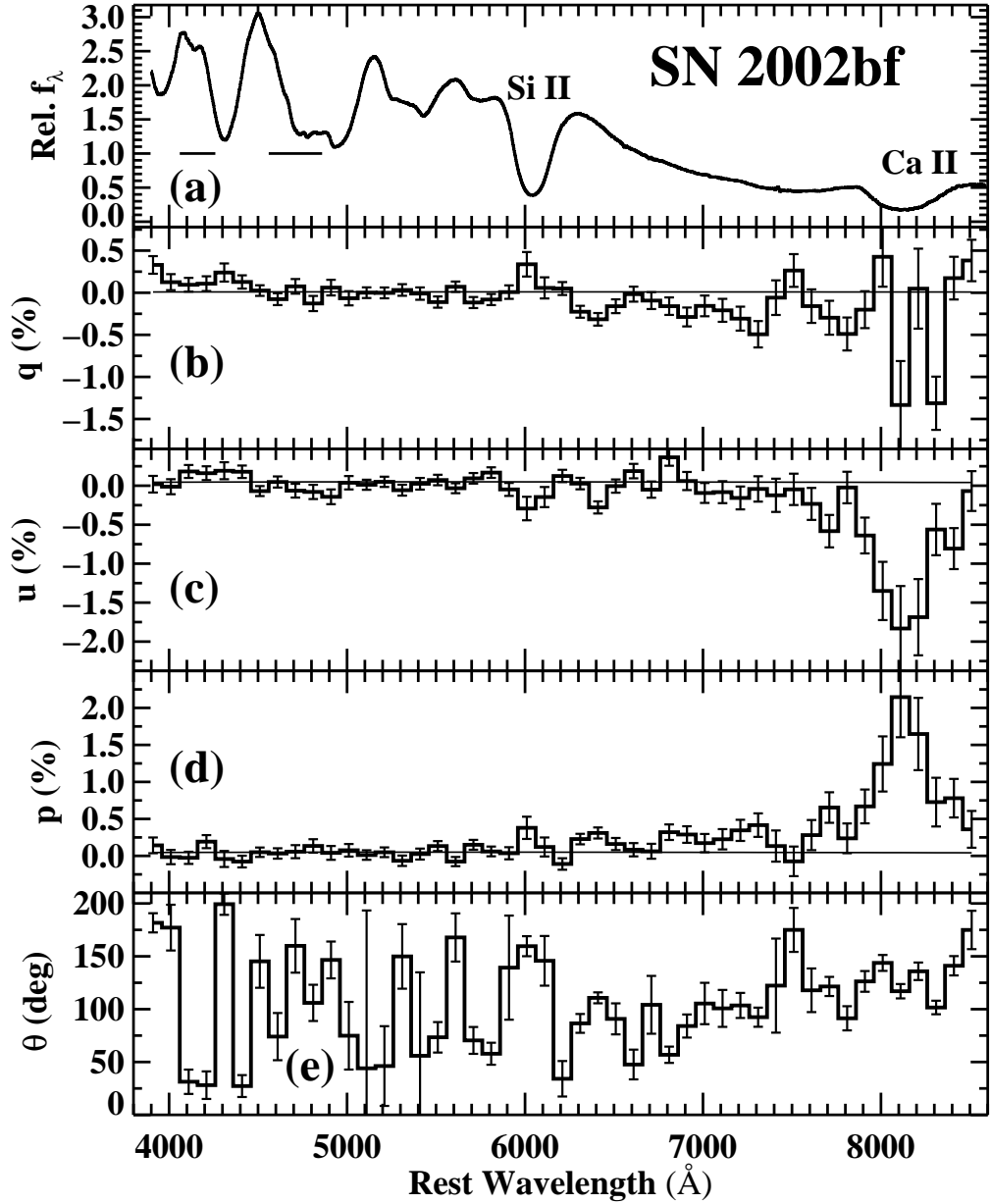}
}}
\vskip 0.2in
\caption{As in Figure~\ref{fig:5}, except for SN~2002bf obtained 2002 Mar. 7,
  when the SN was 3 days past maximum $B$ brightness.  The total-flux spectrum
  is shown at 2~\AA\ bin$^{-1}$, whereas the polarization data are binned at
  100~\AA\ bin$^{-1}$ to improve the S/N.  Note than many of the large jumps
  seen in $\theta$ are an artifact of the data going through the origin in the
  $q$--$u$ plane.  The {\it thin lines} in {\it b, c}, and {\it d}
  show a Serkowski law ISP characterized by the parameters derived in
  \S~\ref{sec:4.4.2}.  The NED recession velocity of $7254\ \Kms$ for
  PGC~029953 has been removed in this and all figures displaying spectra of
  SN~2002bf.  
\label{fig:6} }
\end{figure}

\begin{figure}
\hspace{0.5in}
\rotatebox{0}{
\scalebox{0.9}{
\plotone{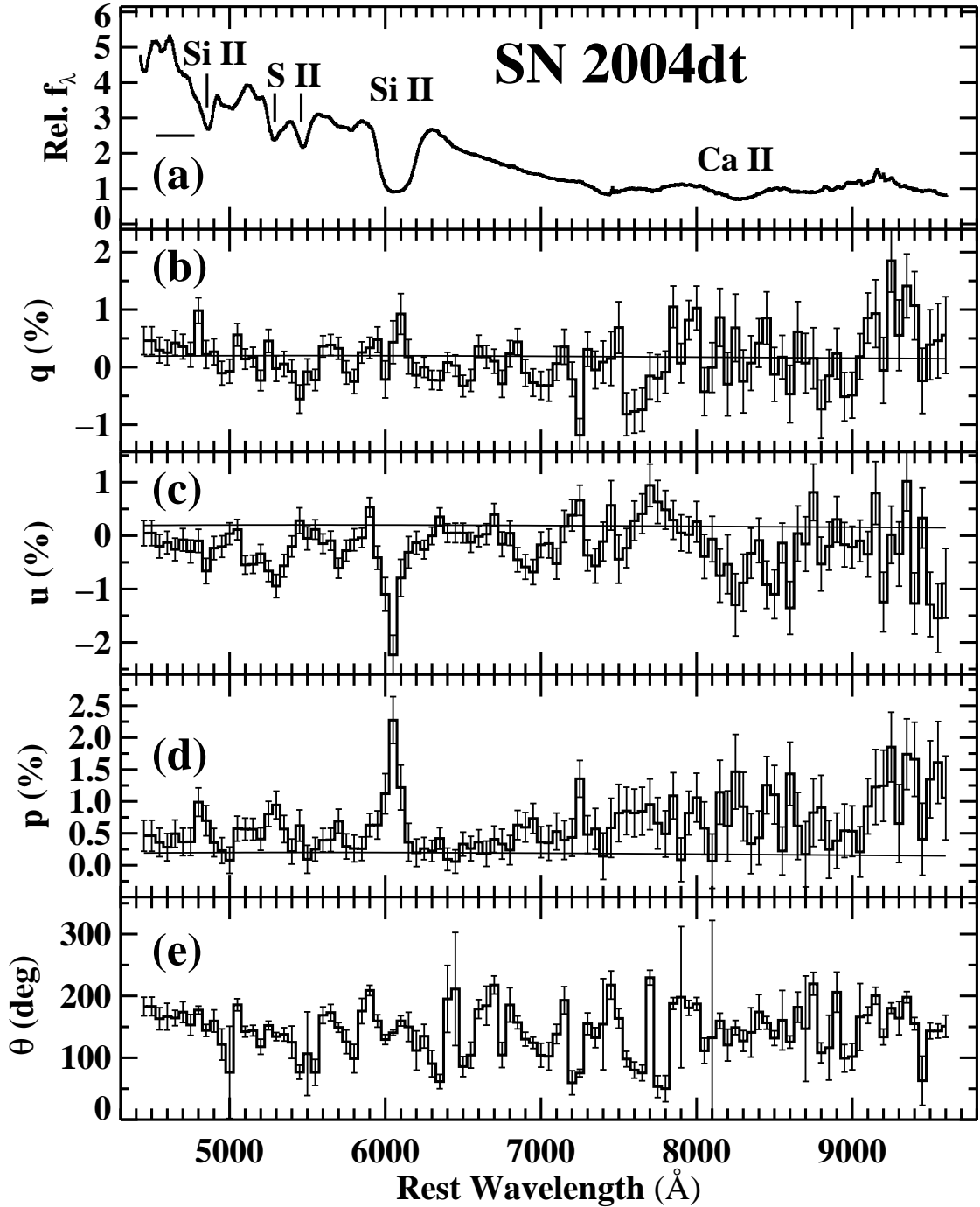}
}}
\vskip 0.2in
\caption{As in Figure~\ref{fig:5}, except for SN~2004dt obtained 2004 Aug. 24,
  about $4$ days after $B_{\rm max}$.  The total-flux spectrum is shown at
  5~\AA\ bin$^{-1}$, whereas the polarization data are binned at 50~\AA\
  bin$^{-1}$ to improve the S/N.  The {\it thin lines} in {\it b, c}, and {\it
  d} show a Serkowski law ISP characterized by the parameters derived in
  \S~\ref{sec:4.4.2}.  The NED recession velocity of $5915\ \Kms$ for NGC 799
  has been removed in this and all figures displaying spectra of SN~2004dt.
\label{fig:8} }
\end{figure}

\begin{figure}
\hspace{0.5in}
\rotatebox{0}{
\scalebox{0.8}{
\plotone{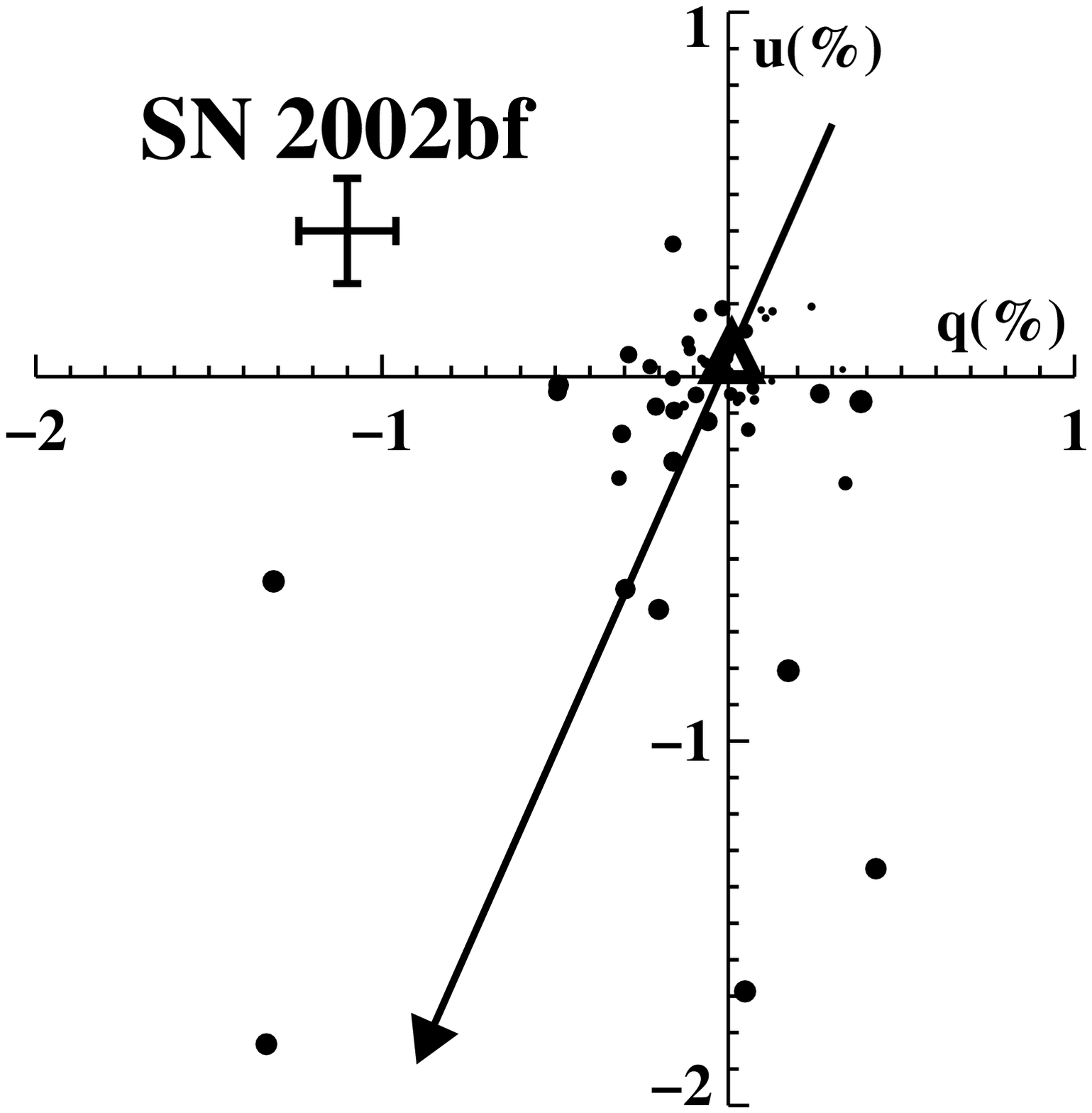}
}}
\vskip 0.2in
\caption{Observed polarization data for SN~2002bf from Figure~\ref{fig:6}
  plotted in the Stokes $q$--$u$ plane.  Each point represents a bin 100~\AA\
  wide, with the symbol size increasing from the blue end of the spectrum
  ($\lambda \approx 3900$~\AA) to the red ($\lambda \approx 8500$~\AA).  The
  {\it error bars} show the typical statistical uncertainty of a data point.
  The large, {\it open triangle} near the origin indicates the adopted ISP, and
  the thick {\it solid line} indicates the axis about which the intrinsic RSP
  and URSP were calculated following ISP removal, with the {\it arrowhead}
  defining the (arbitrarily chosen) direction of increasing RSP.
\label{fig:13} }
\end{figure}

\begin{figure}
\hspace{0.5in}
\rotatebox{0}{
\scalebox{0.8}{
\plotone{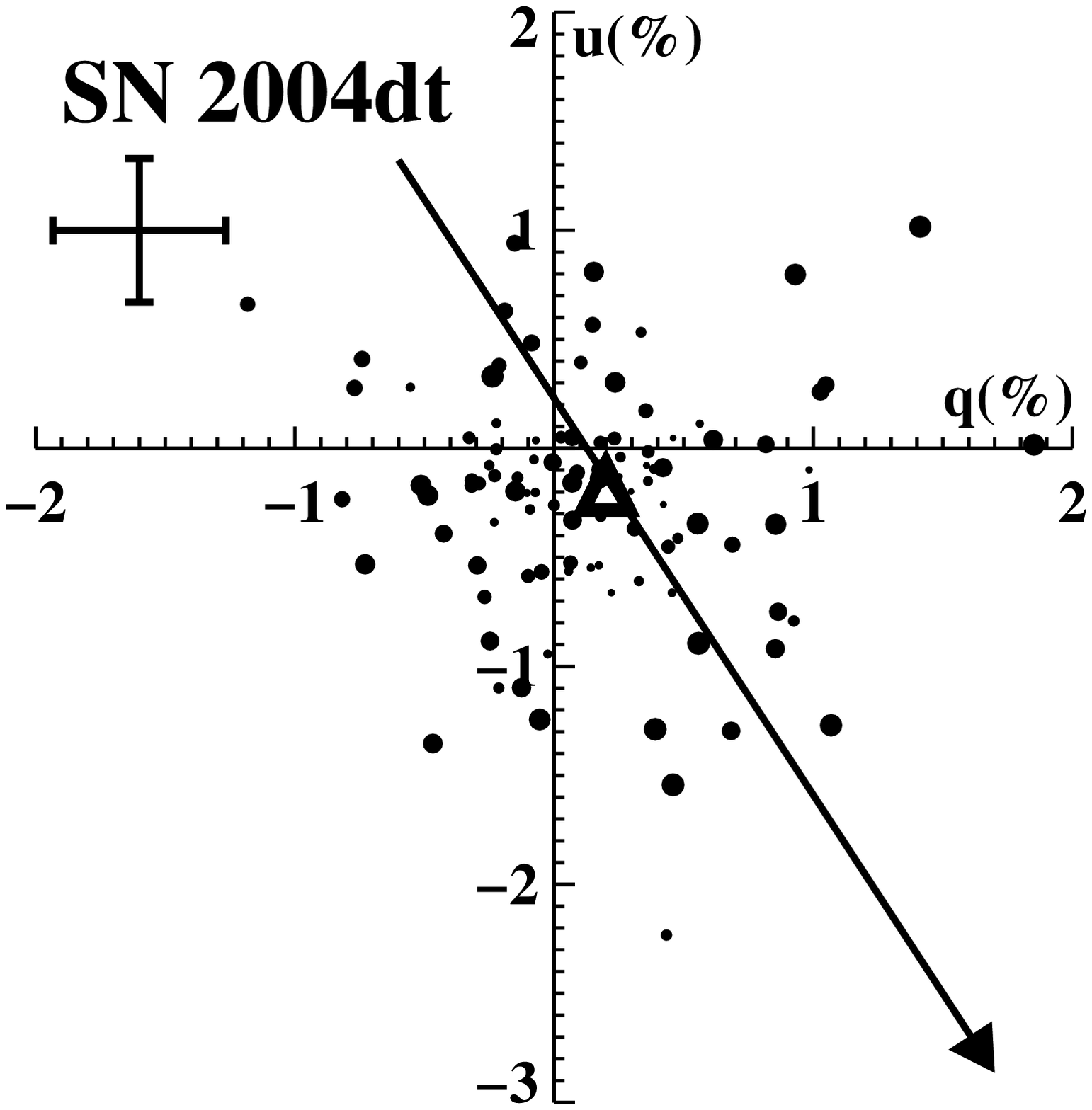}
}}
\vskip 0.2in
\caption{Polarization data for SN~2004dt from Figure~\ref{fig:8} plotted in
  the Stokes $q$--$u$ plane.  Each point represents a bin 50~\AA\ wide, with the
  symbol size increasing from the blue end of the spectrum ($\lambda \approx
  4400$~\AA) to the red ($\lambda \approx 9600$~\AA).  The {\it error bars}
  show the typical statistical uncertainty of a data point.  The large, {\it
  open triangle} indicates the adopted ISP, and the thick {\it solid line}
  indicates the axis about which the intrinsic RSP and URSP were calculated
  following ISP removal, with the {\it arrowhead} defining the (arbitrarily
  chosen) direction of increasing RSP.
\label{fig:14} }
\end{figure}

\begin{figure}
\hspace{0.5in}
\rotatebox{0}{
\scalebox{0.9}{
\plotone{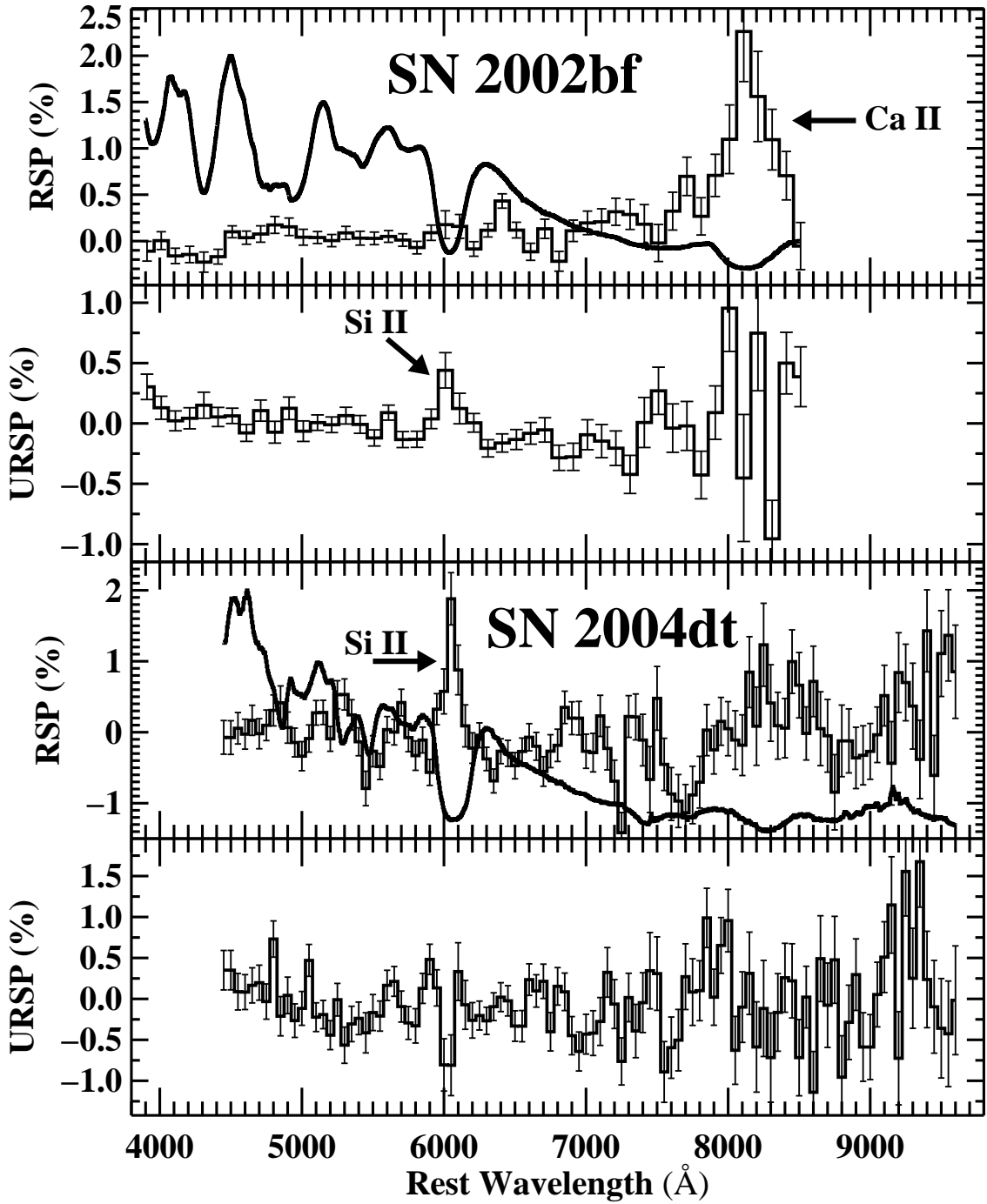}
}}
\vskip 0.4in
\caption{Polarization data for the HV~SNe~Ia SN~2002bf and SN~2004dt following
  ISP removal and calculation of the RSP and URSP with respect to the axes
  indicated in Figures~\ref{fig:13} and \ref{fig:14}, respectively.  The thick,
  {\it smooth lines} without error bars in the RSP plots are the total-flux
  spectra, arbitrarily scaled and shifted for comparison of features.
\label{fig:15} }
\end{figure}

We next turn to the likely subluminous SN~1997dt, which was observed about
three weeks past maximum light.  Figure~\ref{fig:5} reveals an extraordinarily
high level of observed polarization, $p_V = 3.46\%$ at $\theta_V = 112^\circ$.
This is by far the largest polarization yet observed for an SN~Ia.  A distinct
polarization feature is detected in the \ion{Fe}{2} $\lambda 4555$ trough and
probably also in the \ion{Si}{2} $\lambda 5972$ + \ion{Na}{1} D and \ion{Si}{2}
$\lambda 6355$ lines.  The degree of change in the \ion{Fe}{2} $\lambda 4555$
feature reaches nearly $1\%$ in the $q$ parameter.

This amount of observed continuum polarization is surprising, given that our
best total reddening estimate predicts an upper bound on the ISP of only
$1.89\%$ from the \citet{Serkowski75} relation (see \S~\ref{sec:2.3.2}).  If we
trust both the reddening estimate and the upper ISP bound, then an intrinsic SN
polarization of at least $1.57\%$ must exist to explain the observed
polarization, far higher than has been indicated for any previous SN~Ia.
However, the technique of \citet{Wang05} suggests a much higher ISP level:
$(q_{\rm ISP}, u_{\rm ISP}) = (2.53\%, 2.68\%)$, or ISP$_{\rm max} = 3.60\%$,
$\theta_{\rm ISP} = 113\arcdeg$, for $\lambda_{\rm max} = 6500$~\AA, the
wavelength of maximum ISP that yields the most convincing Serkowski-law fit to
the observed polarization (Fig.~\ref{fig:5}d).  If this truly is the ISP, then
it implies an extraordinarily high polarization efficiency for the dust along
the l-o-s to SN~1997dt: ${\rm ISP} / E_{B-V} \approx 18\%\ {\rm mag}^{-1}$,
which is double the observed Galactic limit of $9\%\ {\rm mag}^{-1}$.  Of
course, some of this discrepancy could be removed if the true reddening were
greater than we suspect.  However, even allowing the reddening to equal the
upper limit of $\Ebv_{\rm total} < 0.27$ mag set in \S~\ref{sec:4.3.2} still
requires the dust-polarization efficiency to exceed the Galactic limit.

It thus appears that we face a stark choice: either SN~1997dt has the highest
intrinsic polarization of any SN~Ia yet observed, or the dust along the l-o-s
has an exceptionally high polarization efficiency.  The epoch of our
observation of SN~1997dt is unique for a subluminous SN~Ia, so there is no
empirical database from which to draw expectations and help decide between
these two options.

The situation for SN~1997dt, while perplexing, in fact is not unique: a similar
palette of possibilities presented themselves in our previous study of a single
epoch of spectropolarimetry of SN~1999gi, an SN~II-P that also had a low
reddening and large observed polarization \citep{Leonard4,Leonard6}.  Like
SNe~Ia, SNe~II-P as a group have historically shown very low intrinsic
continuum polarization.  After considering many polarization production
mechanisms, including polarization due to newly formed dust in the SN ejecta
and dust reflection by one or more off-center dust blobs external to the SN, we
concluded that the most likely explanation for the polarization of SN~1999gi
was that the host-galaxy dust along the l-o-s possesses a very high
polarization efficiency, ${\rm ISP}/\EBV = 31^{+22}_{-9}\% {\rm\ mag}^{-1}$,
which remains the largest value yet inferred for a single sight line in either
the MW or an external galaxy \citep{Leonard7}.

There are arguments favoring a similar explanation here.  First, a Serkowski
law reasonably fits the observed continuum polarization of SN~1997dt
(Fig.~\ref{fig:5}d).  Further, at 21 days past maximum, SN~1997dt may be
nearing the end of its photospheric phase, a time when spectropolarimetry may
be losing its efficacy as an asymmetry probe due to a lack of electrons
available to scatter the light.  If we believe this to be the case, then we
should not expect large polarization, even if the SN is highly aspherical.  If
we accept the large ISP, then we naturally would like to know whether it is due
to dust in the MW or NGC~7448.  Unfortunately, we have not observed any distant
Galactic ``probe'' stars (\S~\ref{sec:2.3.2}) near to the l-o-s to estimate the
Galactic ISP.  There are, however, reasons to suspect that it is low.  First,
20 stars within $10\arcdeg$ of the l-o-s are listed in the agglomeration of
stellar polarization catalogs by \citet{Heiles00}, and the greatest observed
polarization is only $0.3\%$.  Second, the great majority of the reddening is
due to host-galaxy dust, since $\Ebv_{\rm Host} = 0.15$ mag while $\Ebv_{\rm
MW} = 0.06$ mag (\S~\ref{sec:4.3.2}).  As was the case with SN~1999gi, we would
again conclude that it is the dust within the host galaxy that must have the
extraordinarily high polarization efficiency.

There are arguments to oppose this, however.  Previous SN~Ia polarization
studies have found intrinsic polarizations increasing toward red wavelengths
(\S~\ref{sec:2.3.3}), which could certainly mimic a Serkowski law over the
limited wavelength band covered by our observations.  Further, the line
polarization features demonstrate that the SN must possess at least some
intrinsic polarization.  Finally, the relatively late phase of the observation,
invoked previously to argue {\it against} high intrinsic polarization, can also
be used to argue {\it in favor} of it: at the stage immediately before an SN
begins the transition to the nebular phase, the deepest layers of the ejecta
are revealed.  Although in need of confirmation by detailed modeling, for this
epoch one can plausibly argue that an optical photosphere still exists with
sufficient optical depth to electron scattering, perhaps even reaching the
single-scattering limit, which is the most polarizing atmosphere
\citep{Hoflich91}.  If the explosion mechanism itself is asymmetric, the
largest imprint of the asymmetry would presumably be in the innermost layers,
which could lead to a very large intrinsic polarization that reveals itself
just at this late phase. 

In fact, such an effect is seen in the
spectropolarimetry of the SN~II-P 2004dj as it transitions to the nebular phase
(D. C. Leonard et al., in preparation). However, in the case of SN~2004dj, the
strong increase is observed primarily in the ``continuum'' region
6800--7800~\AA\ (and in a few strong line troughs), not in the overall level
across the whole spectrum.  Although in need of confirmation by detailed
modeling, depolarizing line blanketing may also be significant for SNe~Ia at
blue wavelengths at this epoch, which would again argue for a large ISP as the
explanation of the high observed polarization.  One possibility that
circumvents this difficulty is that asymmetrically distributed concentrations
of radioactive Ni, recently uncovered in the thinning ejecta at this relatively
late epoch, are responsible for the large polarization.

Curiously, the single, high-polarization observation of SN~1999gi
occurred at a similar stage of its evolution, right at the end of the
optical plateau that characterizes the photospheric phase in SNe~II-P.
It is thus unfortunate that no other spectropolarimetric epochs were
obtained for either event (SN 1999gi and SN 1997dt) to serve as a
basis for comparison.  Multi-epoch data covering the transition from
the photospheric to the nebular phases of SNe of all types will
certainly help reveal more definitively the physical explanation for
such high observed polarizations at these late times.

\begin{figure}
\hspace{0.5in}
\rotatebox{0}{
\scalebox{0.8}{
\plotone{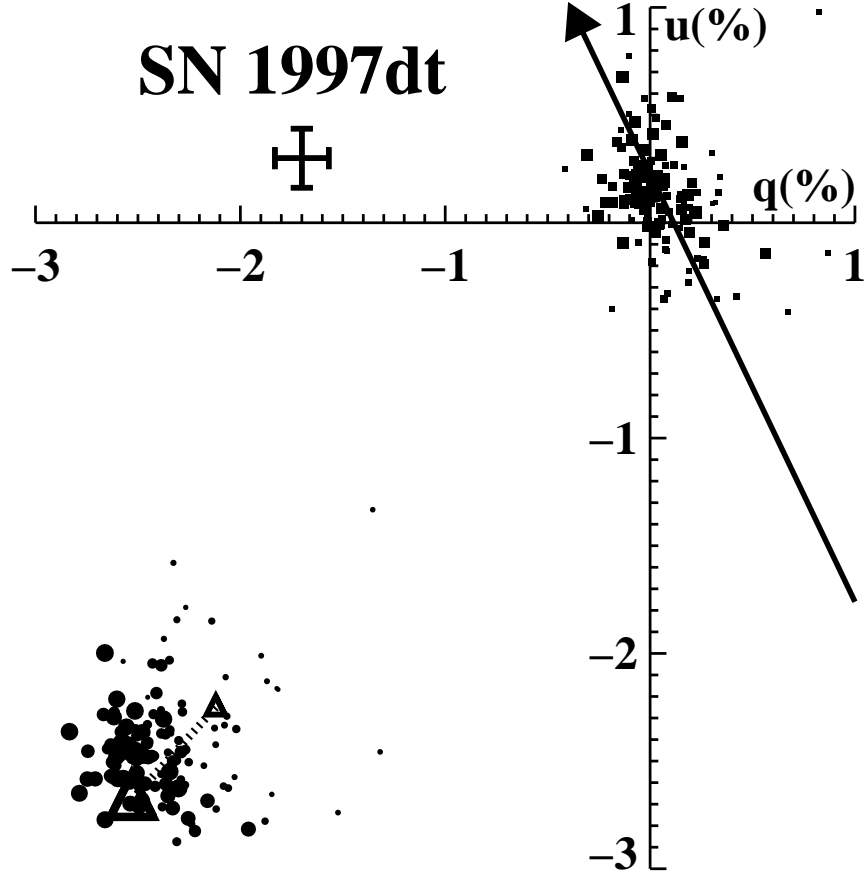}
}}
\vskip 0.2in
\caption{Polarization data for SN~1997dt from Figure~\ref{fig:5} plotted in the
  Stokes $q$--$u$ plane, where {\it filled circles} represent the observed
  polarization, and {\it filled squares} represent the polarization data after
  removal of the ISP derived in \S~\ref{sec:4.4.2}.  Each point represents a
  bin 20~\AA\ wide, with the symbol size increasing from the blue end of the
  spectrum ($\lambda \approx 4300$~\AA) to the red ($\lambda \approx 6750$~\AA).  
  The two {\it open triangles} that are connected by a dotted line
  indicate the range of the ISP for our adopted ISP$_{\rm max}$ of $3.60\%$ at
  $\theta = 113\arcdeg$ for $\lambda_{\rm max} = 6500$~\AA\ over the observed
  wavelength range; the smaller triangle is the ISP at the blue edge of the
  spectrum and the larger triangle is the ISP at $\lambda_{\rm max}$, which is
  very near the red edge of our spectrum.  The {\it error bars} show the
  typical statistical uncertainty of a data point.  The thick {\it solid line}
  indicates the axis about which the intrinsic RSP and URSP were calculated
  following ISP removal.  The (arbitrarily chosen) direction of increasing RSP
  is indicated by the {\it arrowhead}.
\label{fig:16} }
\end{figure}

\begin{figure}
\hspace{0.5in}
\rotatebox{90}{
\scalebox{0.7}{
\plotone{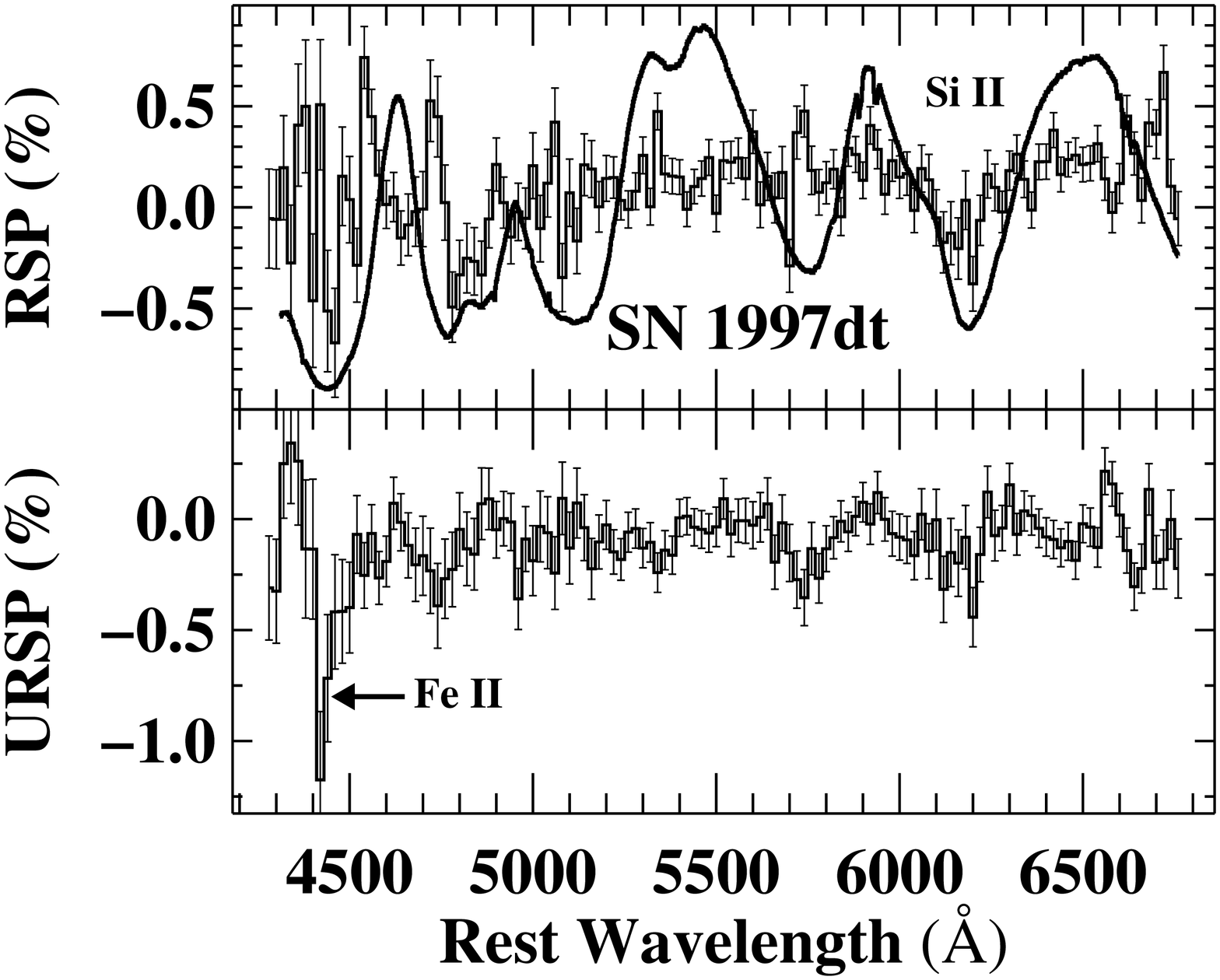}
}}
\vskip 0.4in
\caption{Polarization data for SN~1997dt following ISP removal and calculation
  of the RSP and URSP with respect to the axis indicated in Figure~\ref{fig:16}.
  The thick {\it smooth line} in the RSP plot is the total-flux spectrum,
  arbitrarily scaled and shifted for comparison of features.
\label{fig:17} }
\end{figure}

It may be possible to gain insight into the cause of the observed polarization
from examination of the spectropolarimetry data in the $q$--$u$ plane, shown in
Figure~\ref{fig:16}.  ISP originating from a single source (i.e., characterized
by a single PA) will spread intrinsically unpolarized data points along a line
in a direction that intersects the origin in the $q$--$u$ plane.  Finding that
data lie predominantly along such a line, and exhibit a Serkowski law spectral
shape, provides compelling evidence that a large, single source of ISP
dominates the observed signal.  Such was the case for SN~II-P 1999gi
\citep{Leonard7}.  Similarly, for SN~1997dt, an elongation of the data points
from blue to red wavelengths in a direction that roughly points back toward the
origin also exists. 

When the ISP derived earlier (ISP$_{\rm max} = 3.60\%$, $\theta_{\rm ISP} =
113\arcdeg$, for $\lambda_{\rm max} = 6500$~\AA) is removed, the wavelength
dependence of the polarization largely disappears (Fig.~\ref{fig:16}),
strengthening the argument for a single, dominant ISP source.  The data are
then seen to lie along an axis with a fairly well-defined PA of $58\arcdeg$
(Fig.~\ref{fig:16}), against which we calculate the RSP and URSP shown in
Figure~\ref{fig:17}.  Although of individually low to moderate significance,
the polarization features in the \ion{Fe}{2} $\lambda 4555$ and \ion{Si}{2}
$\lambda 6355$ troughs, seen in both RSP and URSP, do seem to prefer the same
general direction, with most of the modulation occurring along the URSP axis
(e.g., the direction perpendicular to the main axis in the $q$--$u$ plane).
Taken at face value, consistent line-trough polarization changes in a direction
differing from that preferred by the continuum favors an origin in the
selective blocking of photospheric light by clumpy and asymmetrically
distributed intermediate and iron-peak elements overlying the photosphere as
opposed to an ellipsoidal scenario.  

Other plausible ISP choices, however, yield different conclusions.  For
instance, a smaller ISP of $2.6\%$ at the same PA yields a more nearly
spherical constellation of points centered near $(q, u) = (-0.7\%, -0.07\%)$,
with the line excursions now pointing back toward the origin.  Since some
modelers predict that, in SNe~Ia with an ellipsoidal asphericity, polarization
{\it decreases} may exist in absorption troughs (\S~\ref{sec:2.3.1}), the
ellipsoidal model cannot therefore be ruled out in this case as the cause of
the polarization.  Given the marginal significance of the weaker features,
additional interpretation of the line-trough polarization degrees and
directions is probably not warranted with these data.

In conclusion, we find evidence for a large ISP contribution to the observed
polarization of SN~1997dt, probably in the range $2.6\% \lesssim {\rm ISP}
\lesssim 3.6\%$.  This implies a polarization efficiency for the dust along the
l-o-s in NGC~7448 that exceeds the Galactic limit.  We do, however, also find
evidence for polarization intrinsic to the object, most convincingly in
specific line features and, perhaps, in the continuum as well.  A combination
of asymmetrically distributed radioactive Ni and synthesized IMEs overlying the
photosphere may provide the simplest explanation for the line and potential
continuum polarization at this late phase, although the ellipsoidal model 
cannot be definitively ruled out.

\subsubsection{SN 2003du}
\label{sec:4.4.3}

\begin{figure}
\hspace{0.5in}
\rotatebox{0}{
\scalebox{0.9}{
\plotone{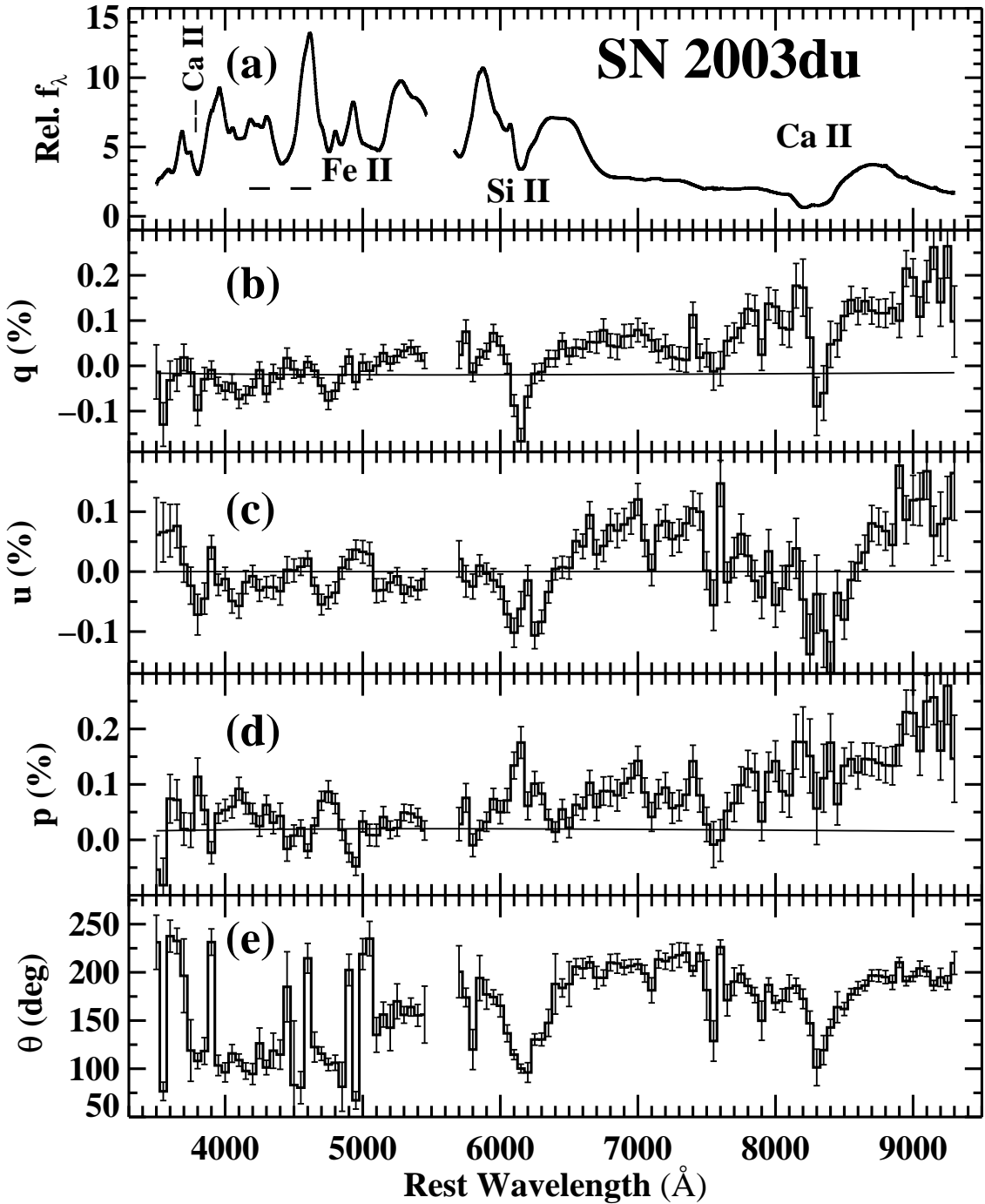}
}}
\vskip 0.2in
\caption{As in Figure~\ref{fig:5}, except for SN~2003du obtained 2003 May 24,
  about $18$ days after $B_{\rm max}$. The total-flux spectrum is shown at
  2~\AA\ bin$^{-1}$, whereas the polarization data are binned at 50~\AA\
  bin$^{-1}$ to improve the S/N.  Note that the region 5400--5700~\AA\ was
  adversely affected by the dichroic used to split the beam, and thus could not
  be calibrated with certainty.  The {\it thin lines} in {\it b, c}, and {\it
  d} show a Serkowski law ISP characterized by the parameters derived in
  \S~\ref{sec:4.4.2}.  The NED recession velocity of $1914\ \Kms$ for UGC 9391
  has been removed in this and all figures displaying spectra of SN~2003du.
\label{fig:7} }
\end{figure}

Our spectropolarimetry of SN~2003du represents the highest S/N data of our
study, and permits a more detailed analysis of line features than has been
possible for the other objects.  SN~2003du presents a low level of observed
polarization across most of the spectrum, with $p_V = 0.04\%$, $\theta_V =
17^\circ$ (Fig.~\ref{fig:7}).  The polarization exhibits an increasing trend
with wavelength, rising from nearly zero at blue wavelengths to $\sim$0.2\% at
the red edge of the spectrum.  There are distinct and significant polarization 
changes across several absorption features, including the \ion{Ca}{2} near-IR 
triplet, the \ion{Si}{2} $\lambda 6355$ line, probably a few weaker lines such 
as \ion{Fe}{2} $\lambda 4924$, and very tentatively the \ion{Ca}{2} H \& K
absorption.

Applying the technique of \citet{Wang05}, and choosing the regions indicated in
Figure~\ref{fig:7}a to estimate the ISP, yields $(q_{\rm ISP}, u_{\rm ISP}) =
(-0.02\%, 0.0\%)$, or ISP$_{\rm max} = 0.02\%$, $\theta_{\rm ISP} = 90\arcdeg$,
for an assumed $\lambda_{\rm max} = 5500$~\AA.  This very low ISP is
consistent with the negligible reddening found earlier.  Furthermore, on the
same night the SN~2003du data were taken, we observed the
distant Galactic star BD~$+50^\circ1593$ (spectral type F8, $V = 10.64$ mag),
located just $0.28^\circ$ from the l-o-s of SN~2003du, and found it to be null
to within $0.1\%$.  From its spectroscopic parallax, we estimate BD~$+50^\circ1593$
to be at least $190$ pc away which, at a Galactic latitude of $53^\circ$,
satisfies the criterion of \citet{Tran95} that a good MW ``probe'' star be more
than 150 pc from the Galactic plane.  We thus have multiple reasons to suspect
little ISP contaminating the data.

\begin{figure}
\hspace{0.5in}
\rotatebox{0}{
\scalebox{0.8}{
\plotone{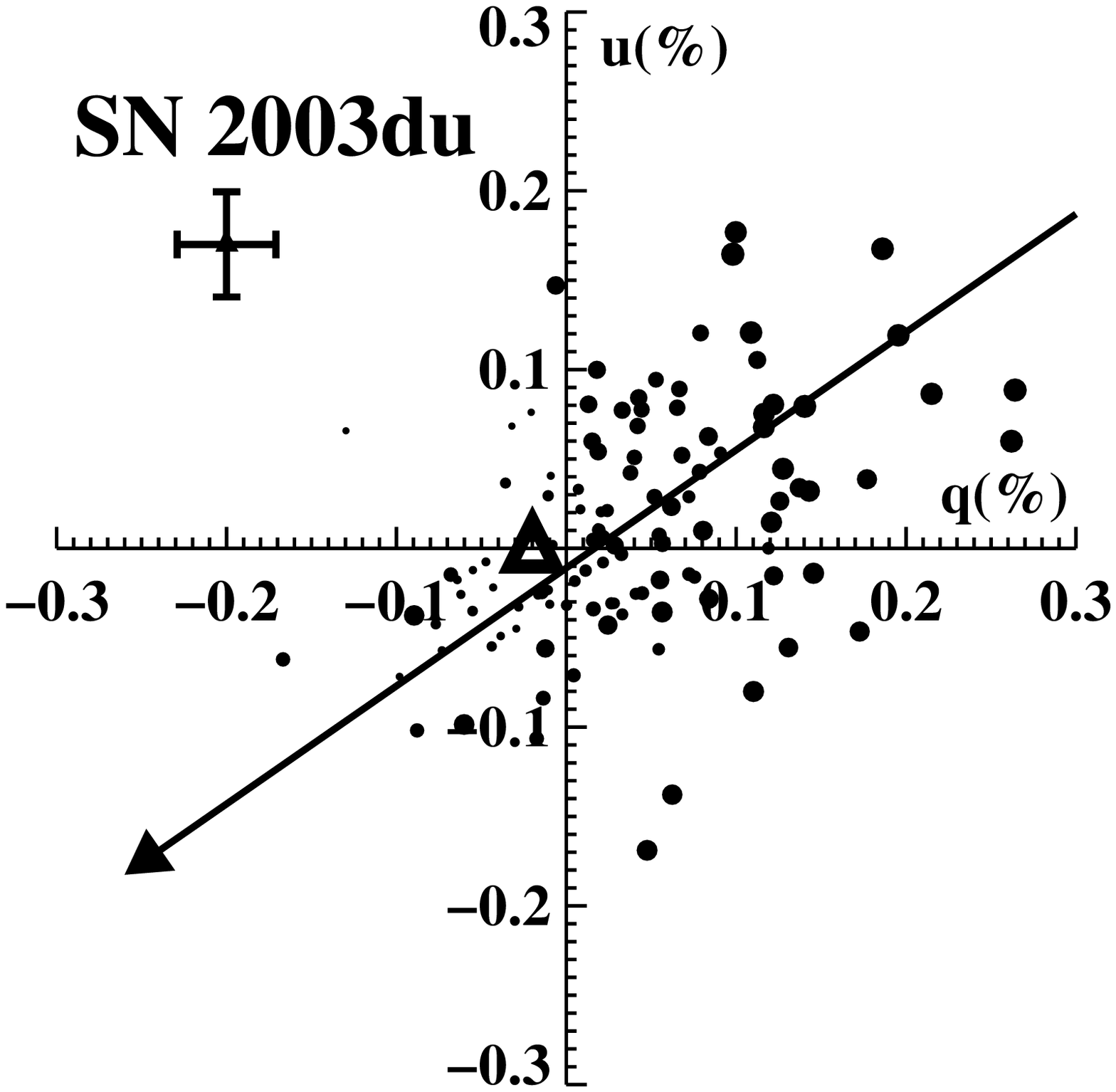}
}}
\vskip 0.2in
\caption{Polarization data for SN~2003du from Figure~\ref{fig:7} plotted in the
  Stokes $q$--$u$ plane.  Each point represents a bin 50~\AA\ wide, with the
  symbol size increasing from the blue end of the spectrum ($\lambda \approx
  3500$~\AA) to the red ($\lambda \approx 9300$~\AA).  The {\it error
  bars} show the typical statistical uncertainty of a data point.  The 
  {\it open triangle} indicates the adopted ISP, and the thick {\it solid line}
  indicates the axis about which the intrinsic RSP and URSP were calculated
  following ISP removal, with the {\it arrowhead} defining the arbitrary
  direction of increasing RSP.
\label{fig:18} }
\end{figure}

\begin{figure}
\hspace{0.5in}
\rotatebox{90}{
\scalebox{0.8}{
\plotone{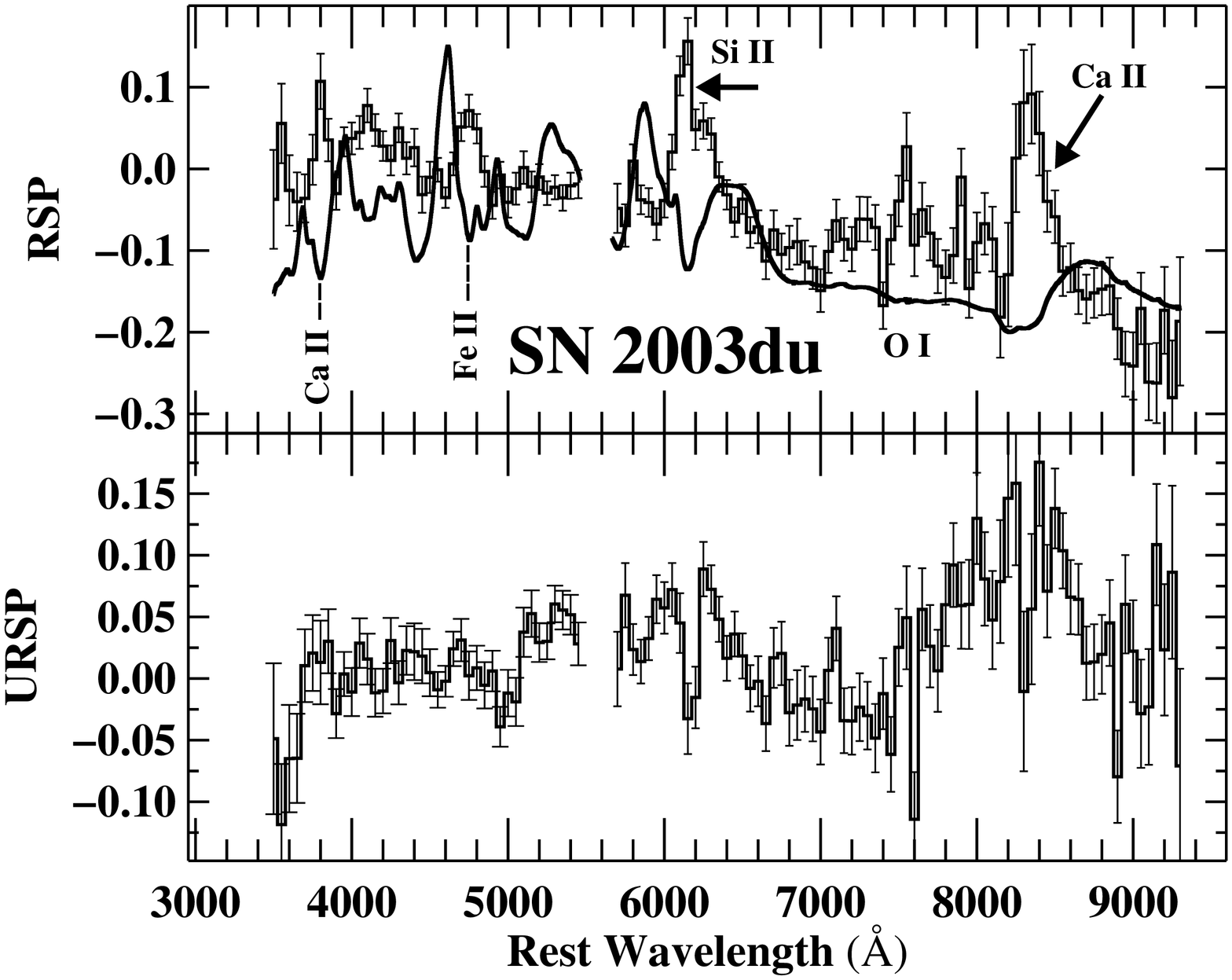}
}}
\vskip 0.2in
\caption{Polarization data for SN~2003du following ISP removal and calculation
  of the RSP and URSP with respect to the axis shown in Figure~\ref{fig:18}.
  The thick {\it smooth line} in the RSP plot is the total-flux spectrum,
  arbitrarily scaled and shifted for comparison of features.
\label{fig:19} }
\end{figure}

After removal of the minimal ISP, a well-defined axis with ${\rm PA\ } =
107\arcdeg$ is derived (Fig.~\ref{fig:18}), about which we calculate the
intrinsic RSP and URSP, shown in Figure~\ref{fig:19}.  Compared with what was
seen in the HV~SNe~Ia, and even SN~1997dt, the line-polarization features are
not large, amounting to no more than $0.3\%$.  However, the very high S/N of
these data makes the detections unequivocal, and establishes intrinsic
polarization in a spectroscopically and photometrically normal SN~Ia at the
latest phase yet observed.

The behavior of the line-trough polarization for the \ion{Ca}{2} near-IR
triplet and \ion{Si}{2} $\lambda 6355$ are very similar.  Both show sharp
increases of $\sim$0.2\% in RSP, as well as overall increases in URSP.  Modest
RSP increases of $\lesssim 0.1\%$ may also be discerned in the \ion{Fe}{2}
$\lambda 4924$ and \ion{Ca}{2} H \& K absorptions.  At this phase, the
\ion{O}{1} $\lambda 7774$ feature in SNe~Ia is quite weak and, coupled with
telluric A-band contamination, makes definitive detection of polarization
modulation in this important region difficult; the observed changes are at
about the level of the statistical noise.

The commonality between the polarization behavior of the Si and Ca lines argues
for similar origins. Given the high S/N of the data, we can examine rather
closely the basic predictions of the ellipsoidal model that (a) the
polarization angle should be independent of wavelength, and (b) the overall
polarization should increase from blue to red wavelengths, with an expectation
of $p \rightarrow 0\%$ at $\lambda \lesssim 5000$ \AA\ (if the line opacity
remains strong in these regions at the epoch of observation;
\S~\ref{sec:2.3.1}).  

With our initial choice of ISP, the first criterion is clearly not satisfied,
as there are obvious excursions in URSP throughout the spectrum, with especially
sharp modulations occurring across the strongest lines.  In fact, for ISP
values constrained to lie along the symmetry axis we are unable to find any
value that convincingly satisfies both criteria of the ellipsoidal model:
values in the upper-right quadrant (e.g., $[q_{\rm ISP}, u_{\rm ISP}] = [0.3\%,
0.18\%]$) produce a polarization {\it decrease} with wavelength across the
spectrum, while ISP values in the lower-left quadrant that satisfy the
criterion of $p \rightarrow 0\%$ at blue wavelengths (e.g., $[q_{\rm ISP},
u_{\rm ISP}] = [-0.1\%, -0.07\%]$) result in strong PA changes across the
spectral lines.  Choosing an ISP point far beyond the constellation of data
points, which may serve to make the PA changes less objectionable (although
still statistically significant), has the unfortunate consequence of straining
the limits implied from the very low reddening, at least for dust with normal
polarizing efficiency.  

In addition, examining the URSP behavior of the \ion{Si}{2} $\lambda 6355$ and
\ion{Ca}{2} near-IR lines in Figure~\ref{fig:19} more closely, we see that the
generally increasing trends in URSP across the lines show sharp decreases right
at the locations of peak RSP modulation, although the statistical significance
of the modulations, especially for \ion{Ca}{2}, is not high.  If the abrupt
changes in URSP are real, such structure is readily explained under the
clumpy-ejecta model by variations in the distribution of the IMEs as a function
of radius in the expanding ejecta.  \citet{Chugai92} also demonstrates that
such sharp changes in line features can be produced by excitation asymmetry.
On the other hand, there is no obvious mechanism to produce such an effect in
the simple ellipsoidal models.  We thus conclude that it is difficult to
reconcile the basic predictions of the ellipsoidal model with the data for
SN~2003du.

We therefore suspect either clumps in the ejecta overlying the photosphere or
ionization asymmetry as the cause of the inferred line and, perhaps, continuum
polarization, and are led again to disfavor the ellipsoidal model as the cause
of the intrinsic polarization of this SN.

\section{Conclusions}
\label{sec:5}

We present post-maximum single-epoch spectropolarimetry of four SNe~Ia,
bringing to six the number of SNe~Ia thus far examined in detail with
spectropolarimetry during the early phases.  The four objects span a range of
spectral and photometric properties, yet all are demonstrated to be
intrinsically polarized.  This suggests that asphericity and/or asymmetry may
be a ubiquitous characteristic of SNe~Ia in the first weeks after maximum
light.  The nature and degree of the polarization varies considerably within
the sample, but in a way that is consistent with, and extends, previously
suspected trends.  Our main spectropolarimetry results are as follows:

\begin{enumerate}

\item SN~2002bf and SN~2004dt, both HV~SNe~Ia observed shortly after maximum
brightness, exhibit the largest polarization features yet seen definitively for
any subtype of SN~Ia, with modulations of up to $\sim$2\% in the troughs of the
strongest lines.  The overall polarization level of both objects is minimal,
at least at blue wavelengths; there is a possible trend of increasing
polarization with wavelength for both objects, though neither shows significant
polarization in the ``continuum'' region 6800--7800~\AA.  The
ISP contamination is not thought to be large in either object.

\item SN~1997dt, believed to be a somewhat subluminous event, has the highest
observed overall polarization of any SN~Ia yet studied, $p_V = 3.46\%$, at 21
days past maximum light.  This demands either an extraordinarily large
polarization efficiency for the dust along the l-o-s in NGC~7448, the largest
intrinsic SN~Ia polarization thus far found, or perhaps some combination of the
two.  The observed polarization rises by about $0.5\%$ from blue ($\lambda =
4300$~\AA) to red ($\lambda = 6700$~\AA) wavelengths, approximating a
Serkowski-law ISP curve rather convincingly, albeit one with a somewhat unusual
peak wavelength ($\lambda_{\rm max} \approx 6500$~\AA).  A polarization
modulation of nearly $1\%$ in the strong \ion{Fe}{2} $\lambda 4555$ absorption,
and a more modest change of $\sim 0.3\%$ in the \ion{Si}{2} $\lambda 6355$
line, demonstrate that the SN does possess intrinsic polarization features.
However, we conclude that ISP is responsible for the bulk of the overall
polarization that is observed, with $2.6\% \lesssim {\rm ISP}
\lesssim 3.6\%$, and that the polarization efficiency of the dust along the
l-o-s in NGC~7448 likely exceeds the empirical Galactic limit.

\item SN~2003du is a slightly overluminous SN~Ia.  Our spectropolarimetry,
  taken 18 days after maximum light, is the highest S/N data obtained for our
  sample of objects.  It reveals a low continuum polarization that increases by
  $\sim$0.3\% from blue to red wavelengths, with distinct changes of $\sim$0.2\% 
  detected in the \ion{Si}{2} $\lambda 6355$ and \ion{Ca}{2} near-IR
  triplet lines; smaller changes may be detected in weaker lines.  The very
  similar behavior of the polarization in the two strongest lines, in both
  magnitude and direction in the $q$--$u$ plane, suggests a common polarization
  origin.  ISP is thought to be minimal.

\end{enumerate}

Ordered by increasing strength of line-polarization features in SNe~Ia, we find
as follows: ordinary/overluminous $<$ subluminous $<$ HV~SNe~Ia, with the
strength of the line-polarization features increasing from $0.2\%$ in the
slightly overluminous SN~2003du to $2\%$ in both HV~SNe~Ia in our study.
Absolute continuum polarization levels are more difficult to establish, due
largely to uncertainties in the ISP, but there are compelling reasons to
believe that at least three of our objects possess very little intrinsic
polarization in spectral regions outside of specific line features.  The
\citet{Howell01} study of SN~1999by and our data on SN~1997dt provide some
evidence that continuum polarization may be higher in subluminous objects than
in other types.

There are a number of alternatives from which to choose for the origin of
polarization in SNe~Ia, including global asphericity (e.g., the ellipsoidal
model), ionization asymmetry, and clumps in the ejecta overlying the
photosphere.  The small, redward rise in the {\it overall} polarization level
that is discerned in at least three of our objects can be reproduced by models
possessing either global asphericity or an ionization asymmetry.  Under the
ellipsoidal model, the levels of continuum polarization for SN~2002bf,
SN~2003du, and SN~2004dt imply minor to major axis ratios of around $0.9$ if
viewed equator-on \citep{Hoflich91,Wang03}.  This level of asphericity would
produce a luminosity dispersion of about 0.1 mag for random viewing
orientations \citep{Hoflich91}, which could explain some of the dispersion seen
in the brightness-decline relation of SNe~Ia.  If the proposed global
asphericity is more complicated, then the luminosity of a single SN~Ia may
depend on viewing angle in a non-trivial way such that, even for a large sample
of objects, an overall bias to slightly higher or lower values may result.

The potential for significantly greater continuum polarization, perhaps of
$\sim$1\%, in the likely subluminous SN~1997dt observed $\sim$21 days after
maximum would imply a more severe distortion, of at least $20\%$, from the
models of \citet{Hoflich91}.  Since most cosmological applications of SNe~Ia
rely on data acquired closer to maximum light, such a late-time asphericity,
even if common, would not seriously affect the utility of SNe~Ia as distance
indicators.

To explain the ubiquitous line polarization, the simple ellipsoidal model is
effectively ruled out for three of our objects, including, most convincingly,
the two HV events.  From a number of lines of reasoning, the most convincing
explanation is partial obscuration of the photosphere by clumps of newly
synthesized IMEs forged in the explosion.

For the HV~SNe~Ia, in particular, the GCD model studied by \citet{Kasen05}
provides a plausible explanation for many of the observed spectral and
spectropolarimetric characteristics.  It predicts the existence of an optically
thick pancake of material with significant radial extent that partially
obscures the optical photosphere, producing larger line velocities and
equivalent widths for many spectral features, and stronger line-trough
polarization than is seen in more typical events.  These qualitative
expectations are borne out by our data: The line features of SN~2002bf and
SN~2004dt possess the strongest polarization modulations and greatest
equivalent widths of our sample.  The astonishing similarity between our epoch
of spectropolarimetry of SN~2004dt, $\sim$4 days after maximum, and that
presented by \citet{Wang05} from 11 days earlier, provides compelling evidence
that the obscuring material also possesses great radial extent in the thinning
ejecta.

That SNe~Ia may be separable into different groups based on their {\it
spectropolarimetric} characteristics yields one more clue to assist in
narrowing down progenitor possibilities and/or models for the physics of the
explosion.  The assertion that some fraction of the IMEs in the ejecta of
SNe~Ia may be confined to bubbles or filaments is, however, a rather blunt
discriminatory tool: At the present stage of theoretical modeling,
deflagration, delayed-detonation, off-center delayed-detonation, and GCD models
can all plausibly be argued to produce clumps in the ejecta \citep[e.g.,][and
references therein]{Wang05}.  The specific predictions of the GCD model in
particular need to be further examined, preferably in full three-dimensional
simulations, to test whether the explosion mechanism itself is viable, and, if
so, whether it can quantitatively reproduce the observed characteristics of at
least some SNe~Ia in detail.  On the observational front, higher S/N data,
preferably obtained at multiple epochs, will help to narrow down the
possibilities as well.  With the steady advances being made in the theoretical
understanding of these events, and the growing rate of SNe~Ia studied in detail
with spectropolarimetry, prospects for improving our understanding of these
events are bright.

\acknowledgments 

We thank Aaron Barth, Louis-Benoit Desroches, Mohan Ganeshalingam, Deborah
Hutchings, Ed Moran, and Karin Sandstrom for assistance with some of the
observations and data reduction, and Saurabh Jha for producing MLCS2k2 fits for
two of our objects. We thank an anonymous referee for useful suggestions that
resulted in an improved manuscript.  This research has made use of the
NASA/IPAC Extragalactic Database (NED), which is operated by the Jet Propulsion
Laboratory, California Institute of Technology, under contract with NASA.  The
work of A.V.F.'s group at UC Berkeley is supported by National Science Foundation
(NSF) grant AST-0307894. D.C.L. is supported by an NSF Astronomy and
Astrophysics Postdoctoral Fellowship under award AST-0401479.  Additional
funding was provided by NASA grants GO-9155, GO-10182, and GO-10272 from the
Space Telescope Science Institute, which is operated by the Association of
Universities for Research in Astronomy, Inc., under NASA contract NAS 5-26555.
A.V.F. is grateful for a Miller Research Professorship at UC Berkeley, during
which part of this work was completed.  Some of the data presented herein were
obtained at the W. M. Keck Observatory, which is operated as a scientific
partnership among the California Institute of Technology, the University of
California, and NASA; the Observatory was made possible by the generous
financial support of the W. M. Keck Foundation.  KAIT was made possible by
generous donations from Sun Microsystems, Inc., the Hewlett-Packard Company,
AutoScope Corporation, Lick Observatory, the NSF, the
University of California, and the Sylvia \& Jim Katzman Foundation. The
assistance of the staffs at Lick and Keck Observatories is greatly appreciated.

\clearpage


\clearpage

\clearpage

\clearpage

\clearpage

\clearpage

\clearpage

\clearpage

\clearpage

\clearpage

\clearpage

\clearpage

\clearpage

\clearpage

\clearpage

\clearpage

\clearpage

\clearpage

\clearpage

\input{tab1.tex}

\input{tab2.tex}
\input{tab3.tex}
\input{tab4.tex}
\input{tab5.tex}

\input{tab6.tex}

\input{tab7.tex}

\end{document}

%% file: tab1.tex


\begin{deluxetable}{lcccc}
\tablewidth{380 pt}
\tablecaption{Magnitudes of Local Standards:  SN 2002bf}
\tablehead{\colhead{Star}  &
\colhead{$B (\sigma_B)$} &
\colhead{$V (\sigma_V)$} &
\colhead{$R (\sigma_R)$} &
\colhead{$I (\sigma_I)$} } 

\startdata

 1 & 12.640 (0.014) &  12.226 (0.011) &  11.986 (0.020)  & 11.709 (0.003)\\
 2 & 17.112 (0.021) &  16.485 (0.010) &  16.116 (0.006)  & 15.776 (0.027)\\
 3 & 17.930 (0.025) &  17.441 (0.021) &  17.143 (0.014)  & 16.792 (0.010 )\\
 4 & 16.305 (0.028) &  14.861 (0.006) &  13.924 (0.016)  & 12.874 (0.024)\\
 5 & 15.624 (0.012) &  14.764 (0.028) &  14.299 (0.025)  & 13.850 (0.003)\\

\enddata
\label{tab:1}

\end{deluxetable}
\clearpage


%% file: tab2.tex
\begin{deluxetable}{lcccc}
\tablewidth{380 pt}
\tablecaption{Magnitudes of Local Standards:  SN 2003du}
\tablehead{\colhead{Star}  &
\colhead{$B (\sigma_B)$} &
\colhead{$V (\sigma_V)$} &
\colhead{$R (\sigma_R)$} &
\colhead{$I (\sigma_I)$} } 

\startdata

 1 & 16.973 (0.013) &  16.291 (0.021) &  15.875 (0.018)  & 15.508 (0.023)\\
 2 & 14.944 (0.025) &  14.329 (0.017) &  13.965 (0.011)  & 13.614 (0.028)\\
 3 & 17.898 (0.022) &  17.025 (0.016) &  16.515 (0.015)  & 16.050 (0.027)\\
 4 & 17.580 (0.032) &  16.903 (0.020) &  16.508 (0.016)  & 16.171 (0.018)\\
 5 & 16.431 (0.026) &  15.813 (0.021) &  15.431 (0.014)  & 15.119 (0.020)\\
 6 & 13.873 (0.025) &  13.351 (0.024) &  13.004 (0.025)  & 12.746 (0.030)\\
 7 & 17.630 (0.028) &  16.295 (0.029) &  15.339 (0.027)  & 14.226 (0.019)\\
 8 & 17.977 (0.031) &  17.508 (0.028) &  17.206 (0.023)  & 16.849 (0.025)\\

\enddata
\label{tab:2}

\end{deluxetable}
\clearpage

%% file: tab3.tex

\clearpage          
\begin{deluxetable}{lccccc}
\tablewidth{450pt}
\tablecaption{Photometric Observations of SN 2002bf}
\tablehead{\colhead{UT Date\tablenotemark{a}}  &
\colhead{Day\tablenotemark{b}}  &
\colhead{$B$ ($\sigma_B$)} &
\colhead{$V$ ($\sigma_V$)} &
\colhead{$R$ ($\sigma_R$)} &
\colhead{$I$ ($\sigma_I$)} }
\startdata
2002-02-22 &-10.07& \nodata       & \nodata       & 17.546(0.452)\tablenotemark{c} & \nodata     \\
2002-02-24 &-8.10& \nodata       & \nodata       & 16.759(0.136)\tablenotemark{c} & \nodata     \\
2002-02-28 &-3.95& \nodata       & \nodata       & 16.303(0.105)\tablenotemark{c} & \nodata     \\
2002-03-09 & 4.95& 16.488(0.116) & 16.185(0.119) & 16.057(0.116) & 16.407(0.123)\\ 
2002-03-12 & 7.95& 16.750(0.048) & 16.285(0.058) & 16.217(0.029) & 16.929(0.101)\\ 
2002-03-15 & 10.93& 16.974(0.044) & 16.553(0.136) & 16.421(0.030) & 17.110(0.096)\\ 
2002-03-19 & 14.92& 17.401(0.060) & 16.691(0.040) & 16.657(0.033) & 17.061(0.145)\\ 
2002-03-25 & 20.90& 18.140(0.102) & 16.998(0.039) & 16.697(0.036) & 16.957(0.076)\\ 
2002-03-29 & 24.99& 18.217(0.571) & 17.019(0.183) & 16.460(0.178) & 16.804(0.048)\\ 
2002-04-01 & 27.84& 18.667(0.054) & 17.481(0.051) & 16.876(0.077) & 16.811(0.072)\\ 
2002-04-04 & 30.95& 18.953(0.034) & 17.760(0.029) & 17.090(0.018) & 17.101(0.090)\\ 
2002-04-07 & 33.89& 19.106(0.064) & 17.856(0.048) & 17.252(0.042) & 17.255(0.172)\\ 
2002-04-13 & 39.86& 19.193(0.072) & 18.042(0.179) & 17.451(0.230) & 17.383(0.414)\\ 
2002-04-22 & 48.86& 19.352(0.056) & 18.083(0.051) & 17.752(0.152) & 17.761(0.115)\\ 
2002-04-30 & 56.81& 19.334(0.079) & 18.400(0.174) & 18.061(0.120) & \nodata\\ 

 \enddata
\tablenotetext{a}{yyyy-mm-dd.}
\tablenotetext{b}{Days since estimated date of maximum $B$ brightness,
2002-03-04.37 UT (HJD 2,452,337.87), taken at the midpoint of the complete 
series of exposures.}
\tablenotetext{c}{Derived from unfiltered image; see text for details.}
\label{tab:3}

\end{deluxetable}

\clearpage


%% file: tab4.tex

\clearpage          
\begin{deluxetable}{llllll}
\tablewidth{450pt}
\tablecaption{Photometric Observations of SN 2003du}
\tablehead{\colhead{UT Date\tablenotemark{a}}  &
\colhead{Day\tablenotemark{b}}  &
\colhead{$B$ ($\sigma_B$)} &
\colhead{$V$ ($\sigma_V$)} &
\colhead{$R$ ($\sigma_R$)} &
\colhead{$I$ ($\sigma_I$)} }
\startdata
2003-05-01           & -4.74&  \nodata      & \nodata       & 13.795(0.085) & 13.938(0.035)\\ 
2003-05-06           &  0.22& 13.508(0.028) & 13.602(0.052) & 13.572(0.035) & 13.920(0.031)\\ 
2003-05-10           &  4.27& 13.613(0.036) & 13.637(0.019) & 13.611(0.015) & 14.058(0.034)\\ 
2003-05-11           &  5.25& 13.650(0.017) & 13.665(0.019) & 13.644(0.016) & 14.105(0.037)\\ 
2003-05-12           &  6.21&  13.698(0.033) & 13.686(0.021) & 13.674(0.021) & 14.164(0.025)\\ 
2003-05-13           &  7.21& 13.752(0.031) & 13.714(0.031) & 13.729(0.024) & 14.220(0.034)\\ 
2003-05-14           &  8.26& 13.852(0.035) & 13.759(0.017) & 13.795(0.033) & 14.280(0.059)\\ 
2003-05-15           &  9.20& 13.886(0.031) & 13.778(0.043) & 13.847(0.017) & 14.336(0.028)\\ 
2003-05-16           &  10.23& 13.981(0.059) & 13.842(0.024) & 13.941(0.023) & 14.427(0.026)\\ 
2003-05-17           &  11.26& 14.105(0.131) & 13.908(0.044) & 14.004(0.036) & 14.524(0.028)\\ 
2003-05-18           &  12.24& 14.141(0.047) & 13.973(0.034) & 14.083(0.020) & 14.572(0.032)\\ 
2003-05-19           &  13.27& 14.246(0.022) & 14.004(0.019) & 14.161(0.015) & 14.638(0.047)\\ 
2003-05-20           &  14.20& 14.327(0.016) & 14.079(0.018) & 14.227(0.023) & 14.681(0.044)\\ 
2003-05-22           &  16.25& 14.570(0.031) & 14.197(0.021) & 14.327(0.028) & 14.710(0.032)\\ 
2003-05-24           &  18.19& 14.764(0.023) & 14.298(0.029) & 14.363(0.026) & 14.671(0.043)\\ 
2003-05-26           &  20.26& 15.018(0.034) & 14.404(0.021) & 14.408(0.020) & 14.632(0.031)\\ 
2003-05-28           &  22.20& 15.207(0.016) & 14.498(0.018) & 14.428(0.012) & 14.578(0.033)\\ 
2003-05-30           &  24.23& 15.404(0.039) & 14.607(0.033) & 14.439(0.026) & 14.529(0.031)\\ 
2003-05-31$^\dagger$ &  25.22& 15.565(0.035) & 14.677(0.014) & 14.437(0.021) & 14.467(0.022)\\ 
2003-06-01           &  26.19& 15.591(0.027) & 14.702(0.028) & 14.475(0.023) & 14.507(0.038)\\ 
2003-06-01$^\dagger$ &  26.20& 15.630(0.032) & 14.722(0.027) & 14.468(0.015) & 14.403(0.022)\\ 
2003-06-04           &  29.21& 15.866(0.018) & 14.868(0.024) & 14.570(0.021) & 14.454(0.032)\\ 
2003-06-07           &  32.15& 16.167(0.090) & 15.054(0.072) & 14.686(0.030) & 14.493(0.065)\\ 
2003-06-10           &  35.17& 16.270(0.025) & 15.206(0.014) & 14.883(0.013) & 14.636(0.037)\\ 
2003-06-13           &  38.19& 16.372(0.055) & 15.388(0.017) & 15.083(0.014) & 14.851(0.038)\\ 
2003-06-16           &  41.13& 16.511(0.031) & 15.526(0.034) & 15.241(0.031) & 15.036(0.047)\\ 
2003-06-19           &  44.10& 16.606(0.015) & 15.639(0.042) & 15.374(0.022) & 15.208(0.051)\\ 
2003-06-22           &  47.11& 16.655(0.023) & 15.748(0.022) & 15.486(0.015) & 15.349(0.039)\\ 
2003-06-25           &  50.09& 16.705(0.025) & 15.841(0.022) & 15.591(0.025) & 15.511(0.039)\\ 
2003-06-26$^\dagger$ &  51.17& 16.772(0.022) & 15.842(0.020) & 15.629(0.037) & 15.495(0.027)\\ 
2003-06-27$^\dagger$ &  52.15& 16.789(0.023) & 15.872(0.019) & 15.664(0.027) & 15.548(0.042)\\ 
2003-06-28           &  53.09& 16.777(0.025) & 15.926(0.025) & 15.697(0.033) & 15.648(0.045)\\ 
2003-07-02           &  57.09& 16.839(0.034) & 16.031(0.017) & 15.824(0.021) & 15.840(0.035)\\ 
2003-07-06           &  61.09& 16.918(0.017) & 16.135(0.016) & 15.953(0.014) & 16.028(0.040)\\ 
2003-07-10           &  65.09& 16.913(0.018) & 16.245(0.017) & 16.088(0.016) & 16.202(0.032)\\ 
2003-07-14           &  69.09& 17.052(0.029) & 16.358(0.014) & 16.192(0.022) & 16.331(0.035)\\ 
2003-07-18           &  73.08& 17.090(0.024) & 16.456(0.030) & 16.322(0.013) & 16.516(0.030)\\ 
2003-07-22           &  77.08& 17.193(0.035) & 16.583(0.019) & 16.461(0.018) & 16.692(0.034)\\ 
2003-08-27$^\dagger$ & 113.06& 17.717(0.014) & 17.367(0.054) & 17.506(0.023) & 17.787(0.038)\\ 
2004-07-15$^\ast$    & 435.94& 22.33(0.03)   & 22.10(0.04)   & 23.44(0.08)   & 21.68(0.07)\\ 
 \enddata
\tablecomments{All photometric observations were made with KAIT,
except those marked with a dagger ($^\dagger$), which
were taken with the 1-m Nickel telescope at Lick Observatory, and those 
marked with an asterisk ($^\ast$), which were taken with {\it HST}'s ACS/HRC.}
\tablenotetext{a}{yyyy-mm-dd.}
\tablenotetext{b}{Days since estimated date of maximum $B$ brightness,
2003-05-06.12 UT (HJD 2,452,765.62), taken at the midpoint of the complete
series of exposures.}

\label{tab:4}
\end{deluxetable}

\clearpage


%% file: tab5.tex
\begin{deluxetable}{lcccclrrrrccccll}
\tabletypesize{\tiny}
\rotate
\tablewidth{650pt}
\tablecaption{Journal of Spectroscopic and Spectropolarimetric Observations}
\tablehead{\colhead{} & 
\colhead{} &
\colhead{} &
\colhead{HJD} &
\colhead{} &
\colhead{Range\tablenotemark{c}}  &
\colhead{Res.\tablenotemark{d}} &
\colhead{P.A.\tablenotemark{e}} &
\colhead{Par. P.A.\tablenotemark{f}} &
\colhead{} & 
\colhead{} &
\colhead{See.\tablenotemark{i}} &
\colhead{Slit} &
\colhead{Exp.} &
\colhead{} \\
\colhead{UT Date} &
\colhead{Object} &
\colhead{Day\tablenotemark{a}} &
\colhead{$-$2,400,000} &
\colhead{Tel.\tablenotemark{b}} &
\colhead{(\AA)} &
\colhead{(\AA)} &
\colhead{(deg)} &
\colhead{(deg)} &
\colhead{Air.\tablenotemark{g}} &
\colhead{Flux Std.\tablenotemark{h}} &
\colhead{(arcsec)} &
\colhead{(arcsec)} &
\colhead{(s)} & 
\colhead{Observers\tablenotemark{j}} }
\startdata

1997-12-20.27 & SN 1997dt & 20.77: & 50802.77 & KIIp & 4310--6820  & 5  & 98  &
77--78    & 1.34--1.57 & HD84 & 1.8 & 1.0 & $4 \times 700$                 &
AF, AB \\
 \\

2002-03-07.41 & SN 2002bf & 3.03   & 52340.91 & KIp  & 3900--8800  & 13 &  98 &  172--178 & 1.24--1.25 & BD26 & 1.5 & 1.5 &  $3 \times 300 + 200$          & AF, DL, EM \\
2002-03-11.44 & SN 2002bf & 7.06   & 52344.94 & L3   & 3300--5400  & 5  &  98 &   98      & 1.26       & F34  & 2.0 & 2.0 & 1078                           & RC, WL        \\
2002-03-11.44 & SN 2002bf & 7.06   & 52344.94 & L3   & 5200--10400 & 11 &  98 &   98      & 1.27       & HD19 & 2.0 & 2.0 & 1161                           & RC, WL        \\
				   
\\

2003-05-24.46 & SN 2003du & 18.34  & 52783.96 & KIp  & 3220--5710  & 9  & 150 & 122--153  & 1.35--1.58 & F34  & 1.5 & 1.5 &  $5 \times 900 + 3 \times 600$ & RC, DL \\
2003-05-24.46 & SN 2003du & 18.34  & 52783.96 & KIp  & 5700--9420  & 9  & 150 & 122--153  & 1.35--1.58 & BD17 & 1.5 & 1.5 &  $5 \times 900 + 3 \times 600$ & RC, DL \\
2003-05-30.40 & SN 2003du & 24.28  & 52789.90 & L3   & 3100--5400  & 5  & 102 & 97--101   & 1.29--1.32 & F34  & 1.5 & 2.0 &  $600 + 500$                   & RF, LD, DH \\
2003-05-30.40 & SN 2003du & 24.28  & 52789.90 & L3   & 5200--10400 & 11 & 102 & 97--101   & 1.29--1.32 & HD84 & 1.5 & 2.0 &  $600 + 500$                   & RF, LD, DH \\
2003-06-07.39 & SN 2003du & 32.27  & 52797.89 & L3   & 3300--5400  & 5  &  96 & 96        & 1.34       & F34  & 2.0 & 3.0 &  700                           & RF, KS    \\
2003-06-07.39 & SN 2003du & 32.27  & 52797.89 & L3   & 5200--10400 & 11 &  96 & 96        & 1.34       & BD17 & 2.0 & 3.0 &  700                           & RF, KS    \\
2003-07-06.34 & SN 2003du & 61.22  & 52826.84 & L3   & 3286--5452  & 5  &  88 & 87        & 1.46       & F34  & 2.0 & 2.0 &  600                           & RF, LD, MG \\
2003-07-06.34 & SN 2003du & 61.22  & 52826.84 & L3   & 5150--10200 & 11 &  88 & 87        & 1.46       & BD26 & 2.0 & 2.0 &  600                           & RF, LD, MG \\
2003-07-27.20 & SN 2003du & 82.08  & 52847.70 & L3   & 3300--5400  & 5  & 120 & 117       & 1.18       & BD28 & 1.0 & 2.0 &  900                           & AF, RF, LD \\
2003-07-27.20 & SN 2003du & 82.08  & 52847.70 & L3   & 5200-10400  & 11 & 120 & 117       & 1.18       & BD17 & 1.0 & 2.0 &  900                           & AF, RF, LD \\

\\
2004-08-24.43 & SN 2004dt & 3.93   & 53241.93 & L3p  & 4500--9800  & 18 & 150 & 145--156  & 1.34--1.48 & BD17 & 4   & 4.0 & $4 \times 900$                 & AF, RF, MG\\
\enddata
\tablenotetext{a}{Days since estimated dates of maximum $B$ brightness, for
  SN~1997dt (1997-11-29; HJD 2,450,782), SN~2002bf (2002-03-04.37; HJD
  2,452,337.87), SN~2003du (2003-05-06.12; HJD 2,452,765.62), and SN~2004dt
  (2004-08-20; HJD 2,453,238).  The epoch of maximum is particularly uncertain
  for SN~1997dt, and thus the phase our our single observation is not well
  known; this uncertainty is indicated by a colon.}
\tablenotetext{b}{L3(p) = Lick 3-m/Kast Double Spectrograph \citep{Miller93}; ``p'' denotes polarimeter attached); KI(II) = Keck-I (II) 10-m/Low-Resolution Imaging Spectrometer \citep[LRIS; ][]{Oke95}; ``p'' denotes polarimeter attached.}
\tablenotetext{c}{Wavelength range of the calibrated flux spectrum.  In some cases, the ends are very noisy and are not shown in the figures.}
\tablenotetext{d}{Approximate spectral resolution derived from night-sky lines.}
\tablenotetext{e}{Position angle of the spectrograph slit.}
\tablenotetext{f}{Parallactic angle \citep{Filippenko82} at midpoint of observation, or PA range for each set of observations.}
\tablenotetext{g}{Airmass at midpoint of observation, or airmass range for each set of observations.  }
\tablenotetext{h}{The standard stars are as follows: F34 = Feige~34, BD28 = BD+28$^\circ$4211---\citet{Stone77}, \citet{Massey90}; HD84 = HD~84937, BD26 = BD+26$^\circ$2606, BD17 = BD+17$^\circ$4711---\citet{Oke83}.}
\tablenotetext{i}{Average value of the full width at half maximum of the spatial profile for each set of observations, rounded to the nearest 0\farcs5.}
\tablenotetext{j}{AB = Aaron Barth; RC = Ryan Chornock; LD = Louis-Benoit Desroches; RF = Ryan Foley; AF = Alex Filippenko; MG = Mohan Ganeshalingam; DH = Deborah Hutchings; WL = Weidong Li; EM = Edward Moran; KS = Karin Sandstrom.}

\label{tab:5}

\end{deluxetable}
\clearpage

%% file: tab6.tex



\clearpage

\begin{deluxetable}{lccclcc}
\tablewidth{0 pt}
\tablecaption{Data on Blueshifts of \ion{Si}{2} $\lambda 6355$}
\tablehead{
\colhead{} &
\colhead{} &
\colhead{Velocity} &
\colhead{} &
\colhead{} &
\colhead{} &
\colhead{Velocity}\\ 
\colhead{Supernova} &
\colhead{Epoch} &
\colhead{\kms} &
\colhead{} &
\colhead{Supernova} &
\colhead{Epoch} &
\colhead{\kms}} 

\startdata

SN 1984A     &  $ -7 $  &   $ 17,040 $  &  &              &  $ +2 $  &   $ 10,940 $ \\
             &  $ -6 $  &   $ 16,840 $  &  &              &  $ +7 $  &   $ 10,400 $ \\
             &  $ -5 $  &   $ 16,370 $  &  &              &  $ +10$  &   $ 10,200 $ \\
             &  $ -3 $  &   $ 15,750 $  &  &              &  $ +14$  &   $ 10,180 $ \\
             &  $ +8 $  &   $ 13,320 $  &  &              &  $ +21$  &   $ 10,190 $ \\
             &  $ +17$  &   $ 12,050 $  &  &              &  $ +28$  &   $ 9,700  $ \\
             &  $ +19$  &   $ 12,290 $  &  & SN 1997bp    &  $ +9 $  &   $ 14,100 $ \\
SN 1991T     &  $ +7 $  &   $  9,730  $  &  & SN 2002bf    &  $ +3 $  &   $ 15,400 $ \\
             &  $ +12$  &   $  9,580  $  &  &              &  $ +7 $  &   $ 15,100 $ \\
             &  $ +13$  &   $  9,580  $  &  & SN 2002bo    &  $ -12$  &   $ 17,310 $ \\
             &  $ +18$  &   $  9,580  $  &  &              &  $ -1 $  &   $ 14,000 $ \\
SN 1991bg    &  $ -2 $  &   $ 10,070 $  &  &              &  $ +16$  &   $ 10,000 $ \\
             &  $ -1 $  &   $ 9,800   $  &  &              &  $ +28$  &   $ 9,600  $ \\
             &  $ +17$  &   $ 7,880   $  &  & SN 2003du    &  $ +18$  &   $ 9,800  $ \\
             &  $ +25$  &   $ 6,920   $  &  &              &  $ +24$  &   $ 9,400  $ \\
SN 1994D     &  $ -12$  &   $ 15,580 $  &  &              &  $ +32$  &   $ 9,500  $ \\
             &  $ -9 $  &   $ 12,530 $  &  & SN 2004dt    &  $ -7$   &   $ 17,200 $ \\
             &  $ -6 $  &   $ 11,100 $  &  &              &  $ +4 $  &   $ 13,500 $ \\
             &  $ -3 $  &   $ 11,160 $  &  &              &          &              \\

\enddata 

\tablecomments{Velocities of absorption minima of the \ion{Si}{2} $\lambda
6355$ line, derived using the relativistic Doppler formula.  All velocities
were measured directly from spectra in our database, except for SN~1984A, for
which the \ion{Si}{2} minima wavelengths reported by \citet{Barbon89} were
used, and the day $-7$ data for SN~2004dt, for which the wavelength of minimum
reported by \citet{Wang05} was used.  All spectra were first deredshifted
according to the values reported in NED.  The adopted recession velocities and
dates of $B$ maximum are as follows: SN~1984A --- $v_{\rm rec} = 261$ \kms,
$B_{\rm max} $ on 1984 Jan. 17
\citep{Barbon89};
SN~1991T --- $v_{\rm rec} = 1736$ \kms, $B_{\rm max} $ on 1991 Apr. 28 \citep{Phillips92};
SN~1991bg --- $v_{\rm rec} =1060 $ \kms, $B_{\rm max} $ on 1991 Dec. 15 \citep{Leibundgut93};
SN~1994D --- $v_{\rm rec} = 850 $ \kms, $B_{\rm max} $ on 1994 Mar. 20 \citep{Richmond95};
SN~1997bp --- $v_{\rm rec} = 2492 $ \kms, $B_{\rm max} $ on 1997 Apr. 7 \citep{Riess98a};
SN~2002bf --- $v_{\rm rec} = 7254$ \kms, $B_{\rm max} $ on 2002 Mar. 4 (this work);
SN~2002bo --- $v_{\rm rec} = 1271$ \kms, $B_{\rm max} $ on 2002 Mar. 23 \citep{Benetti04};
SN~2003du --- $v_{\rm rec} = 1194$ \kms, $B_{\rm max} $ on 2003 May 6 (this work);
SN~2004dt --- $v_{\rm rec} = 5915$ \kms, $B_{\rm max} $ on 2004 Aug. 20 (W. Li,
private communication).}

\label{tab:6}

\end{deluxetable}
\clearpage


%% file: tab7.tex


\clearpage

\begin{deluxetable}{lcccccccc}
\tablewidth{0 pt}
\tablecaption{MLCS2k2 Parameters of SN~2002bf and SN~2003du}
\tabletypesize{\scriptsize}
\tablehead{
\colhead{} &
\colhead{$\mu_0$($\sigma$)\tablenotemark{a}} &
\colhead{$D$($\sigma$)\tablenotemark{a}} &
\colhead{}  &
\colhead{$M_V$($\sigma$)\tablenotemark{c}} &
\colhead{$A_V^0$($\sigma$)\tablenotemark{d}} &
\colhead{} &
\colhead{$B_{\rm max}$\tablenotemark{f}} &
\colhead{$V$(at $B_{\rm max}$)\tablenotemark{f}} \\
\colhead{SN}    &
\colhead{(mag)} &
\colhead{(Mpc)} &
\colhead{$\Delta$($\sigma$)\tablenotemark{b}} &
\colhead{(mag)} &
\colhead{(mag)} &
\colhead{$t_0$($\sigma$)\tablenotemark{e}} &
\colhead{(mag)} &
\colhead{(mag)} }

\startdata

SN 2002bf & 35.47(0.13) & 124(8) & $-$0.11(0.07) & $-$19.58(0.05) & 0.26(0.13) & 52337.87(0.50) & 16.22 & 16.18 \\
SN 2003du & 33.18(0.05) & 43(1)  & $-$0.25(0.03) & $-$19.67(0.02) & 0.04(0.03) & 52765.62(0.50) & 13.56 & 13.58 \\

\enddata

\tablenotetext{a}{Statistical uncertainty only.  }
\tablenotetext{b}{MLCS2k2 light-curve shape parameter, with lower
values of $\Delta$ corresponding to brighter, more slowly declining
light curves, relative to the ``fiducial'' $\Delta = 0$ template, which
has $\Delta m_{15}(B) = 1.07$ mag.}
\tablenotetext{c}{$V$-band absolute magnitude at the time of
peak $B$-band brightness, on the $H_0 = 65\ \kmsmpc$ distance scale
used here. The fiducial $\Delta = 0$ template has the following peak
magnitudes at $B$ maximum light: $M_B^0 = -19.56$ mag, $M_V^0 = -19.50$ mag,
$M_R^0 = -19.50$ mag, and $M_I^0 = -19.20$ mag.}
\tablenotetext{d}{Host-galaxy extinction, assuming an $R_V = 3.1$ extinction law.}
\tablenotetext{e}{Heliocentric Julian date minus 2,400,000 for the time of $B$-band
peak.  The time of $V$-band maximum for both SNe is about 1 day after the time
of $B$-band peak.}
\tablenotetext{f}{Observer frame, no $k$-correction, no extinction
correction.  $V$-band peak at $t = t(V_{\rm max})$ is 0.01 mag brighter than
$V$(at $t_0$) for both objects.}
\label{tab:7}

\end{deluxetable}
\clearpage
